\pdfoutput=1
\documentclass[fleqn,usenatbib]{mnras}

\usepackage{type1cm}
\usepackage[T1]{fontenc}
\usepackage{newtxtext,newtxmath}
\usepackage{graphicx}
\usepackage{amsmath}
\usepackage{booktabs}
\usepackage{subcaption}
\usepackage{multirow}
\usepackage{needspace}
\usepackage{natbib}
\usepackage{orcidlink}
\usepackage{float}
\usepackage{enumitem}
\usepackage{hyperref}
\usepackage{xcolor}
\title[Role of Inner Halo Spin in Shaping DM Bars]
{The Role of Inner Halo Angular Momentum (Spin) in Shaping Dark Matter Bars in Milky Way Analogs}

\author[Shouvik Ghosh and S.K. Kataria]{
Shouvik Ghosh\,\orcidlink{0009-0008-7578-1952}$^{1}$\thanks{E-mail: shouvikghosh.nit@gmail.com},
Sandeep Kumar Kataria\,\orcidlink{0000-0003-3657-0200}$^{2}$\thanks{E-mail: skkataria.iit@gmail.com},
\\
$^{1}$Department of Physics, Sardar Vallabhbhai National Institute of Technology, Surat 395007, India\\
$^{2}$Department of Space, Planetary \& Astronomical Sciences and Engineering, Indian Institute of Technology, Kanpur 208016, India
}

\pubyear{2026}

\begin{document}

\maketitle


\begin{abstract}
Studies of galactic bars have primarily focused on stellar bars, since they can be directly observed through ultraviolet to infrared wavebands. Cosmological as well as idealised simulations  reveal that the dark matter (DM) halos interact with baryonic matter, primarily the stellar bars, dynamically by means of the exchange of angular momentum. In these simulations, the spherical DM halo dynamically responds to interaction with the stellar bar by reshaping its orbital structure in the proximity of the stellar bar, forming a bar-like configuration, called the Dark Matter (DM) bar. Using N-body simulations of Milky Way analogs we discuss the role of inner halo angular momentum, measured as halo spin parameter $(\lambda)$ of the dark matter halo, on formation and evolutionary characteristics of the DM bars. Our systematic study involves halos with initial spin configurations ranging from $\lambda = 0-0.1$. The result conveys that DM bar formation and its characteristics are extensively dependent on the initial spin parameter $(\lambda)$ of the DM halo. We demonstrate that the strength of the dark matter bar gradually increases with an increase in halo spin in long-term evolution, with a significant impact of stellar bar buckling on dark matter bar strength. The evolutionary characteristics of the DM bar are strongly influenced by the initial spin of the host halo.

\end{abstract}



\section{Introduction}
The baryonic matter of a galaxy is embedded within the dark matter halo, together forming large-scale structures such as galaxies. In the case of disk galaxies, these two components form a disk-halo system and co-evolve dynamically over time. \\
\indent Observations reveal that stellar bars are commonly found in the universe. According to the observed data from optical and near infrared (NIR) surveys, nearly two-thirds of the nearby disk galaxies exhibit stellar bars \citep{Eskridge2000,2008ApJ...675.1194B,2009A&A...495..491A,2010ApJ...714L.260N,2015ApJS..217...32B,2018MNRAS.474.5372E}. Their presence plays a crucial role in shaping the secular evolution of the host galaxy by redistributing the stars, gas and angular momentum within them \citep{2002MNRAS.330...35A,2003MNRAS.341.1179A,2005ApJ...632..217S,2006ApJ...645..209D,Athanassoula_2013,2013A&A...553A.102D,Kataria.Das.2018,Kataria.Das.2019,Kataria_etal_2020,Kumar.Kataria.2022,2014RvMP...86....1S,2023MNRAS.518.1002R, Kataria.Vivek.2024, Tahmasebzadeh_2024,Kataria.2024}. \\
\indent Bars in disk galaxies originated through dynamical instability, first identified in N-Body simulations \citep{1969JCoPh...4..306H,1970ApJ...161..903M}. The stellar bars resonantly interact with the dark matter halo, causing the bar to slow down over time \citep{1981A&A....96..164C,1992ApJ...400...80H,2000ApJ...543..704D,2002ApJ...569L..83A,2002MNRAS.330...35A,2003MNRAS.341.1179A,2003MNRAS.346..251O,2005MNRAS.363..991H,2006ApJ...637..214M,2007MNRAS.375..460W,2009ApJ...697..293D,KumarA.et.al.2022,Chiba2024-bh,kataria_effects_2022,Kataria2024}. The mode of this interaction is through the transfer of angular momentum between the bar and the dark matter halo \citep{Ansar.et.al.2023}.\\
\indent The stellar bars tend to undergo an event of buckling instability during their evolution \citep{1981A&A....96..164C,1991A&A...252...75P,1991Natur.352..411R,2019A&A...624A..37L}. The bars thicken substantially, become increasingly centrally concentrated, and develop a characteristic peanut/boxy morphology when viewed edge-on \citep{1990A&A...233...82C,1991A&A...252...75P,1991Natur.352..411R,1998MNRAS.300...49B,2002MNRAS.337..578P}. These peanut/boxy morphologies are similar to bulge shapes observed in edge-on galaxies \citep{1986AJ.....91...65J,1987MNRAS.229..691S,1999AJ....118..126B,1999A&A...345L..47M}.\\
\indent The vertical thickening in bars arises from three processes:  vertical buckling\citep{1991Natur.352..411R},  vertical heating of 2:1 resonance \citep{1991A&A...252...75P,10.1093/mnras/stt1972}  and gradual trapping of 2:1 orbits resonances. The peanut/boxy morphology has also been shown to depend on bar angular momentum \citep{https://doi.org/10.48550/arxiv.2409.03746}. \\
\indent The instability may occur multiple times over the lifetime of a bar, with the second episode typically lasting longer \citep{MartinezValpuesta2006}. Furthermore, the buckling event tends to weaken the bar.\\
\indent Dark matter(DM) bars are the integral structural feature of the dark matter halo and are embedded within the parent DM halo \citep{Ash2024}. The DM bar is an intrinsic property of the halo and is formed as a dynamic response of the DM halo, when it interacts and co-evolves with its baryonic counterparts, particularly the stellar bar. The DM bar transforms the orbital structure of the halo in the vicinity of the bar.
\\
\indent Dark matter (DM) halos play a pivotal role, providing the structural framework for galaxy formation and evolution  \citep{2005Natur.435..629S,2015MNRAS.446..521S}. Recent studies reveal that a DM halo spin has a profound effect on stellar bar evolution and also affects the angular momentum redistribution in disk-halo system \citep{Long2014,Collier2018,kataria_effects_2022,Kataria2024,Chen_2025, Kataria2025}.\\
\indent The key parameter that characterises the dynamical state of the halo is the dimensionless spin parameter, $\lambda$. It is defined as:
\begin{equation}
\lambda = \frac{J}{\sqrt{2GM R_{\mathrm{vir}}}},
\end{equation}
where $J$ is the specific angular momentum, $M$ is the virial mass, and $R_{\mathrm{vir}}$ is the virial radius. In the standard $\Lambda$CDM cosmological model, $\lambda$ follows a lognormal distribution, with a typical value around 0.035 \citep{2001ApJ...555..240B}. This parameter quantifies the rotation of the DM halo and is known to significantly affect disk morphology by regulating the exchange of angular momentum between the disk and the halo.\\
\indent The structural integrity and density of the inner region of the parent DM halo change in response to the dynamical evolution with the stellar bars. The stellar bars have been observed to trap DM in `bar' like orbits and consequently form DM bars \citep{2005NYASA1045..168A,2006ApJ...648..807B,2006ApJ...644..687C,2007MNRAS.377.1569A,2013MNRAS.429.1949A,2021ApJ...915...23C,2024Galax..12...27M}.\\
\indent The dark matter bar is shorter than its stellar counterpart \citep{2007MNRAS.377.1569A}. The DM bars are also significantly weaker in amplitude than the stellar bar, although the stellar and DM bar co-evolve synchronously and exhibit coupled dynamical behaviour. The halos with an initial spin parameter $\lambda$, appear to have a greater acceleration on the bar instability than non-rotating halos \citep{Long2014,Saha2013}.
\\ 
\indent Previous studies demonstrated that with increasing the spin parameter of the halo, the triggering of bar formation happens earlier in time than low-rotating or non-rotating halos \citep{Saha2013,collier_dark_2019,kataria_effects_2022}. The retrograde halo displays late triggering of bars. The DM bar exhibits similar behaviour to the stellar bar in both prograde and retrograde halos across all the models, as it forms as a dynamic response to the co-evolution of halo and baryonic matter. Over time the barred galaxy has been shown to lose angular momentum to host halos \citep{Athanassoula2003,MartinezValpuesta2006,Berentzen2007,VillaVargas2009,VillaVargas2010} and performs it mainly by resonance interaction \citep{Athanassoula2003,MartinezValpuesta2006,Dubinski2009, Chiba2024-bh}. The angular momentum transfer between halo and disk is affected by different initial halo spins, which subsequently influences not only the bar formation in the stellar disk but also the DM bar.\\
\indent DM bars remain relatively less explored in the context of galaxy dynamics compared to stellar bars. Studying the evolution of the DM bar allows us to investigate previously unexplored dynamical processes and helps to bridge the gaps in our existing knowledge of galaxy dynamics; mainly about the dark matter distribution in the galaxy. 
In this study, we try to focus on the development of the DM bar in halos with varying spin parameter, both in prograde and retrograde rotation with respect to the stellar disk.\\
\indent Here, we try to investigate (1) How does the spin parameter affect the onset of the DM bar formation, its strength, and morphological evolution? (2) As we have seen, buckling affects the strength of the stellar bar, does it also affect the DM bar and how it evolves after the event? (3) How does the initial spin parameter of the parent DM halo influence the evolutionary characteristics of the DM bar, such as its pattern speed and angular evolution with the stellar bar over time? 

We study the evolution of the DM bar in a disk halo system as they co-evolve through the transfer of angular momentum. As the disk-halo co-evolves, we explore how changes in the spin parameter, $\lambda$, affect the formation and evolution of the DM bar throughout its evolutionary phase.
\\
\indent In this paper, we study a galaxy model with increasing halo spin $(\lambda=0-0.1 )$ in prograde direction. We also studied a model with $(\lambda=0.1)$ in the retrograde direction with respect to the disk. In these models, the spin of halos is controlled by varying inner halo angular momentum only. We discuss the model setup in section \ref{model} and the results in section \ref{results}. Discussion is done in section \ref{discusion} and we summarise and conclude our key results in section \ref{conclusion}.


\section{Model Setup}
\label{model}
The detailed description of the model setup, including the numerical setup and the initial condition, are provided in \citep{kataria_effects_2022}.
All the initial galaxy models were constructed using the \textsc{GalIC} code.  Each model comprises $10^{6}$ dark matter halo particles and $10^{6}$ stellar disk particles. We performed convergence tests, by doubling the number of particles, which confirmed the numerical stability of the results. The total mass of each galaxy is $6.38 \times 10^{11}\,M_{\odot}$, and the circular velocity near the solar radius is approximately $250~\mathrm{km\,s^{-1}}$. The stellar disks remain locally stable with a Toomre parameter $Q > 1$, which varies radially as shown in the equation:

\begin{equation}
    Q(R) = \frac{\sigma(R)\,\kappa(R)}
    {3.36\,G\,\Sigma(R)}.
\end{equation}

where $\sigma(R)$, $\kappa(R)$, and $\Sigma(R)$ denote the radial velocity dispersion, epicyclic frequency, and surface mass density, respectively. 

The dark matter halo follows a spherically symmetric Hernquist density profile, as shown in the equation:

\begin{equation}
    \rho_{h}(r) = \frac{M_{halo}\,a}{2\pi\,r\,(r + a)^{3}},
\end{equation}

where ${M_{halo}}={M_{200}}$\citep{2005MNRAS.361..776S}. ${M_{200}}$ and ${R_{200}}$ are the virial mass and virial radius for the NFW halo, respectively. $M_{200} = 5.68 \times 10^{11}\,M_{\odot}$ and $R_{200} = 140~\mathrm{kpc}$. The stellar disk has an exponential radial and $\mathrm{sech}^{2}$ vertical density distribution, characterised by scale lengths of $R_{d} = 2.9~\mathrm{kpc}$ and $z_{0} = 0.58~\mathrm{kpc}$, contributing approximately $10\%$ of the total galactic mass.

To study the effect of halo spin, five galaxy models were constructed with increasing spin parameters. The variation in spin was achieved by reversing the direction of retrograde orbits within the central $30~\mathrm{kpc}$, which results in enhancing the prograde orbital fraction. We follow the probabilistic spin-reversal scheme of \citep{Kataria2024}. Instead of reversing a fixed fraction of retrograde particles, which introduces an artificial discontinuity in the angular momentum distribution around $J_z=0$, the reversal is performed probabilistically using a smooth angular-momentum-dependent function. This procedure generates the desired halo spin while maintaining a smooth $J_z$ distribution and the equilibrium properties of the halo. We have verified the equilibrium of dark matter halo for all of the models in \hyperref[app:appendix]{Appendix}. This modification preserves the equilibrium configuration of the halo, in accordance with Jeans’ theorem. For completeness, a counter-rotating configuration with negative halo spin was also constructed for comparative analysis.

\indent To assess the numerical stability of the simulations, we monitored the virial ratio and the total angular momentum throughout the full simulation duration ($0 \leq t \leq 9.78$\, Gyr). The virial ratio remains close to unity, indicating that the galaxy remains in dynamical equilibrium, while the total angular momentum is conserved to within numerical precision. Both the virial ratio and the total angular momentum remain stable, indicating that no fast-growing numerical instability develops in the core.

 We denoted our models with spin parameter $\lambda = 0, 0.025, 0.050, 0.075, 0.1$ in prograde  and $0.1$ in retrograde as $ S000,S025,S050,S075,S100,SM100$ respectively, consistent with the nomenclature of \citep{kataria_effects_2022}. Our simulations were run for approximately 9.78 Gyr, providing a comprehensive view of long-term evolutionary processes.


\section{Results}
\label{results}
The evolutionary pathways of our models differ from one another and are governed by the initial halo spin parameter. Following the initial formation of asymmetric disk structures, the system evolves under self-gravity, leading to the development of non-axisymmetric bar structures in both the stellar and dark matter components. We analysed the stellar bar within the range of $R\leq 20$ kpc, while the dark matter bar is studied within $R\leq 10$ kpc. The vertical extent of bars was limited to $|z| \leq 3$ kpc. The limits were chosen based on direct inspection of the surface density maps of the S100 model, which exhibits the most prominent bar morphology in both the stellar and dark matter components. The isodensity contours clearly show that the elongated bar structure remains well within  $R \leq 20$\, kpc for the stellar bar as shown in \citep{kataria_effects_2022} and 
$R \leq 10$\,kpc for the dark matter bar throughout the full simulation duration ($0 \leq t \leq 9.78$\, Gyr). The measurements are performed using the mass-weighted method, ensuring that the signal is dominated by the coherent inner bar region, and their characteristic features are insensitive to the exact choice of outer radial limit.
 


\begin{figure}
    \centering
    \includegraphics[width=0.48\textwidth]{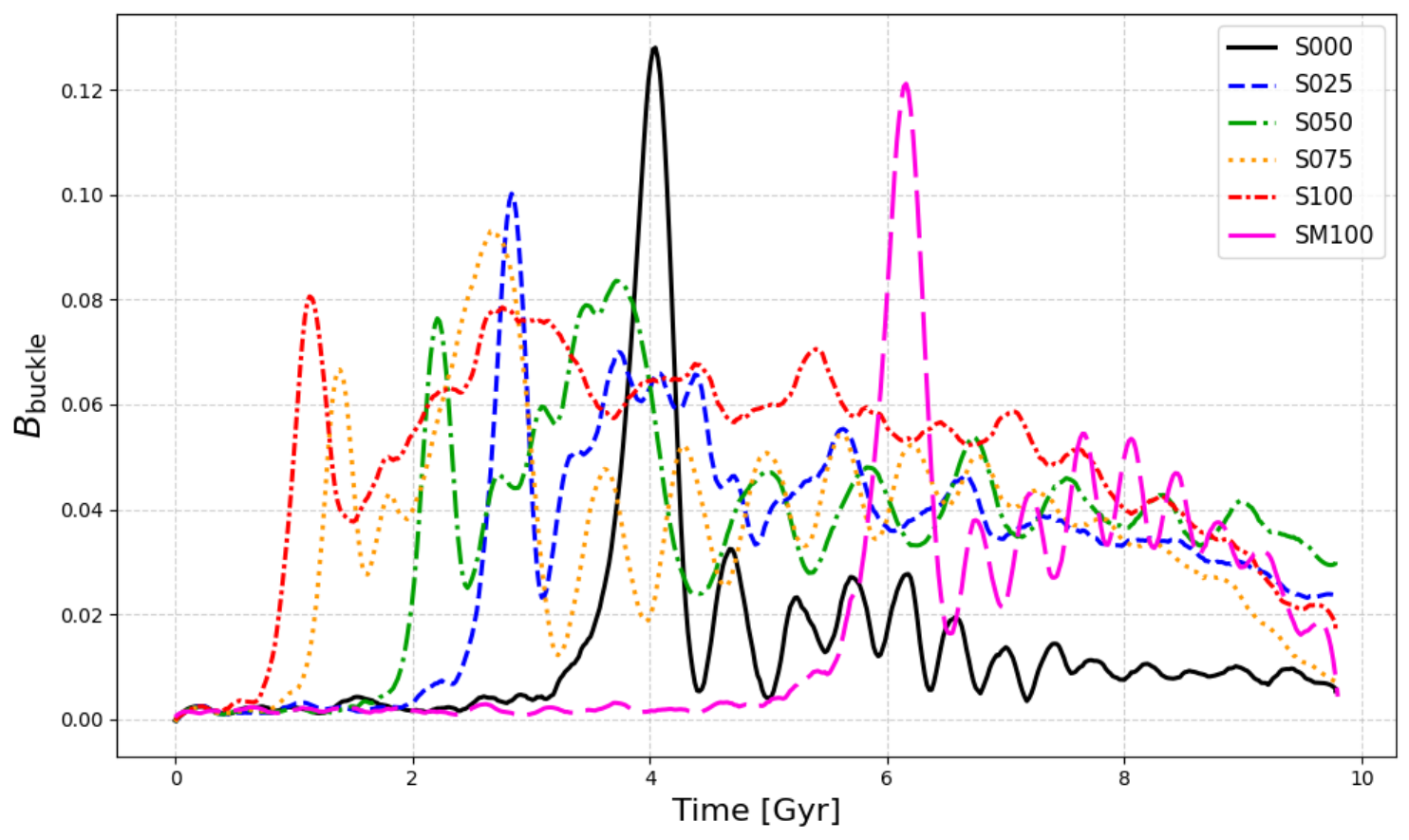}
    
    \caption{Stellar buckling strength over time for increasing halo spin parameter.}
    \label{fig: Buckling Strength}
\end{figure}
\begin{figure}
    \centering
    \includegraphics[width=0.48\textwidth]{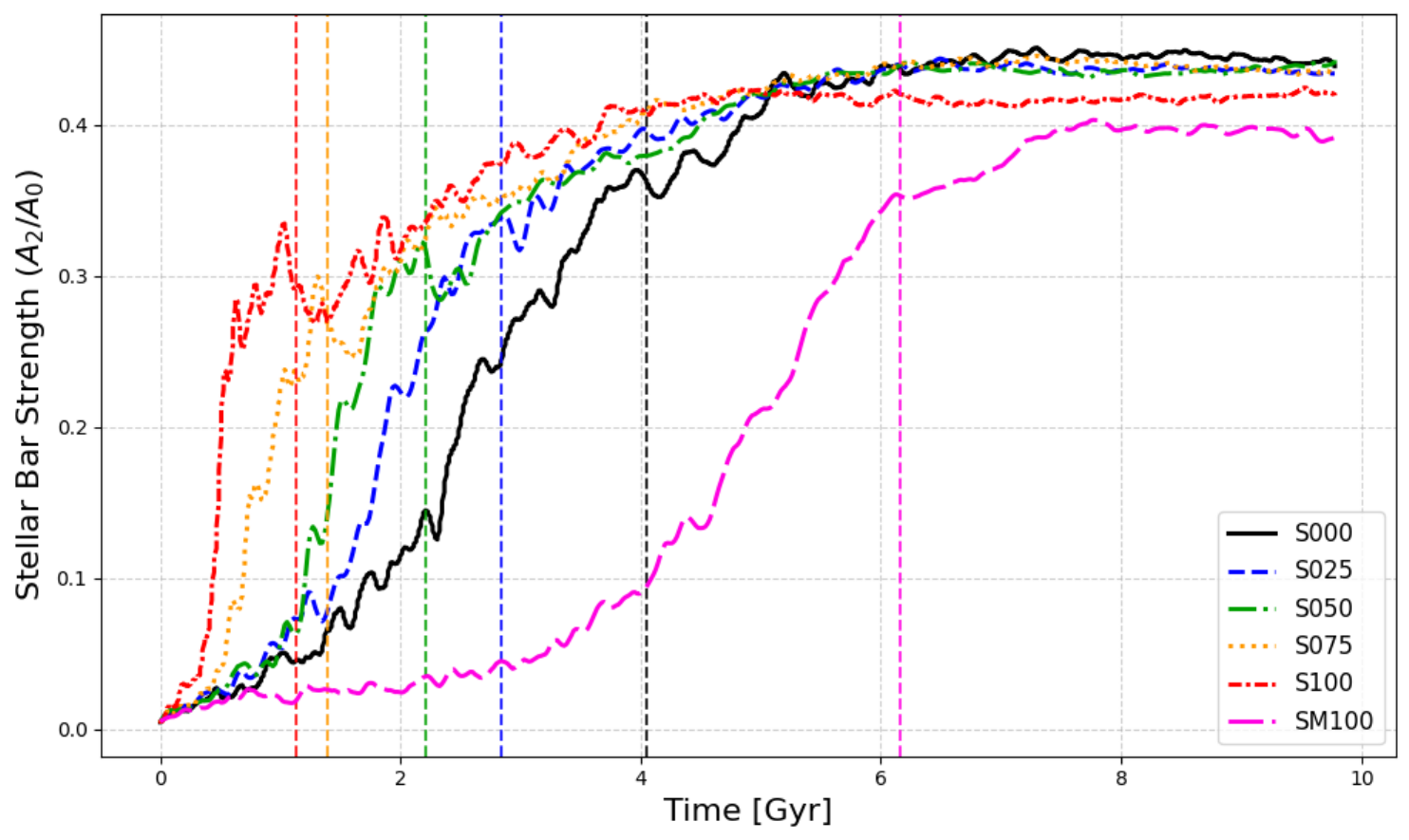} 
    \caption{Evolution of stellar bar strength $A_2/A_0$ over time for increasing halo spin parameter. The dashed vertical lines represents the epoch at which the first buckling occurred.}
    \label{fig stellar bar strength}
\end{figure}
\begin{figure}
    \centering
    \includegraphics[width=0.48\textwidth]{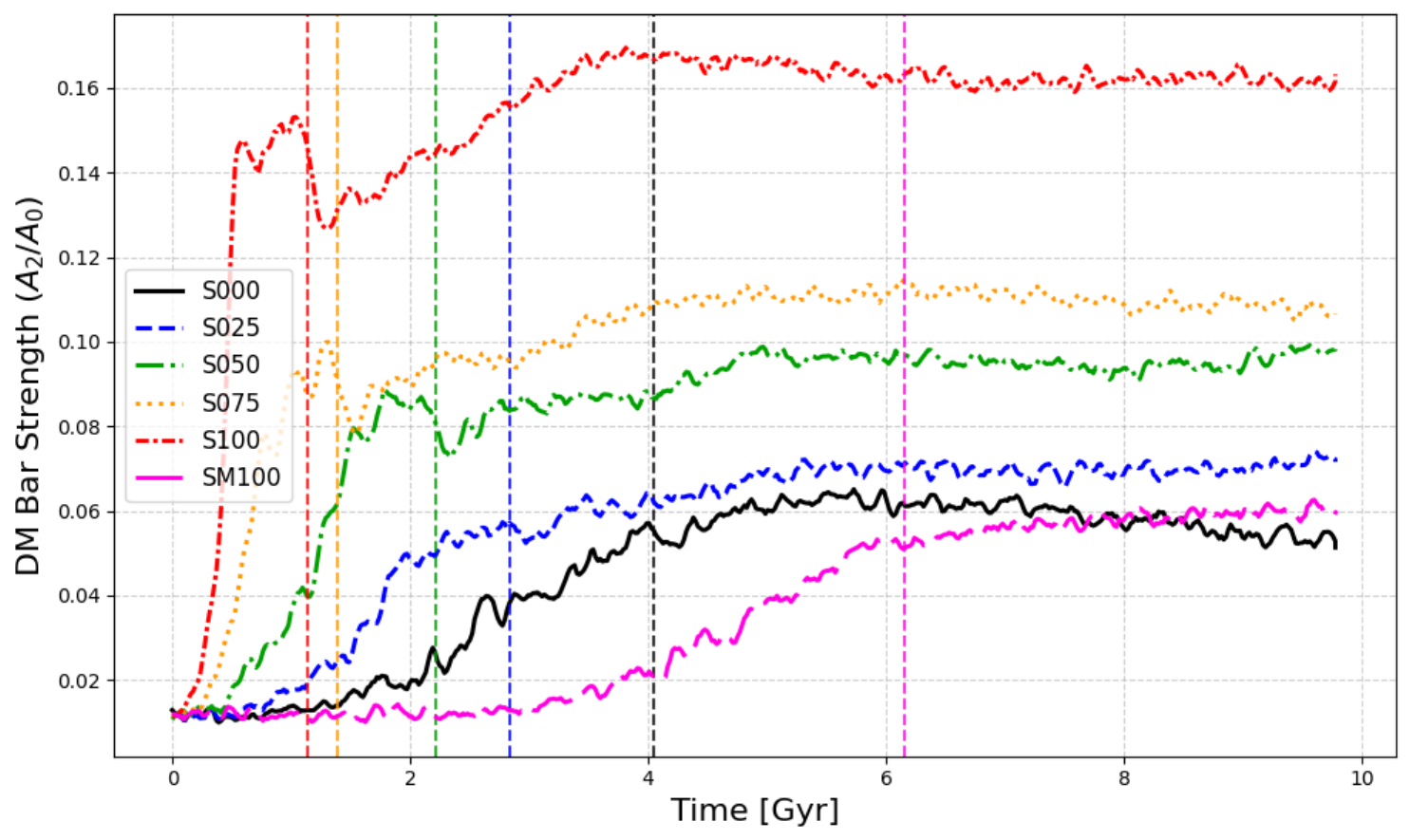} 
    \caption{Evolution of dark matter bar strength $A_2/A_0$ over time for increasing halo spin parameter. The dashed vertical lines represents the epoch at which the first buckling occurred.}
    \label{fig dark matter bar strength}
\end{figure}

\begin{figure}
    \centering
    \includegraphics[width=0.48\textwidth]{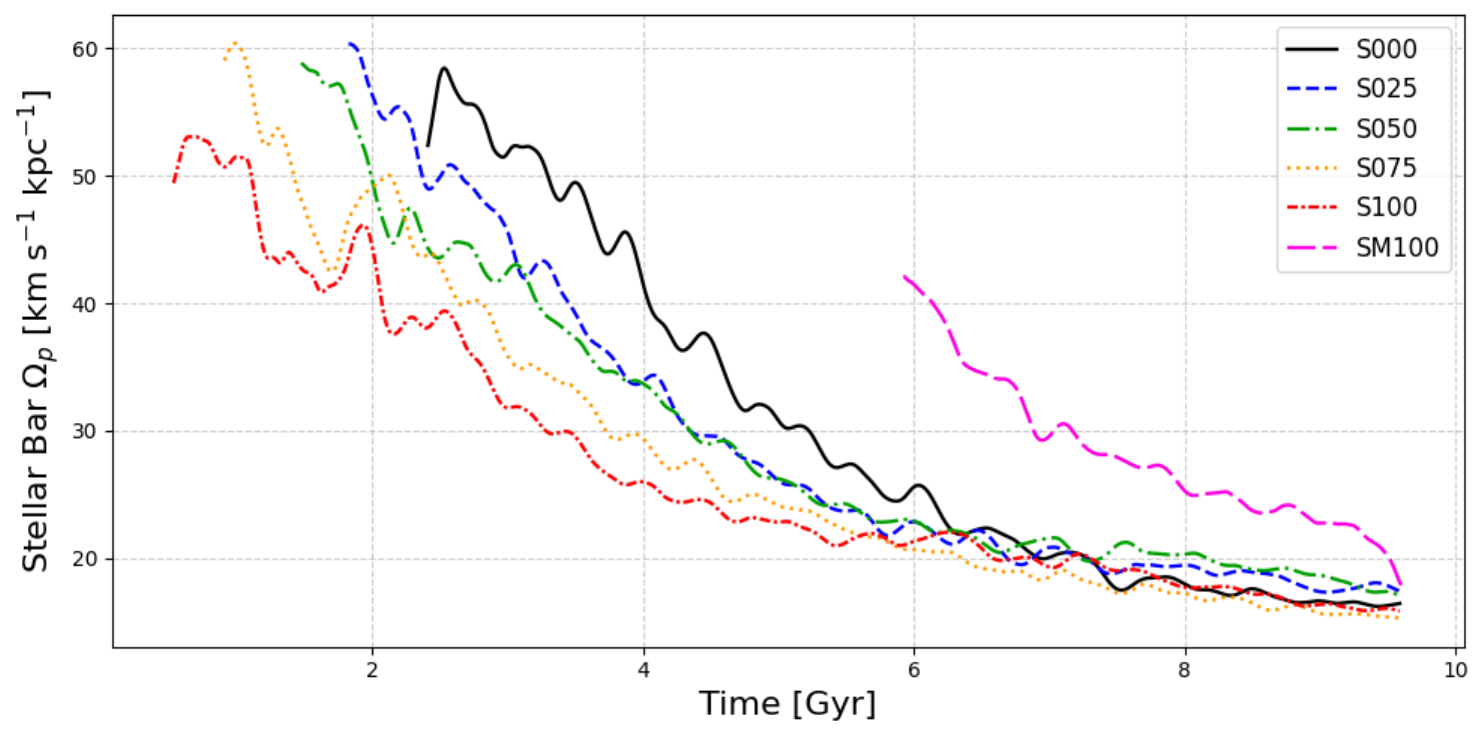} 
    \caption{Evolution of stellar bar pattern speed over time for varying halo
spin parameter.}
    \label{fig stellar pattern speed}
\end{figure}

\begin{figure}
    \centering
    \includegraphics[width=0.48\textwidth]{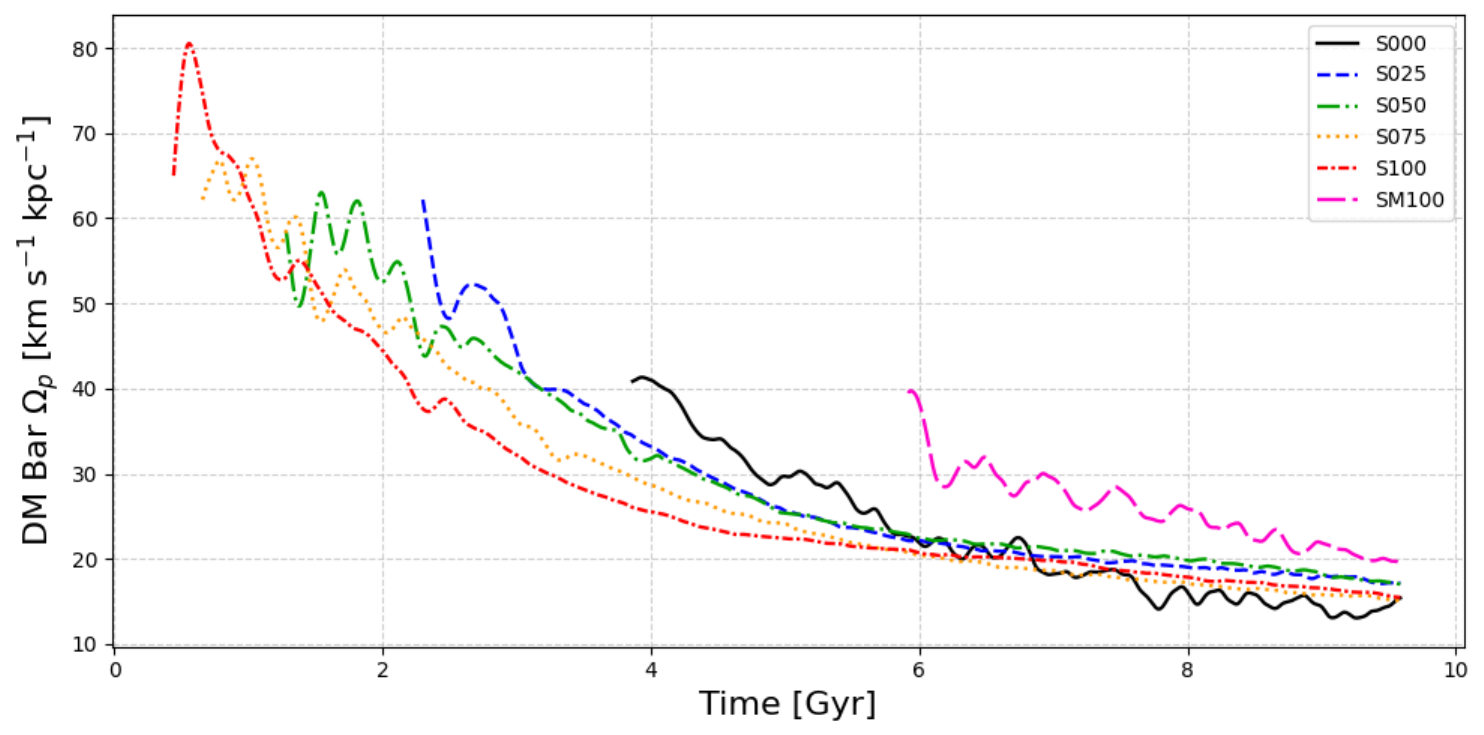} 
    \caption{Evolution of dark matter bar pattern speed over time for varying halo
spin parameter.}
    \label{fig pattern speed}
\end{figure}

\subsection{Evolution of Bar Strength}
\label{evolution of bar strength}
The bar strength is calculated using the $m=2$ mode, where $A_2$, represents the bi-symmetric structure, while $A_0$ corresponds to the axisymmetric component. The ratio $A_2/A_0$, provides a dimensionless parameter to measure the strength of the bar. The parameter is applicable to both stellar and dark matter bar.

The Fourier coefficients for the \(m = 2\) mode are measured as  
\begin{equation}
a_2(R) = \sum_{i=1}^{N} m_i \cos\left( 2\phi_i \right), \quad
b_2(R) = \sum_{i=1}^{N} m_i \sin\left( 2\phi_i \right),
\end{equation}
where \(a_2\) and \(b_2\) are calculated for all disk particles, \(m_i\) is the mass of the \(i\)-th particle, and \(\phi_i\) is its azimuthal angle.  
The bar strength is then defined as  
\begin{equation}
\frac{A_2}{A_0} = 
\frac{\sqrt{a_2^2 + b_2^2}}{\sum_{i=1}^{N} m_i},
\end{equation}
\indent Furthermore, the bar strength was measured using the mass-weighted Fourier decomposition ($A_2/A_0$) method across radial bins within $R \leq 20$\,kpc for the stellar component and $R \leq 10$\,kpc for the dark matter component. The mass weighting ensures that the characteristic evolution remains consistent across different radial ranges, with the bar region dominating the final $A_2/A_0$ signal. Contributions from the outer regions beyond the bar are negligible. \\
\indent The dashed vertical line in Figure \ref{fig stellar bar strength} and Figure \ref{fig dark matter bar strength} represents the epoch at which the first buckling occurred.

 \subsubsection{Stellar Bar Strength}
 \label{stellar bar strength}
 
 Figure \ref {fig stellar bar strength} shows the time evolution of stellar bar strength. We observe that the bar formation timescale decreases with increasing initial halo spin, as supported by prior work \citep{Saha2013,Long2014,Collier2018}. In contrast, the $SM100$ model represents a significantly delayed formation of the bar. As discussed earlier, $S050,S075, S100$ models demonstrate two subsequent buckling events in their evolutionary phase, where we notice the first buckling duration was for a very short period of time, which has an impact on the strength of the bar, and we see a sudden drop in the strength of the stellar bar during that phase. The drop in strength becomes more significant with the increase in $\lambda$, in the models.\\
\indent From Figure \ref{fig stellar bar strength}, we conclude that after the conclusion of the first buckling phase, the bar regains its strength. Furthermore, the bar gains strength more than its pre-buckling phase and eventually saturates.

 The stellar bar strength in all our models eventually saturates at a similar level irrespective of the halo spin parameter $\lambda$. \\

\subsubsection{Dark Matter Bar Strength}
\label{dark matter bar strength}

We measure the dark matter bar strength using the same methodology used for stellar bar strength, which is illustrated in Figure \ref{fig dark matter bar strength}. The dark matter bar strength demonstrates a similar trend relative to the stellar bar, where the bar's triggering timescale decreases with an increase in halo spin. Resembling the stellar bar, the model $SM100$ with retrograde halo displays a delayed formation of dark matter bar. The strength of dark matter bars is comparatively lower than their stellar counterpart, and with increasing halo spin in the prograde direction, the bar formation becomes more prominent.

\begin{figure*}
\centering

\begin{minipage}{0.48\textwidth}
    \centering
    \includegraphics[width=\linewidth]{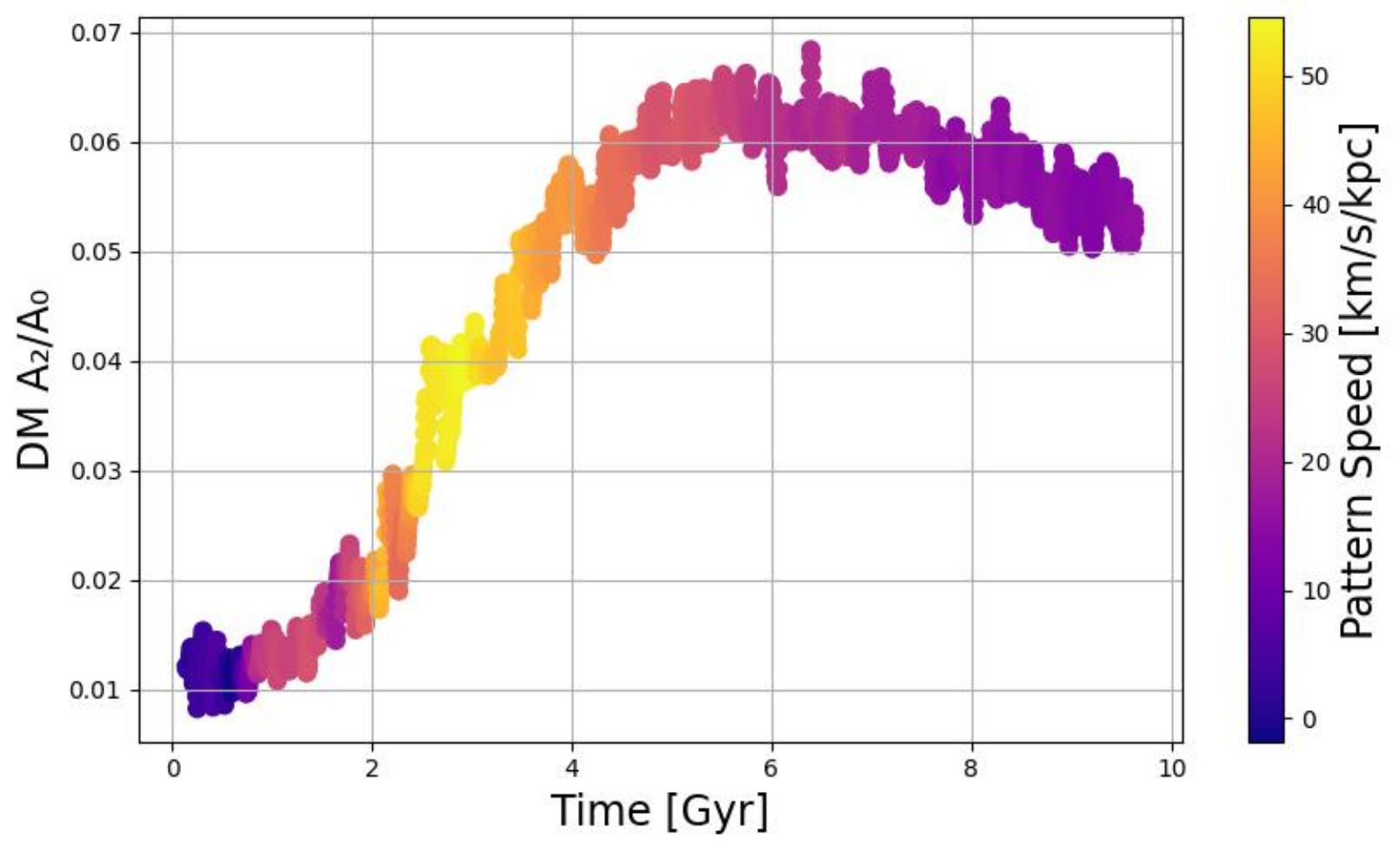}
    
    (a) $S000$
\end{minipage}
\hfill
\begin{minipage}{0.48\textwidth}
    \centering
    \includegraphics[width=\linewidth]{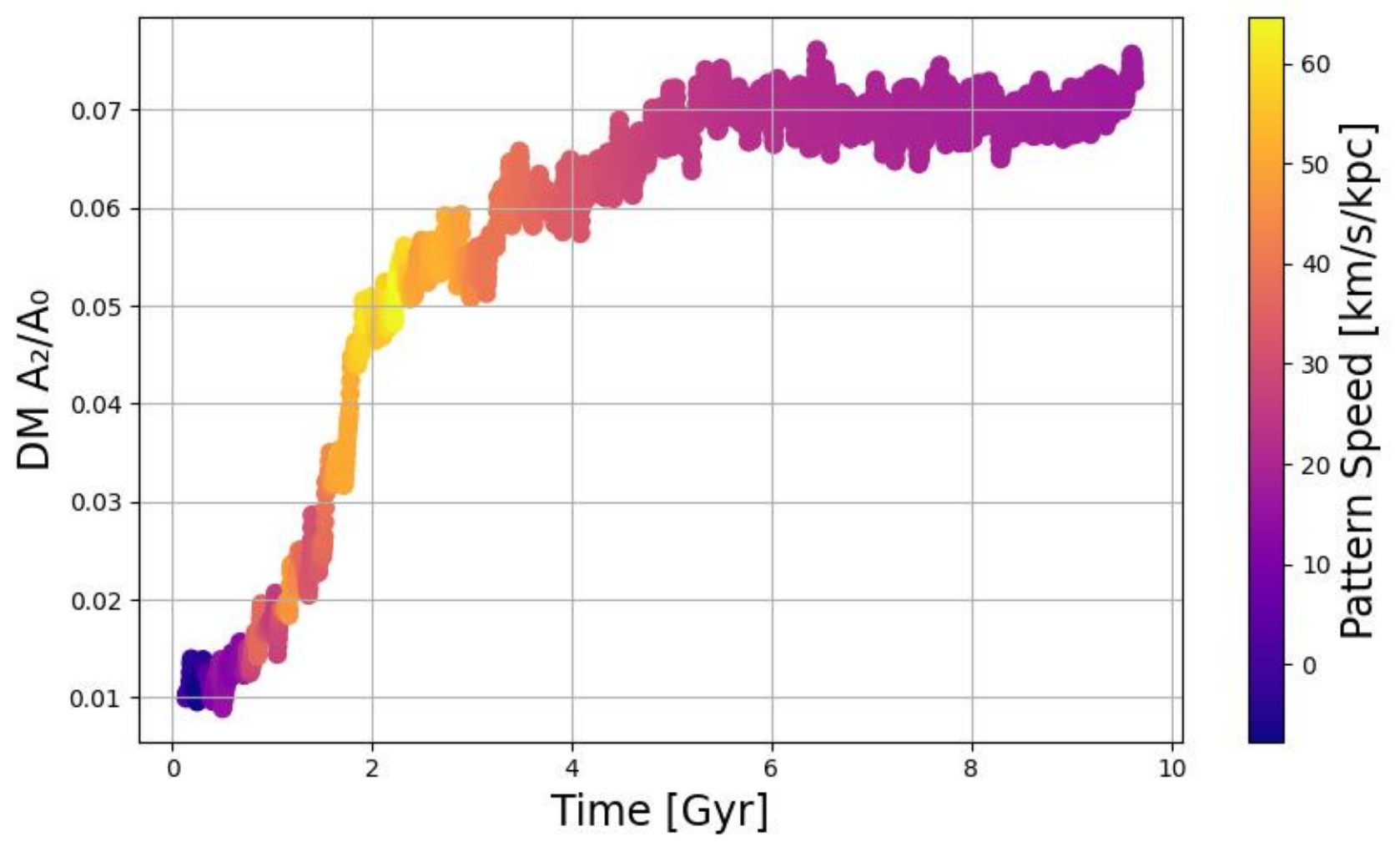}
    
    (b) $S025$
\end{minipage}

\vspace{0.3cm}

\begin{minipage}{0.48\textwidth}
    \centering
    \includegraphics[width=\linewidth]{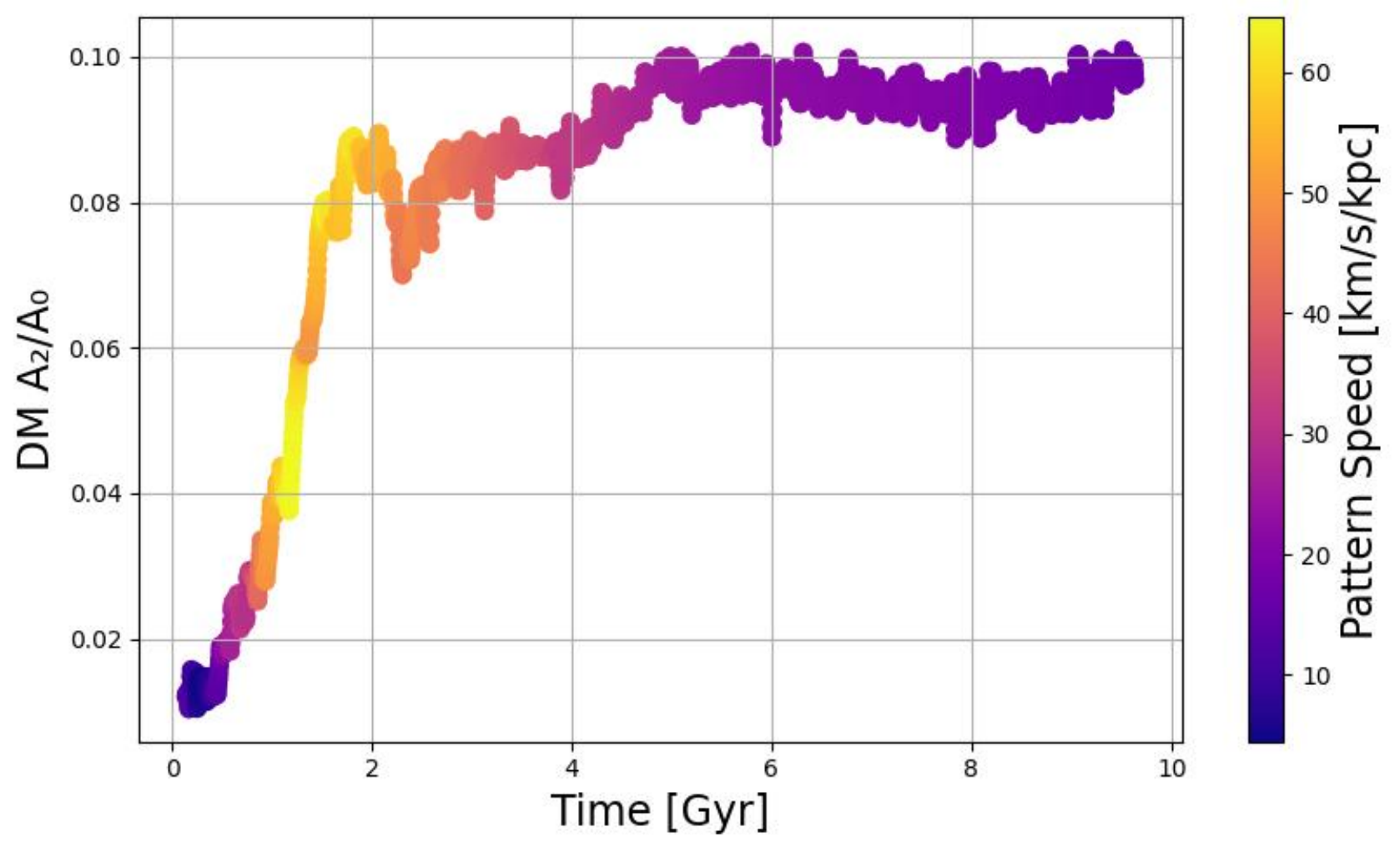}
    
    (c) $S050$
\end{minipage}
\hfill
\begin{minipage}{0.48\textwidth}
    \centering
    \includegraphics[width=\linewidth]{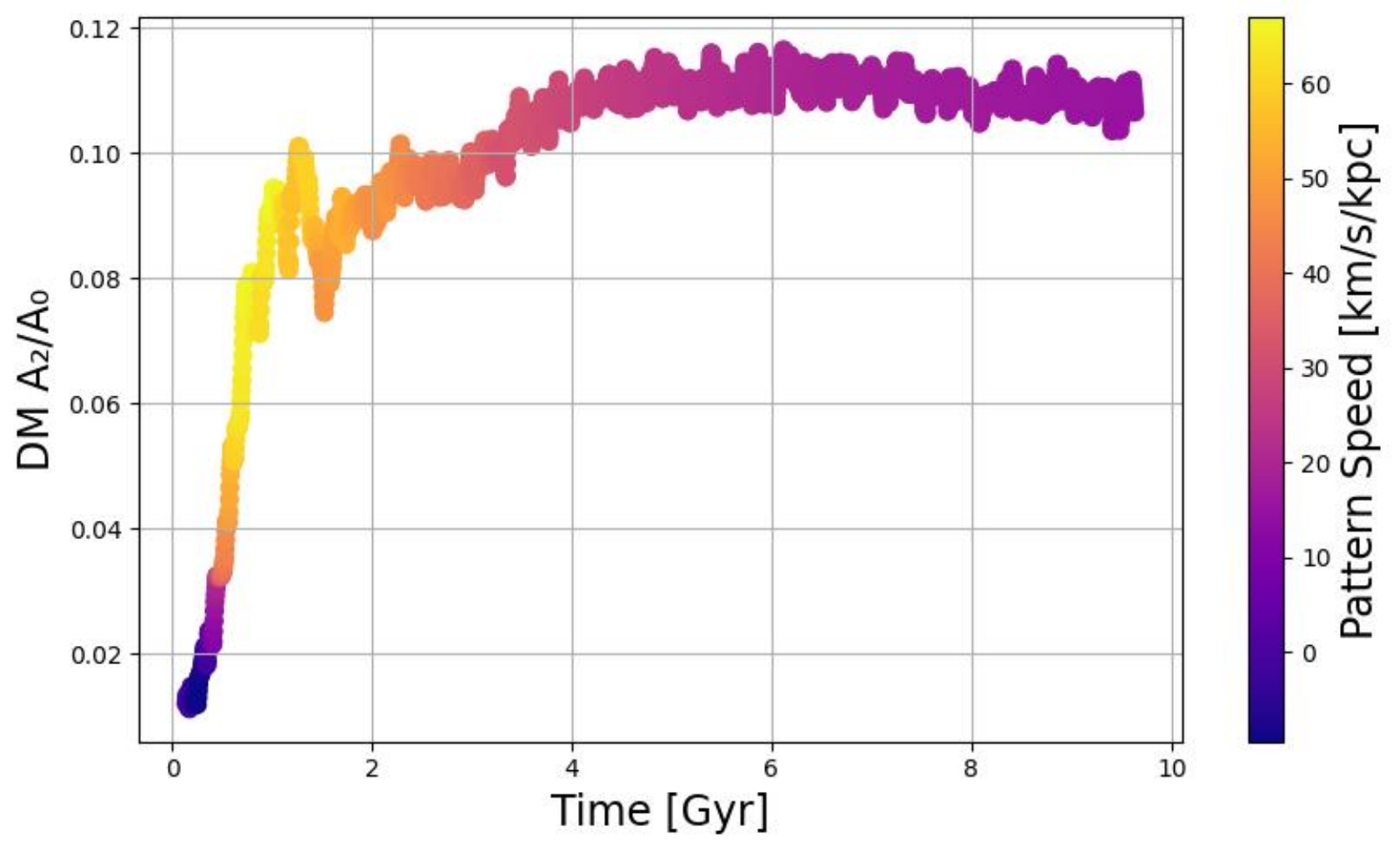}
    
    (d) $S075$
\end{minipage}

\vspace{0.3cm}

\begin{minipage}{0.48\textwidth}
    \centering
    \includegraphics[width=\linewidth]{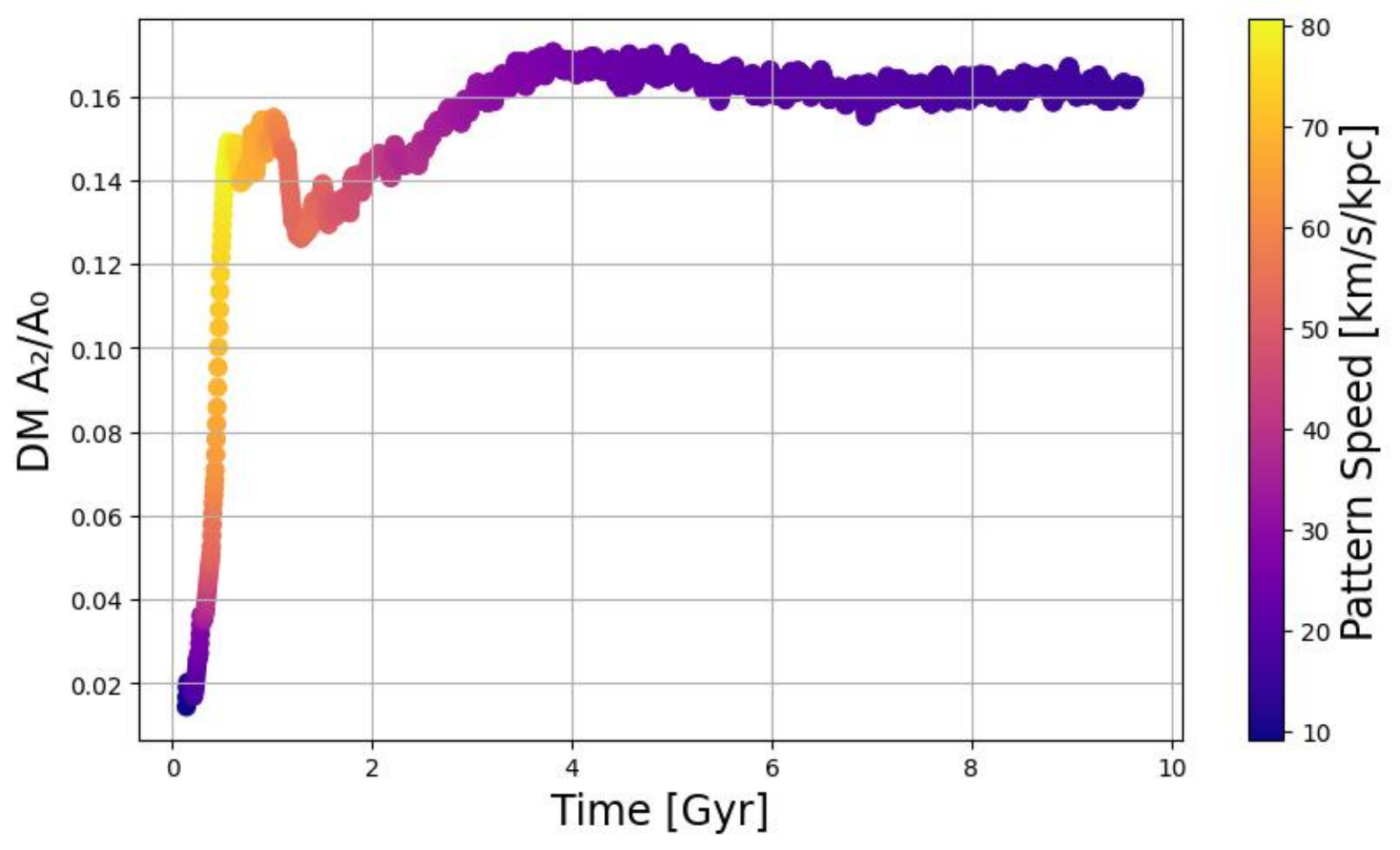}
    
    (e) $S100$
\end{minipage}
\hfill
\begin{minipage}{0.48\textwidth}
    \centering
    \includegraphics[width=\linewidth]{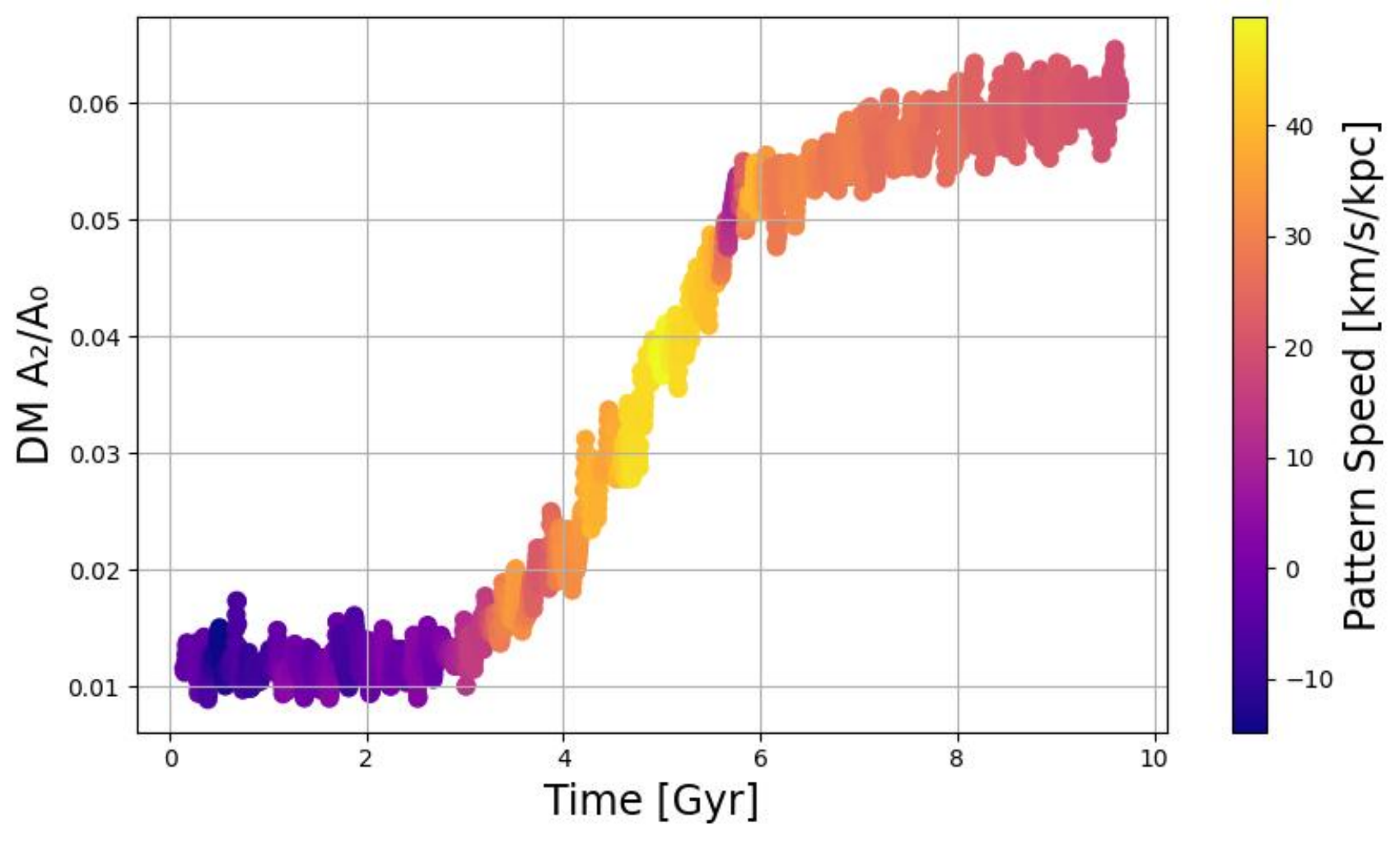}
    
    (f) $SM100$
\end{minipage}

\caption{
Co-evolution of bar strength ($A_2/A_0$) and pattern speed ($\Omega_{\mathrm{bar}}$) over time for dark matter bars in halos with varying spin parameters.
}

\label{fig:co_evolution_bar_strength_pattern_speed}

\end{figure*}
\begin{figure*}
\centering

\begin{minipage}{0.48\textwidth}
    \centering
    \includegraphics[width=\linewidth]{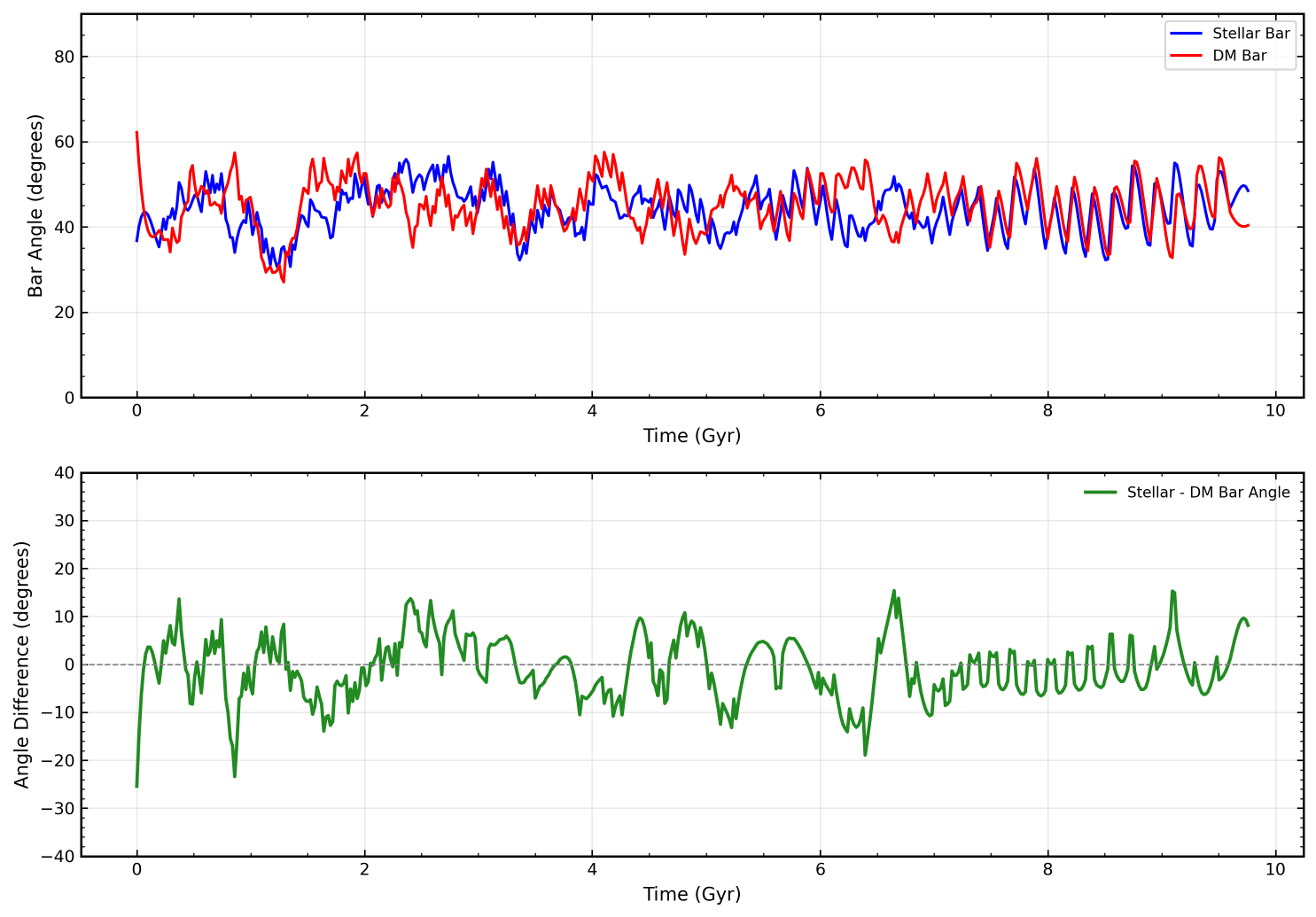}
    
    (a) $S000$
\end{minipage}
\hfill
\begin{minipage}{0.48\textwidth}
    \centering
    \includegraphics[width=\linewidth]{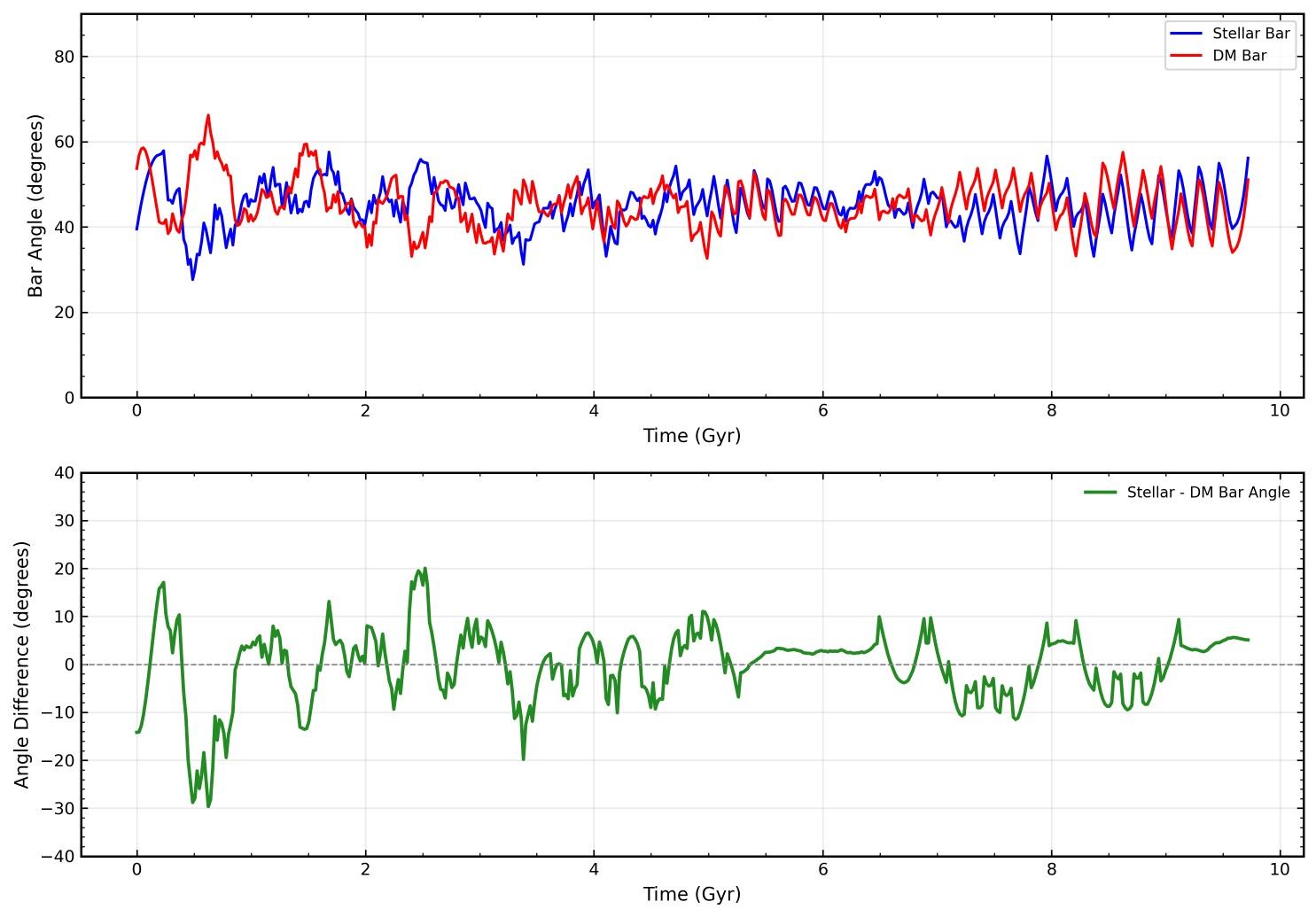}
    
    (b) $S025$
\end{minipage}

\vspace{0.3cm}

\begin{minipage}{0.48\textwidth}
    \centering
    \includegraphics[width=\linewidth]{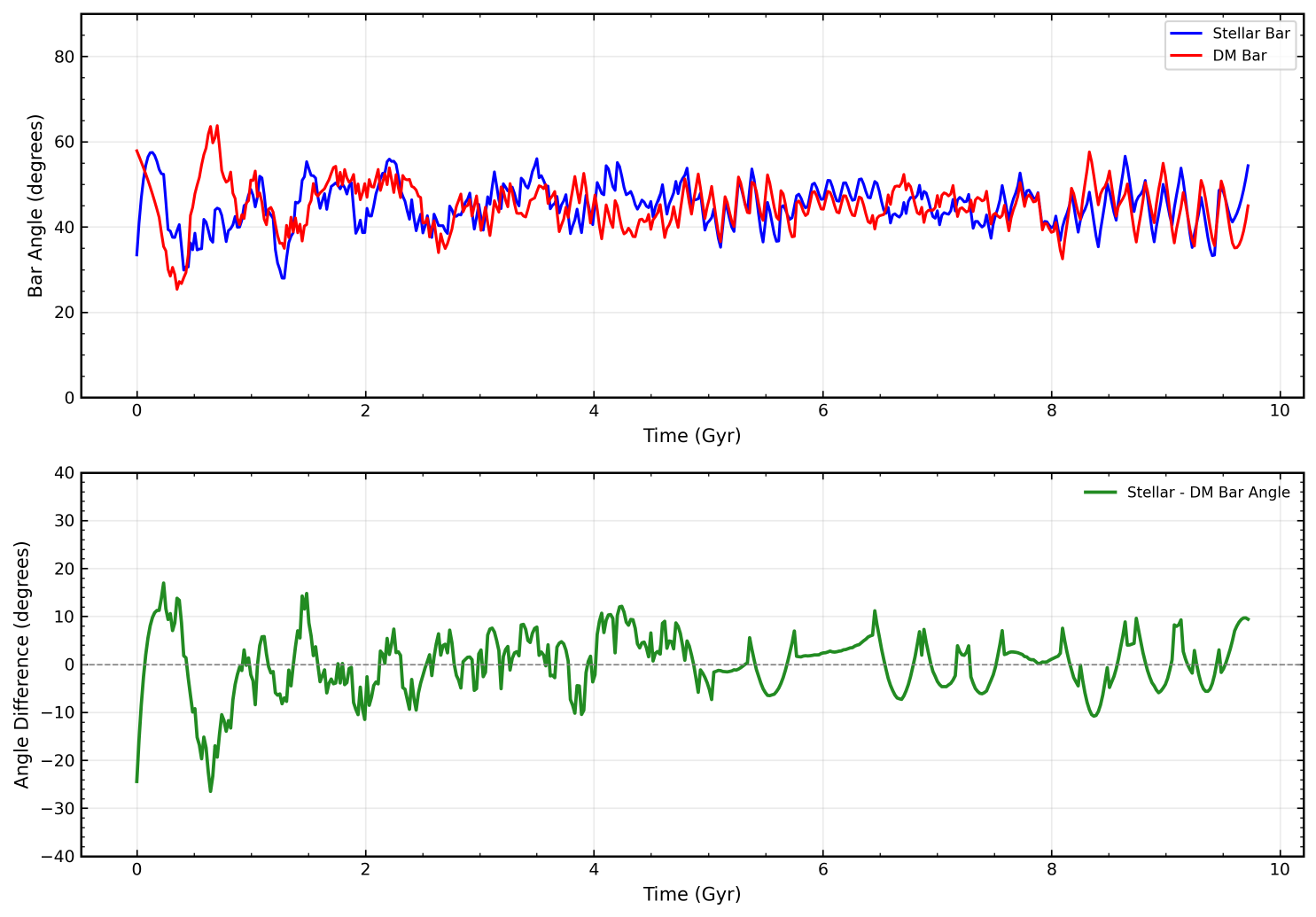}
    
    (c) $S050$
\end{minipage}
\hfill
\begin{minipage}{0.48\textwidth}
    \centering
    \includegraphics[width=\linewidth]{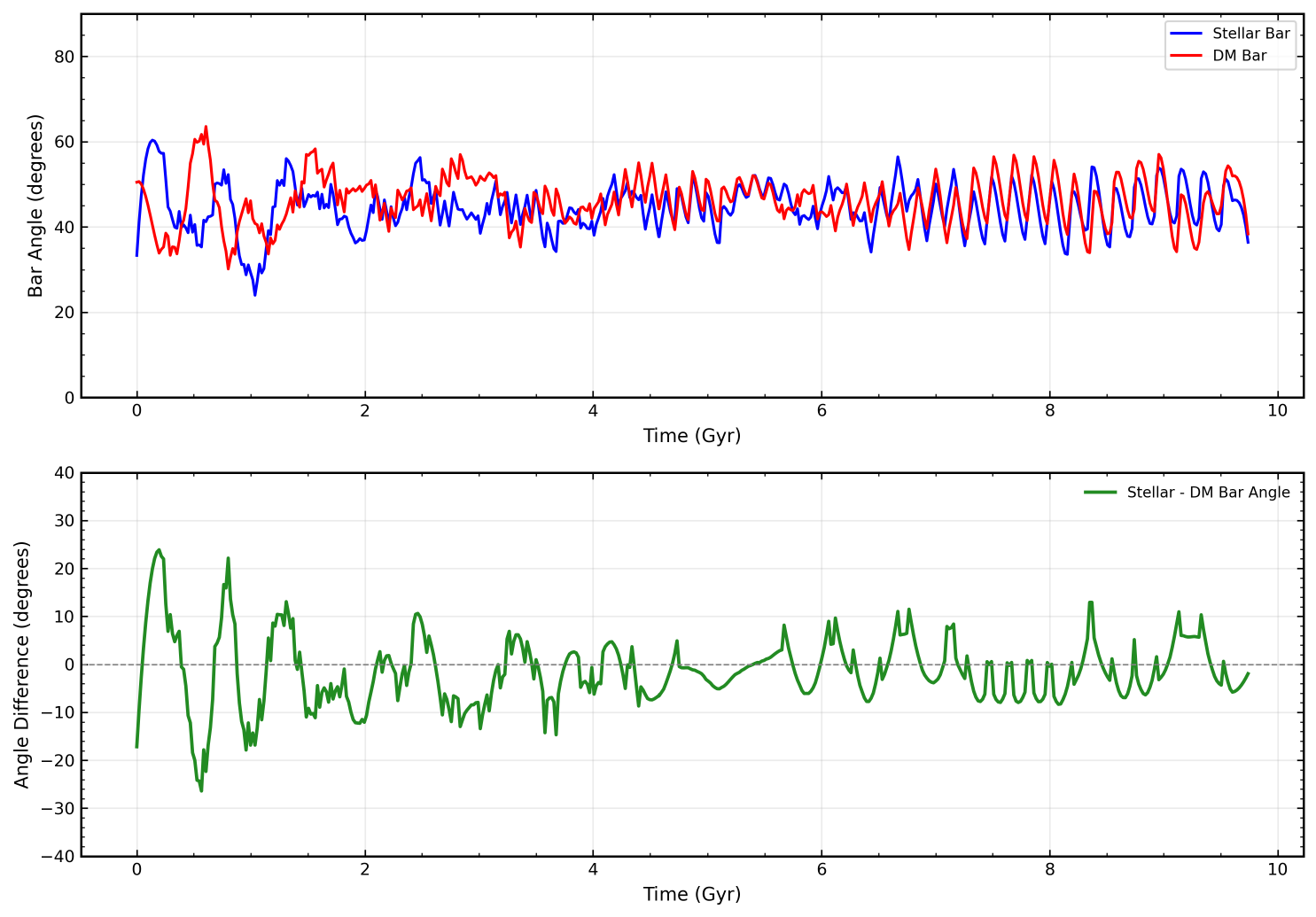}
    
    (d) $S075$
\end{minipage}

\vspace{0.3cm}

\begin{minipage}{0.48\textwidth}
    \centering
    \includegraphics[width=\linewidth]{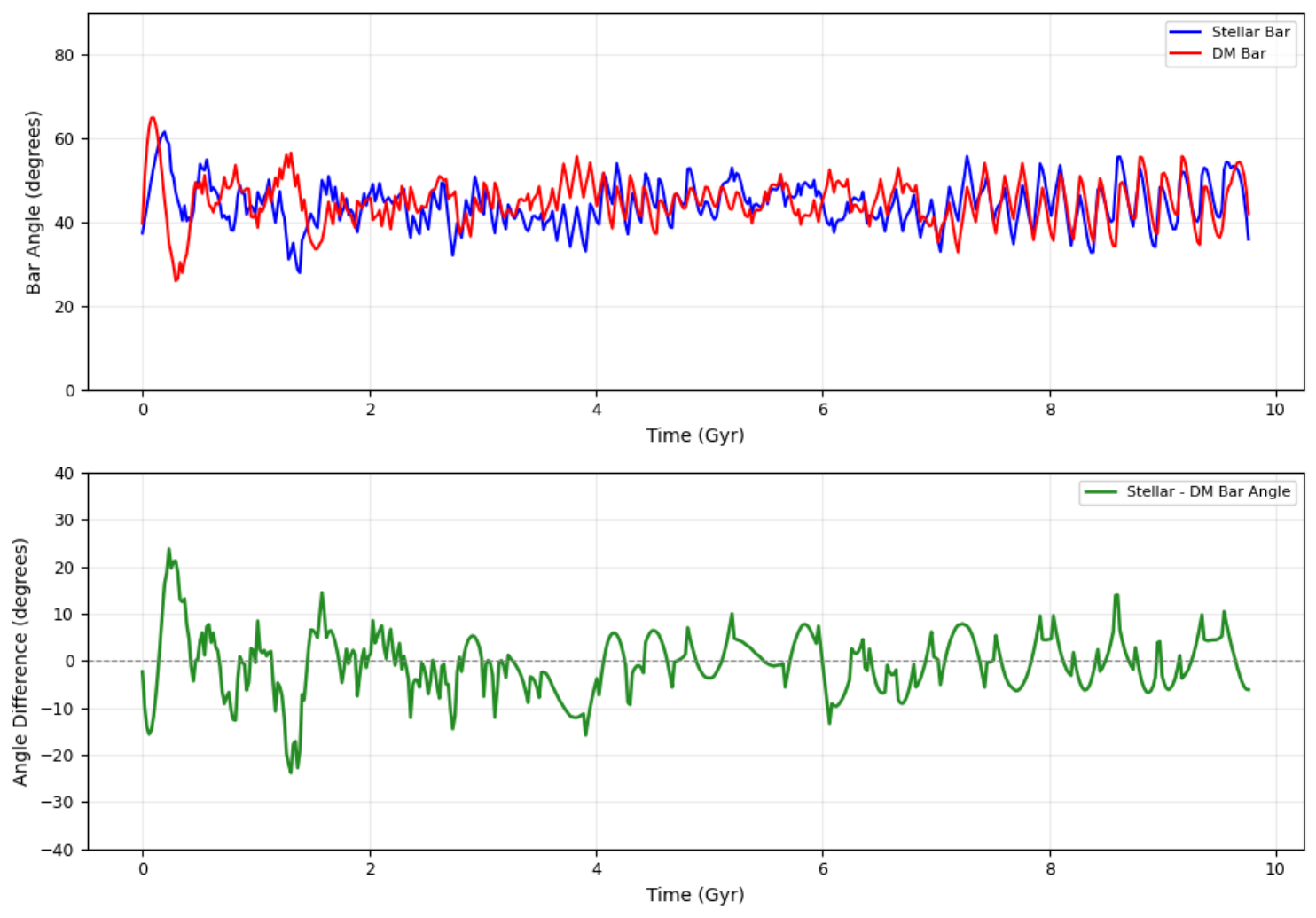}
    
    (e) $S100$
\end{minipage}
\hfill
\begin{minipage}{0.48\textwidth}
    \centering
    \includegraphics[width=\linewidth]{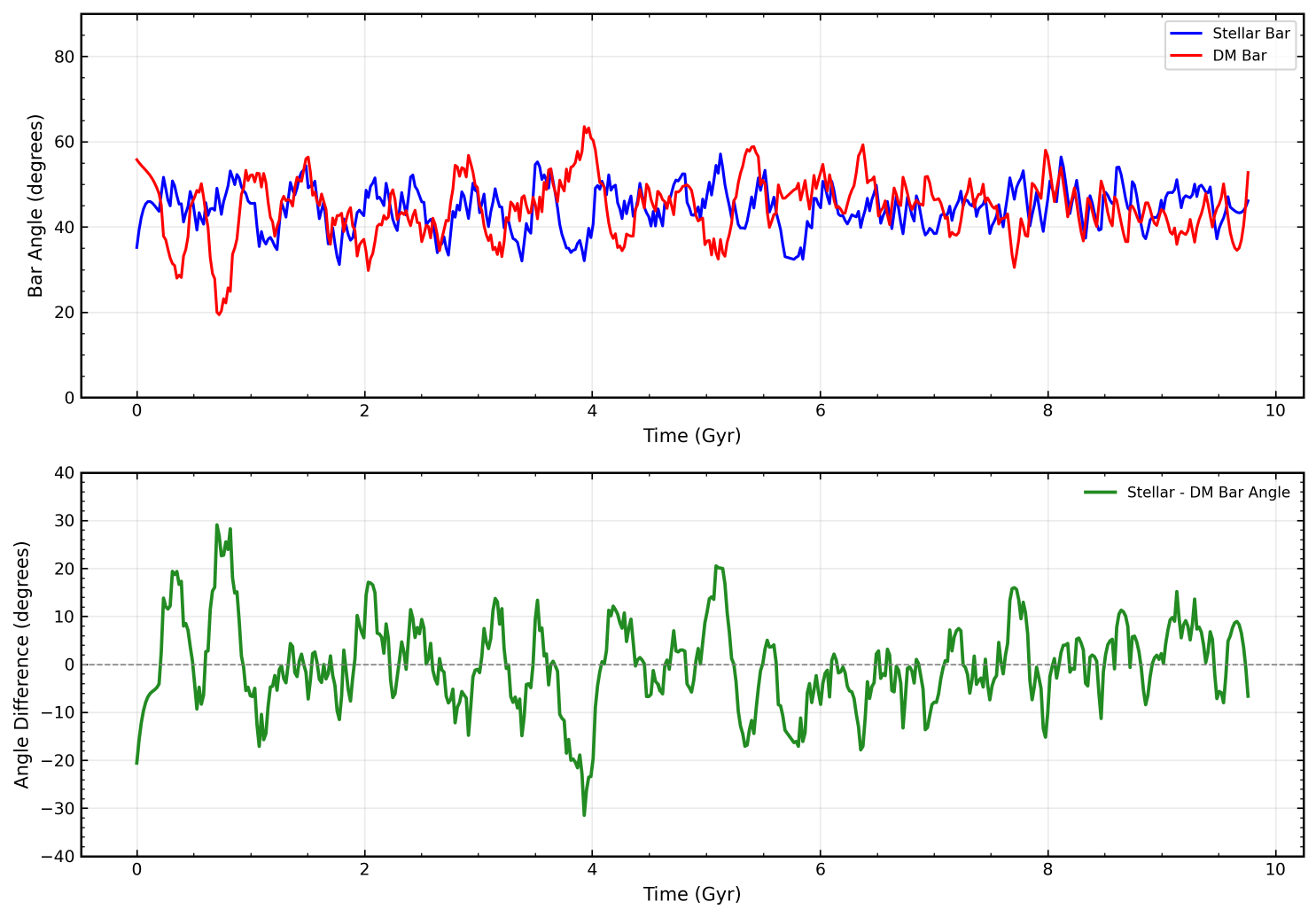}
    
    (f) $SM100$
\end{minipage}

\caption{
Evolution of stellar bar and dark matter bar angle over time for varying halo spin parameter.}

\label{fig: Bar Angle}

\end{figure*}

 In contrast to the stellar bar, where the bar strength saturates at a comparable level, the dark matter bar strength has a strong dependence on its initial spin parameter, exhibiting an increase in the magnitude of the bar strength with the increase in $\lambda$ in the prograde direction. At the end of its evolutionary phase, the DM bar strength saturates at a different magnitude, depending upon its initial $\lambda$. The $S100$ model with the highest initial spin in prograde direction displayed the highest magnitude, while the $SM100$ model with the retrograde halo showed the least magnitude in bar strength.\\
\indent As discussed earlier, in the case of stellar bars, we also notice a drop in the strength of DM bars. The characteristic is prominent in $S050, S075, S100$, where they undergoes two buckling events in their evolutionary phase. The first buckling episode, which is short-lived, affects not only the stellar bar but also the dark matter bar by causing a drop in its strength. A slight drop in strength is also visible in the $S025$ model, where it experiences only one buckling event in its evolutionary phase. In contrast, the models experiencing two buckling events exhibit a more pronounced decrease in their bar strength. The first buckling event in these models is short-lived and therefore the brief impulsive phase causes a sharper and more abrupt drop in $A_2/A_0$ compared to models undergoing a single prolonged buckling episode. After the completion of this buckling phase, the dark matter bar regains its strength, and its magnitude exceeds its pre-buckling phase, which differs from the previous studies \citep{collier_dark_2019}, where the bar failed to recover its strength after the buckling event. The discrepancy arises due to radial variations in the fraction of retrograde halo orbits in the inner regions. Our model contains a higher fraction of retrograde halo orbits at inner radii, as shown in Figure 2 of \citep{kataria_effects_2022}, with no discontinuity at the halo angular momentum at $\left| L_z \right|=0$ \citep{Kataria2024}, leading to a more efficient transfer of angular momentum to the halo. The difference in the mechanism, arising due to differences in the retrograde orbit fraction, is demonstrated in Figure 9 of \citep{kataria_effects_2022}.
The DM bar strength in all our models eventually saturates at distinct magnitudes, depending upon its initial halo spin parameter $\lambda$. The saturation magnitude of the bar strength increases with the increment in $\lambda$.

\subsection{Stellar Buckling Strength}
\label{Stellar Buckling strength}
The stellar buckling is demonstrated by all the bars as they evolve, irrespective of the initial halo spin. The buckling amplitude is calculated using the following equation as given by \citep{2006ApJ...645..209D}
\begin{equation}
B_{\text{buckle}} =
\left|
\frac{\sum\limits_{i=1}^{N} m_i \, z_i \, e^{2i\phi_i}}
{\sum\limits_{i=1}^{N} m_i}
\right|.
\label{eq:buckling_strength}
\end{equation}
where $z_i$ is the vertical position, $m_i$ is the mass, and $\phi_i$ is the azimuthal angle of the $i$-th stellar particle.

The timescale associated with the buckling event decreases with increasing halo spin, as shown in Figure \ref{fig: Buckling Strength}, differing from the previous findings \citep{10.1093/mnras/stad2799}, where the buckling event is delayed with increasing halo spin. Therefore, the stellar bars have different buckling time periods depending upon their initial halo spin parameter $\lambda$.\\
\indent All the bars in our simulation undergo the buckling phase irrespective of $\lambda$, but we witness that in some of our models, $S050, S075, S100$, there is a demonstration of the second buckling phase.  The second buckling duration occurs for a longer period of time than the first buckling phase. In models $S050, S075, S100$, where there is an occurrence of the second buckling phase, we observe that the first phase of buckling is very violent or happens for a very short period of time, which also causes a drop in bar strength as discussed in \ref{stellar bar strength} and \ref{dark matter bar strength}. In models $S000, S025, SM100$ with a single buckling event in their evolutionary phase, they have an intermediate stellar buckling time period. $SM100$ model shows a suppressed buckling with the event taking place at a later stage.\\

\subsection{Pattern Speed}
\label{pattern speed}

The pattern speed $\Omega_p$ of the stellar and dark matter bars is computed using the D-matrix method. The complex $m=2$, fourier phase is computed within $R \leq 20$\,kpc for the stellar bar and $R \leq 10$\,kpc for the DM bar and $|z| \leq 3$\,kpc, interpolated onto a uniform time grid, and the pattern speed is extracted by minimising the phase residual within a sliding Gaussian-weighted window.

The time evolution of both the stellar and DM bar pattern speed $(\Omega_p)$ are shown in Figure \ref{fig stellar pattern speed} and Figure \ref{fig pattern speed}, respectively. The initial dark matter bar pattern speed in the higher prograde halo model is greater than its stellar counterpart, as shown in Figure \ref{fig stellar pattern speed} and Figure \ref{fig pattern speed}.
The pattern speeds of all our models are similar across both stellar and dark matter components at the later stages of its evolutionary phase, irrespective of its initial halo spin parameter $\lambda$. However, the halo spin strongly influences the initial pattern speed of both bars, higher prograde halo spin leads to a higher initial pattern speed in the DM bar, while the retrograde model exhibits the lowest initial value. As the bar evolves, the pattern speed of both the bar decreases over time.\\
\indent The $SM100$ model with the retrograde halo has the slowest initial pattern speed compared to the prograde halo models. Despite the delayed onset of bar formation, this model also shows the steepest subsequent decline in pattern speed.
$S100$ model with the highest prograde halo spin demonstrates the early triggering of bar formation and also the highest initial pattern speed.\\
\indent The bar strength and the pattern speed exhibit an inverse relationship, as the bar gains strength and develops its bar structural morphology over time, its pattern speed gradually decreases.

    





\begin{figure*}
\centering

\begin{subfigure}[b]{0.48\textwidth}
    \centering
    \includegraphics[width=\linewidth]{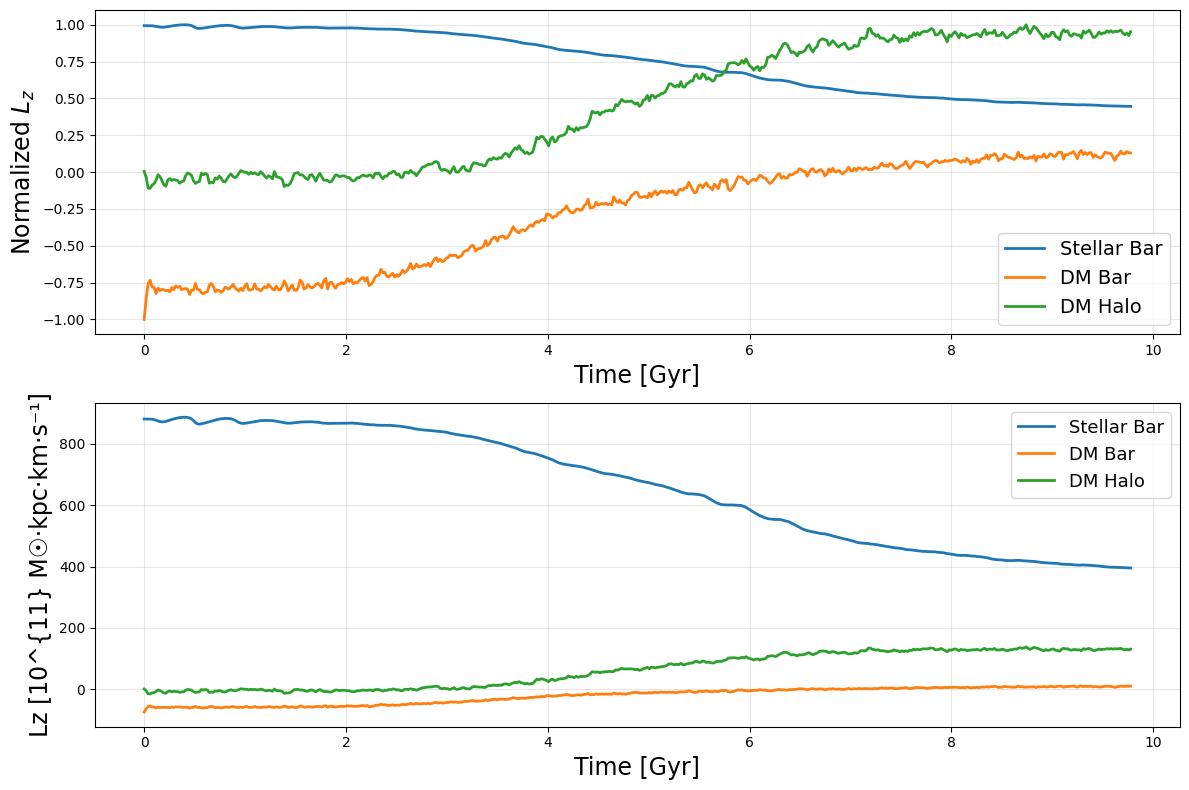}
    \caption{$S000$}
    \label{fig:S000}
\end{subfigure}
\hfill
\begin{subfigure}[b]{0.48\textwidth}
    \centering
    \includegraphics[width=\linewidth]{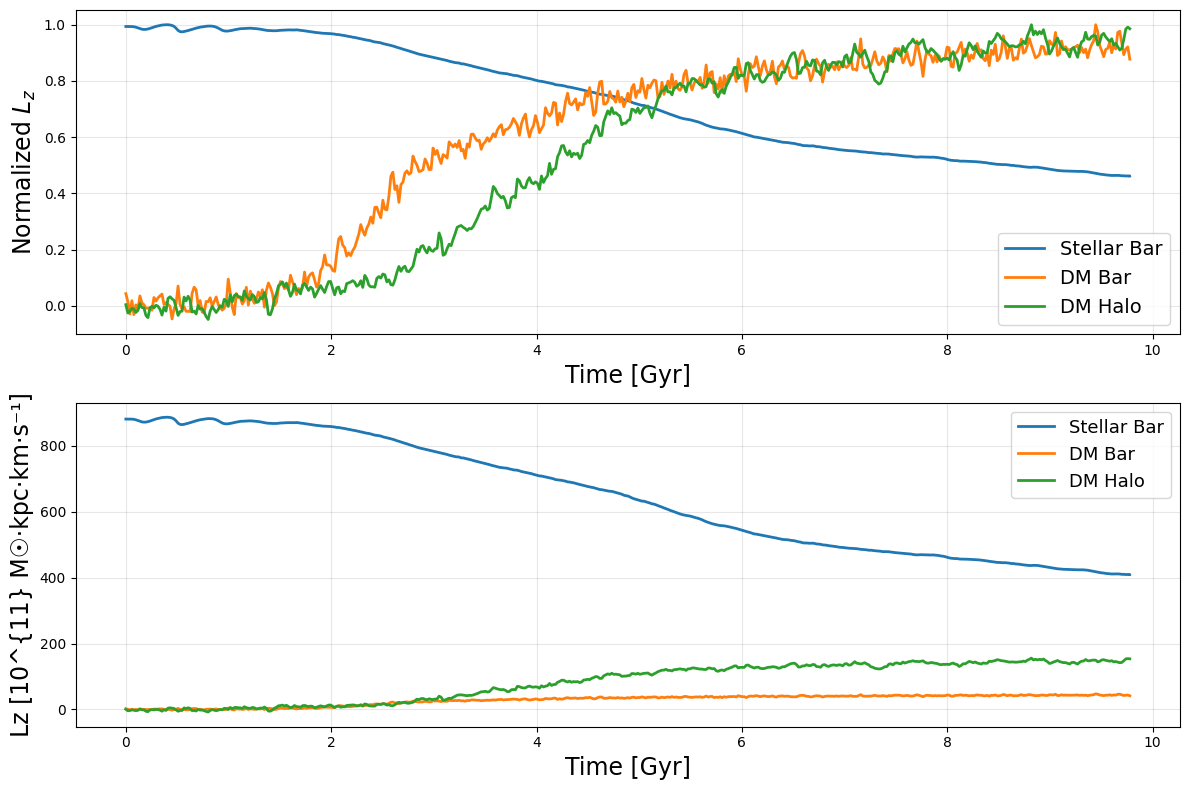}
    \caption{$S025$}
    \label{fig:S025}
\end{subfigure}

\vspace{0.3cm}

\begin{subfigure}[b]{0.48\textwidth}
    \centering
    \includegraphics[width=\linewidth]{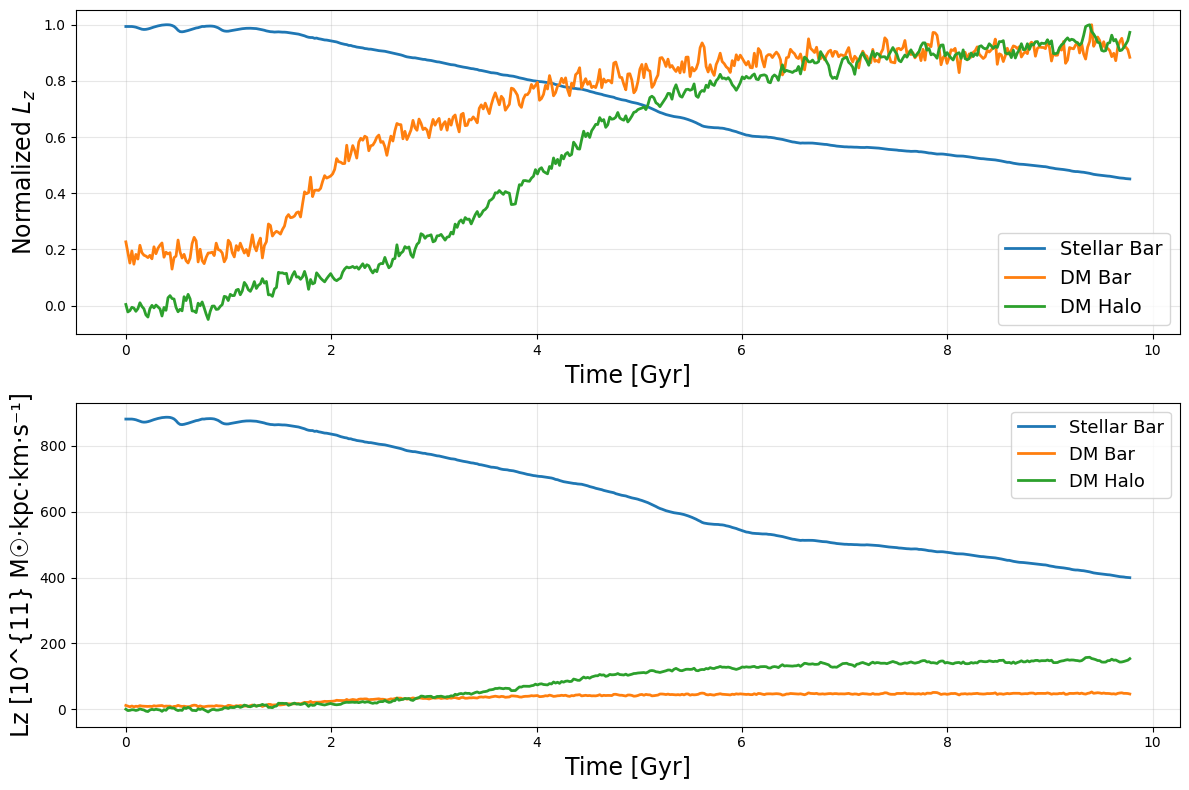}
    \caption{$S050$}
    \label{fig:S050}
\end{subfigure}
\hfill
\begin{subfigure}[b]{0.48\textwidth}
    \centering
    \includegraphics[width=\linewidth]{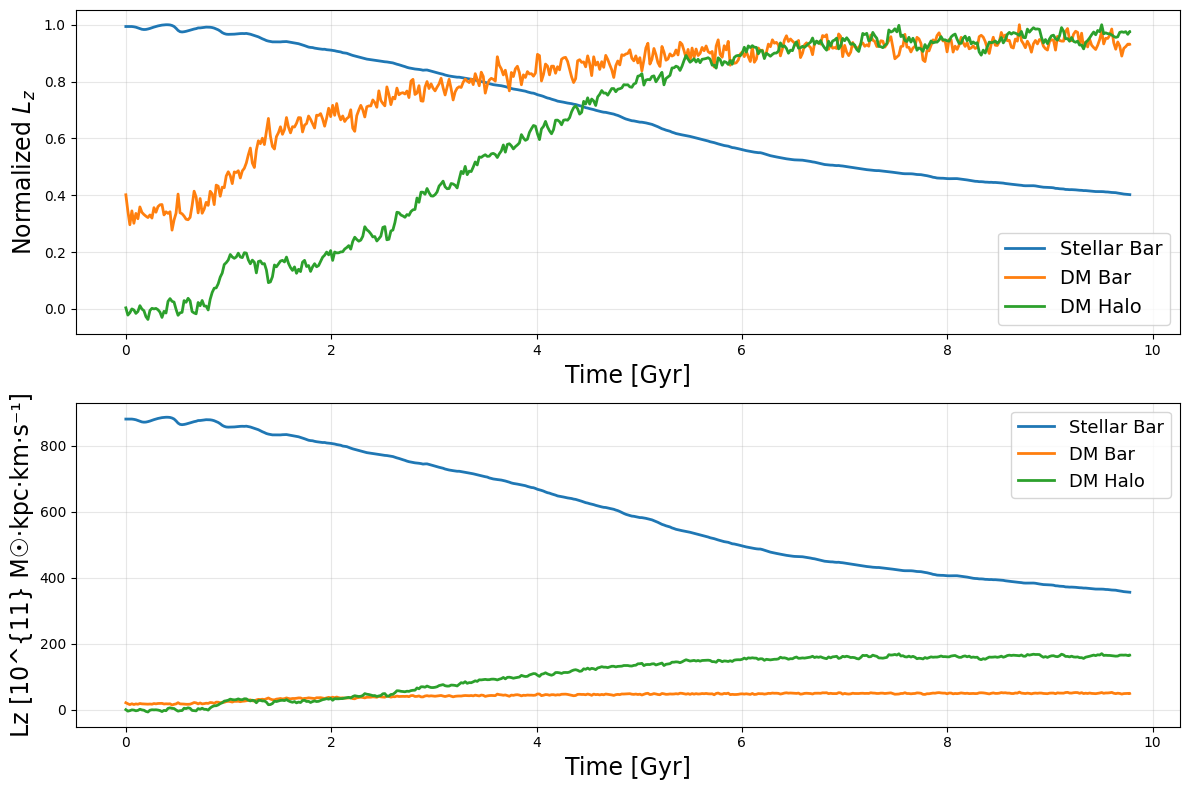}
    \caption{$S075$}
    \label{fig:S075}
\end{subfigure}

\vspace{0.3cm}

\begin{subfigure}[b]{0.48\textwidth}
    \centering
    \includegraphics[width=\linewidth]{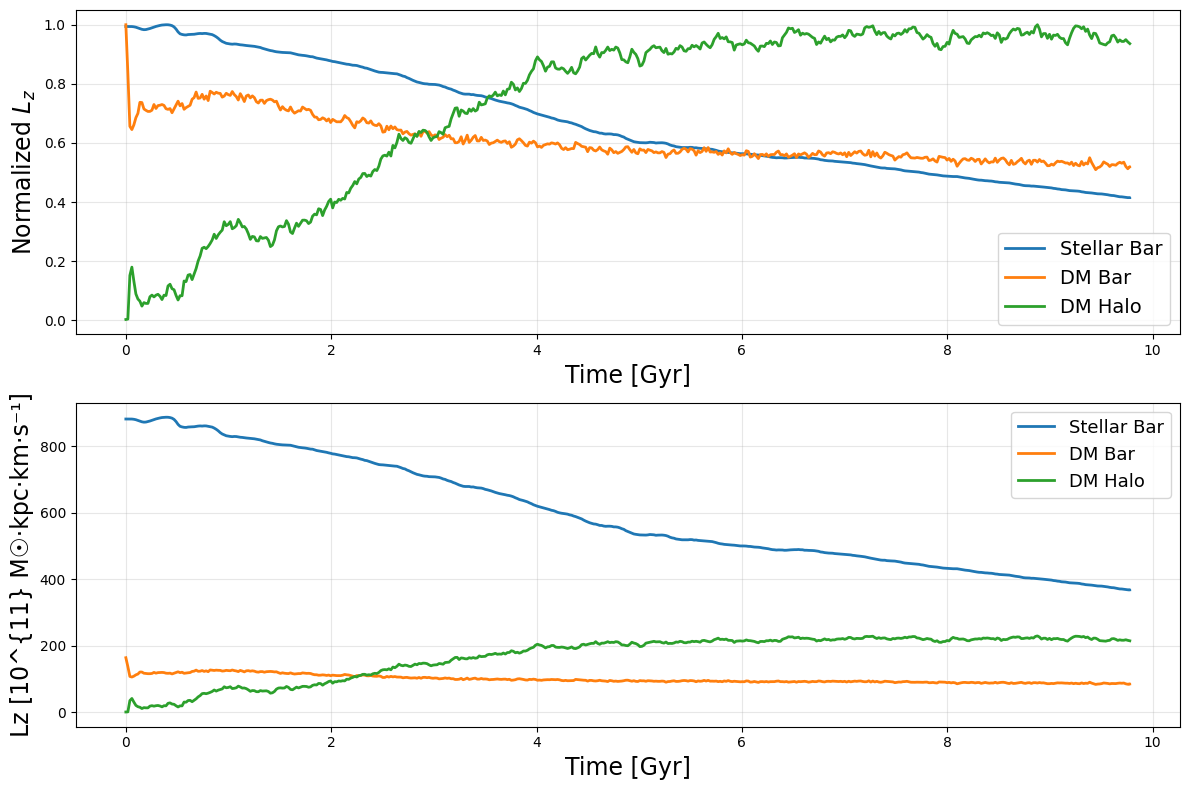}
    \caption{$S100$}
    \label{fig:S100}
\end{subfigure}
\hfill
\begin{subfigure}[b]{0.48\textwidth}
    \centering
    \includegraphics[width=\linewidth]{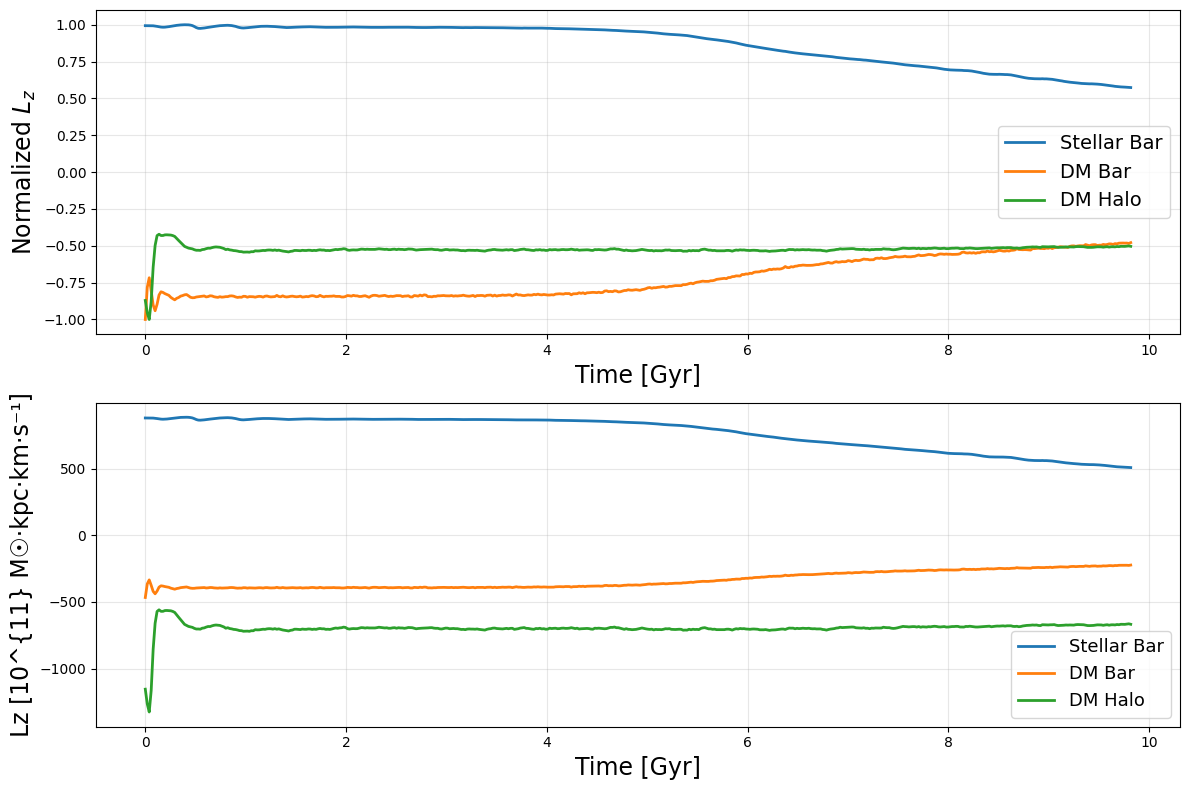}
    \caption{$SM100$}
    \label{fig:SM100}
\end{subfigure}

\caption{
Evolution of angular momentum for the stellar bar, dark matter bar, and the dark matter halo over time, shown for increasing initial halo spin parameters.
}
\label{fig: Angular Momentum}

\end{figure*}

\subsection{Co-evolution of Bar Strength and Pattern Speed Over Time}
\label{co evolution }

\indent Figure \ref{fig:co_evolution_bar_strength_pattern_speed} presents the combined evolution of the DM bar strength and pattern speed, illustrating their dynamical relationship.
 In the primordial bar triggering phase, when the bar starts forming, the initial pattern speed is highest, which gradually decreases over the course of its evolution. \\
\indent The dark matter bar pattern speed starts decreasing after the sole buckling events in the models $S000,S025, SM100$, and after the first buckling event in the models $S050,S075, S100$ and eventually saturates at a similar level.\\
\indent Figure \ref{fig:co_evolution_bar_strength_pattern_speed} represents that as the DM bar starts to evolve further and starts gaining strength, the pattern speed in an inverse relationship with the bar strength starts to reduce its magnitude. As the bar-like morphology becomes prominent for the DM bar in our models, the pattern speed magnitude declines.\\
\indent We conclude that the pattern speed of DM bar eventually reaches a comparable saturation level, as shown in Figure \ref{fig pattern speed} and Figure \ref{fig:co_evolution_bar_strength_pattern_speed}, irrespective of its initial halo spin parameter $\lambda$ and the final bar strength attained at the end of its evolutionary phase.

\subsection{Stellar and Dark Matter Bar Angle}
\label{bar angle}
The stellar and dark matter bar angles, along with their respective differences between them are illustrated in Figure \ref{fig: Bar Angle}. The orientation of both the bars remained mostly independent of the initial spin parameter $\lambda$. \\
\indent The model with retrograde halo displayed the maximum angular difference between the bars throughout its evolutionary phase compared to the bars in the prograde halo. The bar angles displayed minimal variation across models with different prograde halo spins, differing from the previous findings \citep{10.1093/mnras/stad2799}, where the lower-spin models exhibit a larger angular difference between the bars, which decreases significantly with increasing DM halo spin. From Figure \ref{fig: Bar Angle}, we observe a relatively large angular difference between the bars during the initial years, which arises due to differences in their pattern speed as shown in Figure \ref{fig stellar pattern speed} and Figure \ref{fig pattern speed}. As time progresses, both bars evolve and develop prominent bar morphology and therefore, the pattern speed of both bars decreases, which, in turn, gradually reduces the angular difference between them and eventually leads to near coupling as shown in Figure \ref{fig: Bar Angle}. This occurs as the pattern speed of both stellar and dark matter bar saturates at comparable values at a later stage of their evolution as depicted in Figure \ref{fig stellar pattern speed} and \ref{fig pattern speed}, consistent with \citep{Ash2024}. The strong coupling alignment facilitates the transfer of angular momentum between the various components. The alignment of stellar and DM bars has been confirmed in the TNG50 simulations \citep{Ash2024}.



\subsection{Angular Momentum Over Time }
\label{angular momentum result}
Angular momentum transfer plays a pivotal role in galaxy dynamics, governing the co-evolution of bars and their surrounding halo. Figure \ref{fig: Angular Momentum} displays the dynamics of the angular momentum transfer over time by the dark matter bar, halo and the stellar bar. The angular momentum flows from the barred stellar disk to the DM halo as discussed in prior works \citep{Athanassoula2003,Athanassoula2005,Berentzen2007,MartinezValpuesta2006,Dubinski2009}  The transfer of angular momentum is driven by orbital resonances, and primarily by the corotation resonance (CR), inner Lindblad resonance (ILR) and outer Lindblad resonance (OLR) as shown in previous studies \citep{Athanassoula2003,MartinezValpuesta2006,collier_dark_2019,Chiba2024-bh}. \\
\indent For the stellar bar, we considered the region $R \leq 20$kpc and $|z| \leq 3$kpc while for the dark matter bar, we analysed the region $ R \leq 10$kpc and $|z| \leq 3$kpc. The dark matter halo was analysed within the radial range $ 10$kpc $\leq R \leq 25$kpc and $|z| \leq 10$kpc. \\
\indent The stellar bar, as it evolves, interacts dynamically with the DM halo, and it transfers angular momentum to the halo and consequently loses it. The DM halo gains angular momentum over time, and the DM bar, being embedded in the halo, also exhibits a similar trend except for $S100$. The $S100$ model displays that the DM bar in correspondence with its stellar counterpart loses angular momentum over its evolutionary phase. This trend is displayed since the initial angular momentum of DM bar is much larger than the outer DM halo in $S100$ model.\\
\indent As discussed, the $SM100$ model exhibits delayed triggering of the stellar bar, as shown in Figure \ref{fig:SM100}, which displayed the transfer of angular momentum occurs primarily in the later period of its evolutionary phase.\\
\indent The upper panel of Figure \ref{fig: Angular Momentum} shows the angular momentum $L_z$ of each 
component normalised by its own maximum absolute value over the 
simulation duration, i.e.:
\begin{equation}
    \tilde{L}_z = \frac{L_z(t)}{\max|L_z(t)|}
\end{equation}
This normalisation is applied independently to each component 
(stellar bar, DM bar, and DM halo) to facilitate a
comparison of the relative temporal evolution and exchange of 
angular momentum between components, independent of their absolute 
magnitudes. While the lower panel of Figure \ref{fig: Angular Momentum} conveys 
the absolute angular momentum, the upper normalised panel 
reveals the relative timing and phase of angular momentum exchange 
between the stellar bar, DM bar, and DM halo.\\
\indent The normalised profiles highlight the impact of buckling on angular momentum transfer. Figure \ref{fig: Angular Momentum} clearly shows that, during the buckling phase (Figure \ref{fig: Buckling Strength}), all models undergo a rapid exchange of angular momentum between the stellar and dark matter components.\\
\indent The DM halo exhibits a larger absolute gain in angular momentum, as shown in the bottom panel of each model in the Figure \ref{fig: Angular Momentum}, whereas the normalised plots reveal that the DM bar experiences a greater relative growth.


\begin{figure}
    \centering
    \includegraphics[width=0.48\textwidth]{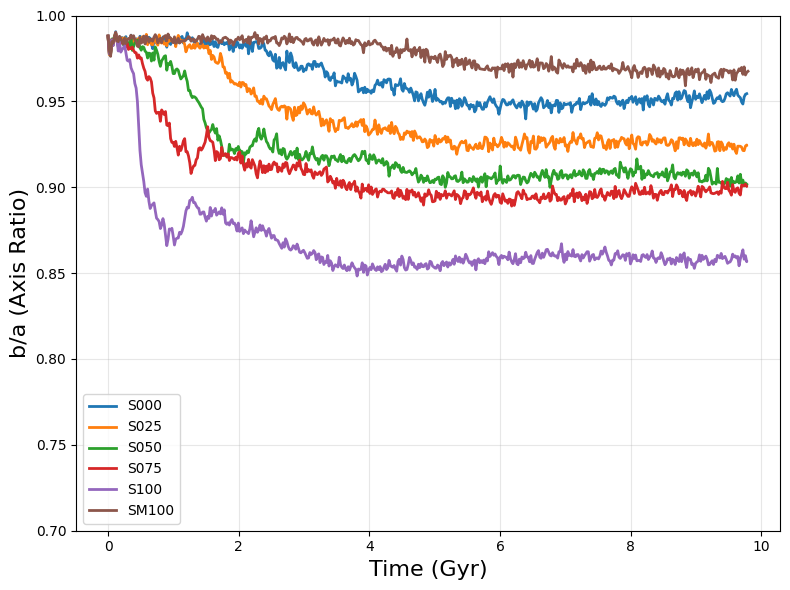} 
    \caption{The time evolution of the $b/a $ axis ratio for all models, where $b/a$ represents structural morphology of the DM bar.}
    \label{fig b/a}
\end{figure}
\begin{figure}
    \centering
    \includegraphics[width=0.48\textwidth]{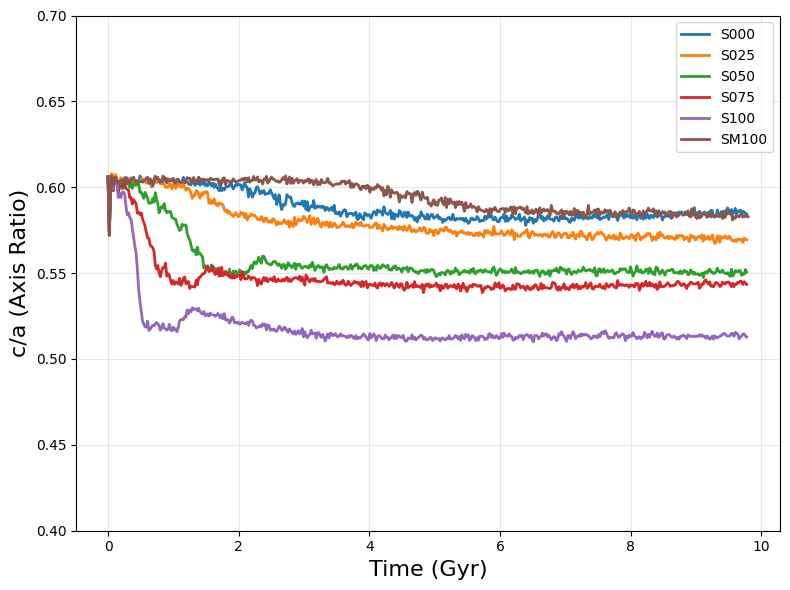} 
    \caption{The time evolution of the $c/a$ axis ratio for all models,  where $c/a$ represents vertical thickness of the DM bar.}
    \label{fig c/a}
\end{figure}

\subsection{Morphological Evolution of DM Bar Structure}
\label{axis ratio}
\indent The evolution of the axial ratios $b/a $ and $c/a$, representing the intermediate: major and minor: major axis ratios, respectively, is shown in Figures \ref{fig b/a} and \ref{fig c/a}, which represent the evolution of the structural morphology of the DM bar and its vertical flattening. We observe that the dark matter bar in all our models undergoes progressive morphological evolution, with the higher values of the initial halo spin parameter $\lambda$, in prograde direction leading to a stronger and more prominent bar in the later stages of its evolution as summarised in the Table \ref{tab:axis_ratios}.\\
\indent As the halo spin parameter $\lambda$ increases in our prograde models, the $b/a$ ratio decreases, indicating that the DM bar develops a prominent bar-like structural morphology with increasing prograde halo spin. The $S100$ model with the lowest $b/a$ ratio has the most prominent DM bar structure among other models, while the $SM100$ model has the least distinct bar morphology at the final stage of its evolutionary phase.\\
\indent
Similar to the $b/a$ axis ratio, the $c/a$ ratio also decreases with increasing $\lambda$, indicating a more enhanced vertical flattening in our prograde models, with increasing the $\lambda$. The DM bar in the $S100$ model, showed the most enhanced flattening, while the $ SM100$  model demonstrated the least significant vertical flattening, as shown in Table~\ref{tab:axis_ratios}. The $c/a$ ratio of the DM bar in each model stabilises after the initial period of bar formation and remains constant throughout the rest of its evolution.\\ 
\indent We observe that the $b/a$ ratio decreases and the $c/a$ ratio remains constant over time, following the establishment of the non-axisymmetric morphology as previously shown in\citep{Ash2024}, in all the models irrespective of the spin parameter $\lambda$ of the halo.

\begin{table}
\centering
\caption{
Axis ratios for dark matter bars across different spin models at the end of their evolutionary period.
\label{tab:axis_ratios}
}

\begin{tabular}{lcc}
\hline
Model & DM $b/a$ & DM $c/a$ \\
\hline
S000  & 0.9545 & 0.5844 \\
S025  & 0.9246 & 0.5694 \\
S050  & 0.9018 & 0.5504 \\
S075  & 0.9004 & 0.5432 \\
S100  & 0.8566 & 0.5127 \\
SM100 & 0.9675 & 0.5828 \\
\hline
\end{tabular}

\end{table}


\section{Discussion}
\label{discusion}

In contrast to stellar bars, the Dark Matter (DM) bars remain comparatively less explored in the context of galactic dynamics. The study of the dark matter bar is important to get a complete picture of the evolutionary galaxy dynamics. In the recent literature, the non-axisymmetric dark matter structure that develops in the inner halo is commonly referred to as a DM bar, despite exhibiting an axis ratio $(b/a)$ that is significantly higher than that of a typical stellar bar \citep{Ash2024}. In this work, we adopt the terminology DM bar to remain consistent with the current literature. This study aims to bridge the gap in our current literature knowledge about the DM bar, particularly how it co-evolves with its baryonic counterpart and what its evolutionary traits are under the influence of the different initial halo spin parameters of the parent DM halo. We also investigated how stellar disturbances shape the evolutionary growth of the DM bar. \\
\indent We investigated the formation and evolution of DM bars in the disk halo system under varying initial halo spin parameter $\lambda$ using the N-body simulation evolved for 9.78 Gyr. We analysed the evolutionary characteristics of DM bar in the halo with the spin parameter $\lambda = 0-0.1$ in prograde and $\lambda=0.1$ in retrograde, in order to obtain a conclusive result on how the dynamics of the disk-halo system and primarily the properties of DM bar are impacted by both prograde and retrograde halo spin.\\
\indent As discussed, DM bars are formed and evolve as a response to the parent DM halo during its co-evolution with the baryonic disk, with the stellar bar playing the most significant role.

    






\begin{figure*}
\centering

\begin{minipage}{0.48\textwidth}
    \centering
    \includegraphics[width=\linewidth]{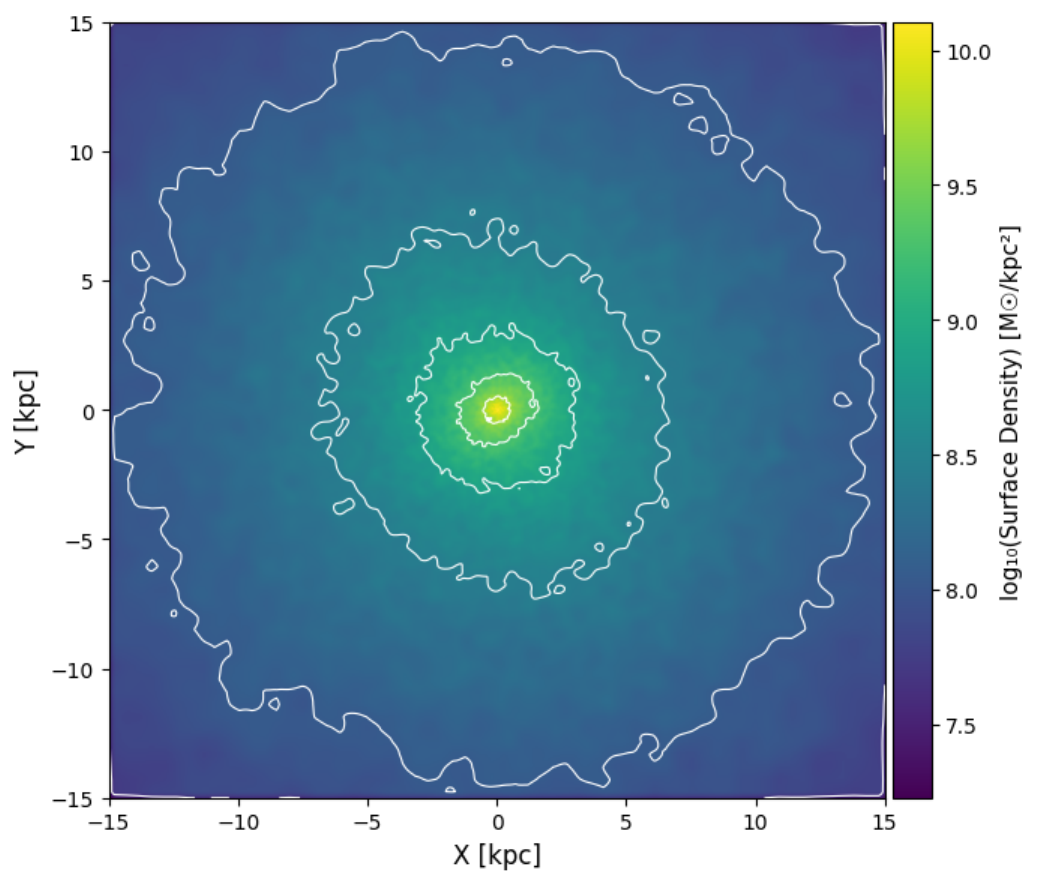}
    
    (a) $S000$
\end{minipage}
\hfill
\begin{minipage}{0.48\textwidth}
    \centering
    \includegraphics[width=\linewidth]{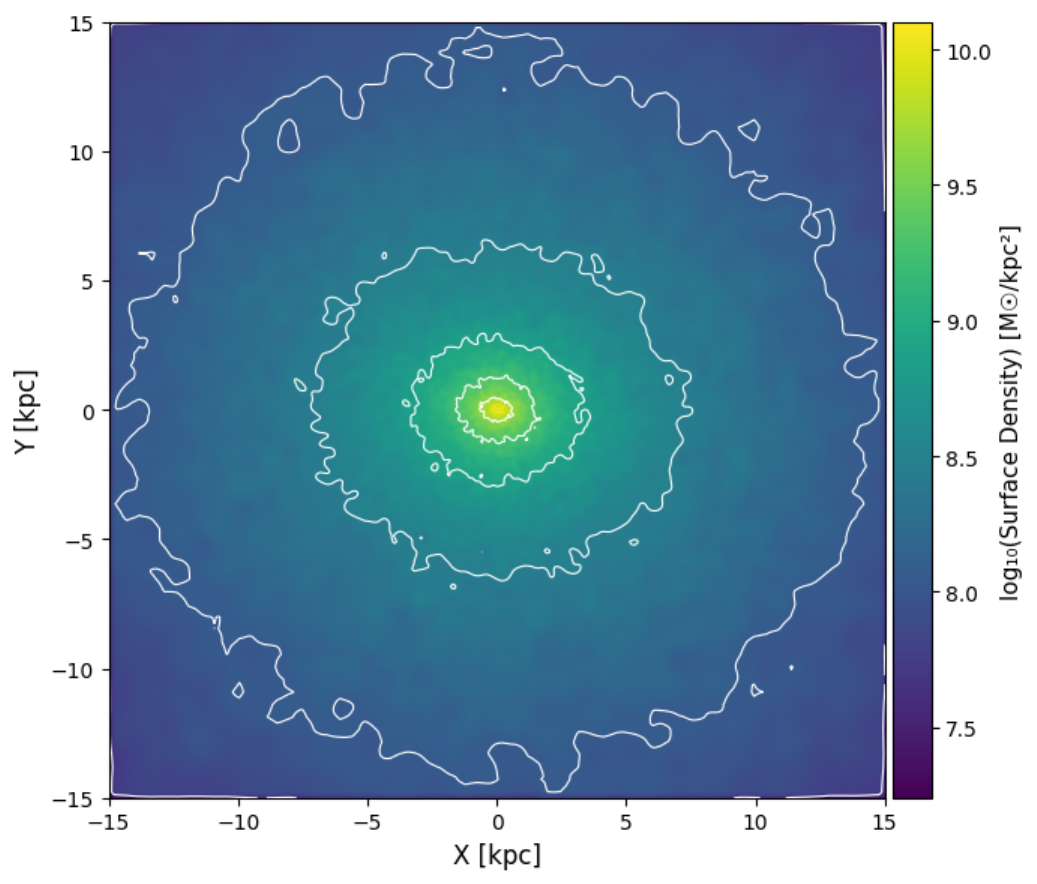}
    
    (b) $S025$
\end{minipage}

\vspace{0.3cm}

\begin{minipage}{0.48\textwidth}
    \centering
    \includegraphics[width=\linewidth]{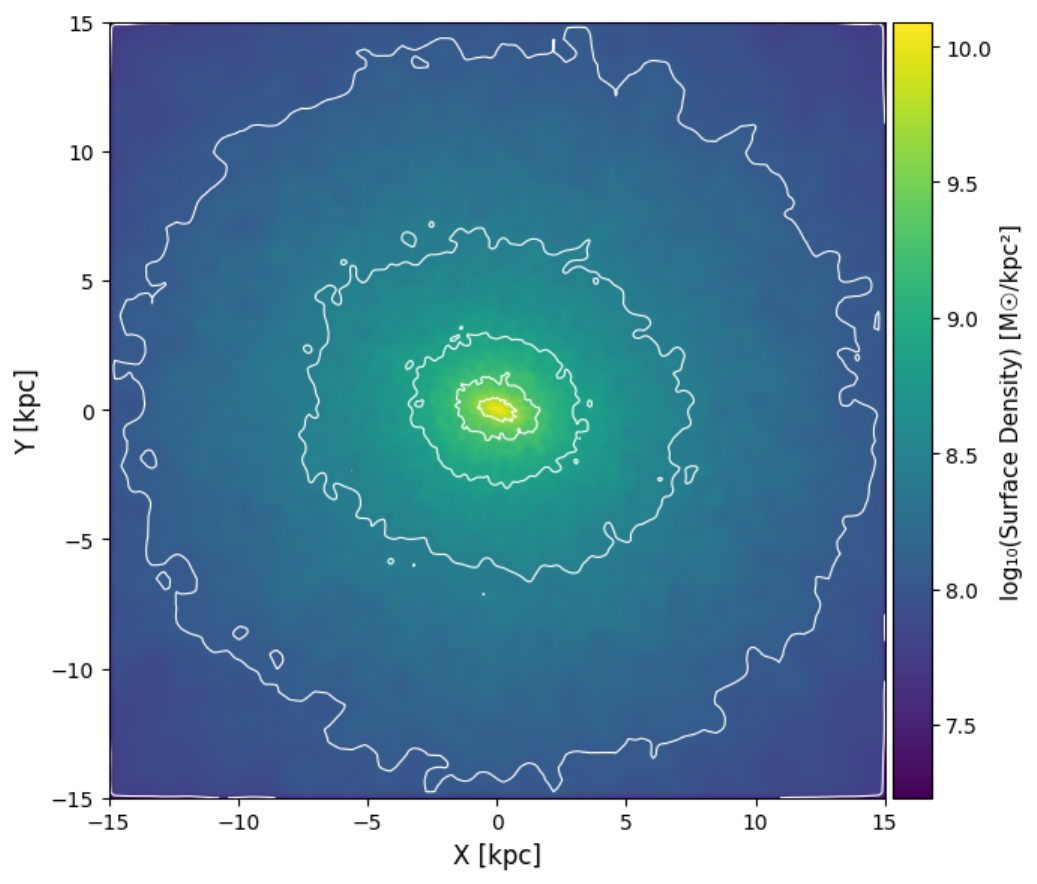}
    
    (c) $S050$
\end{minipage}
\hfill
\begin{minipage}{0.48\textwidth}
    \centering
    \includegraphics[width=\linewidth]{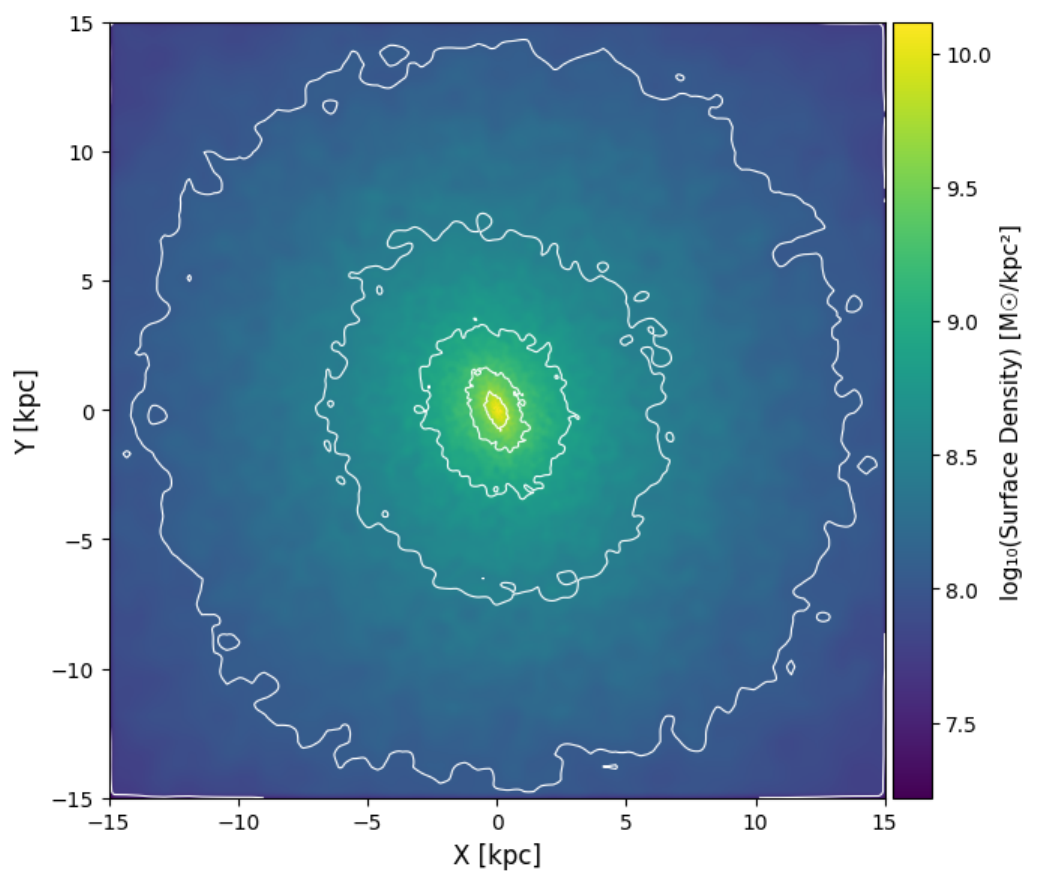}
    
    (d) $S075$
\end{minipage}

\vspace{0.3cm}

\begin{minipage}{0.48\textwidth}
    \centering
    \includegraphics[width=\linewidth]{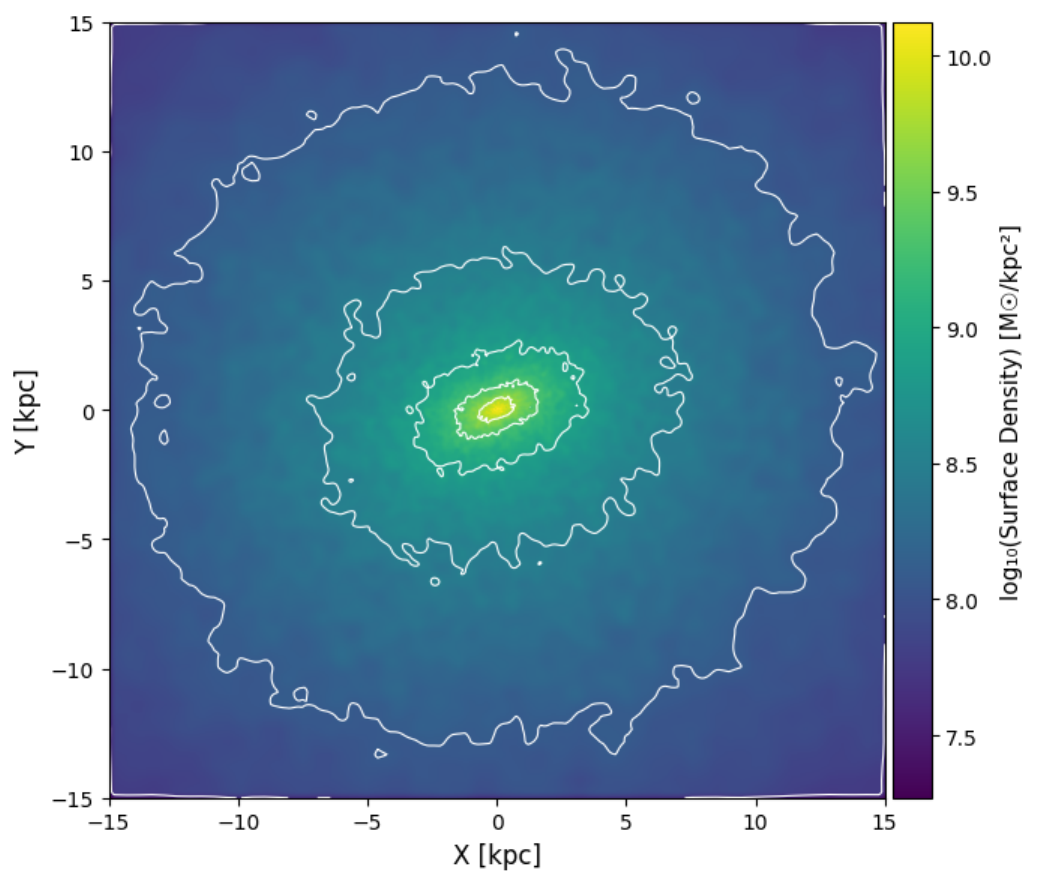}
    
    (e) $S100$
\end{minipage}
\hfill
\begin{minipage}{0.48\textwidth}
    \centering
    \includegraphics[width=\linewidth]{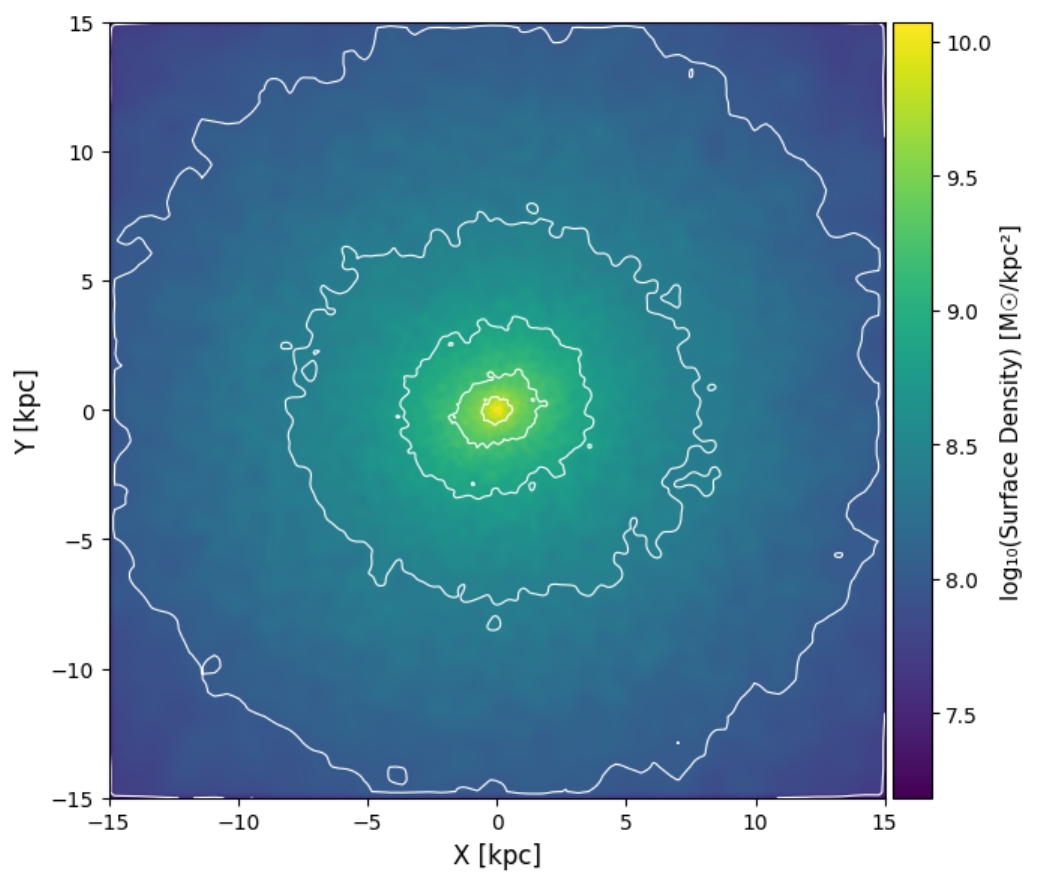}
    
    (f) $SM100$
\end{minipage}

\caption{
Dark matter surface density for models with varying spin parameter at $t=9.78$ Gyr.
}

\label{dark matter surface density}

\end{figure*}

\indent First, we measured the bars' instability, and observed that the higher prograde spin leads to an earlier onset of the buckling event while the model with retrograde halo spin displayed a delayed occurrence of the buckling event. The models with higher prograde spin parameter displayed a second buckling event in their evolutionary phase, in agreement with \citep{MartinezValpuesta2006}, which revealed that recurrent buckling is a key characteristic of long-term bar evolution. We see that co-rotating DM halos facilitate bar formation through resonant gravitational interactions as shown in prior study \citep{Saha2013}. The results also convey that the stellar bar formation timescale decreases with an increase in halo spin, agreeing with previous studies \citep{Saha2013,Long2014,Collier2018}, facilitated by efficient angular momentum transfer. The dark matter bar mirrors the stellar bar in this trend. \citep{Kataria2025} further revealed that spinning halos are approximately eight times more efficient in transporting angular momentum from the disk to the halo compared to non-spinning halos.\\
\indent Our results are consistent with recent findings from cosmological simulations. The study from TNG50 simulations \citep{Ash2024,ansar2024stellarbardark} supports that DM bars commonly occur in barred disk galaxies in realistic cosmological contexts, with properties consistent with isolated simulations. The results of our work help us to understand how different initial spin configurations will shape the evolutionary characteristics of the system through differential angular momentum exchange.\\
\indent We observe that the DM bar exhibits lower strength in correspondence to the stellar bar, aligning with earlier results \citep{collier_dark_2019}. The stellar bar strength, irrespective of the initial spin parameter $\lambda$, saturates at a similar level, whereas the dark matter bar strength at the end of its evolutionary period varies with $\lambda$, showing clear dependence on it. The DM bars display a higher saturation level with higher spin in prograde, while both the bars in the retrograde halo displayed lower bar strength in comparison to their prograde counterparts at the end of their evolutionary phase.\\
\indent Figure \ref{dark matter surface density}, also reflects that the Dark Matter halo gradually develops a non-axisymmetrical structure in the inner region of the DM halo, losing its initial symmetry as the spin increases in the prograde direction, therefore developing a DM bar in the inner region of the DM halo. The geometry of the non-axisymmetric structure of the inner DM halo becomes more pronounced as the spin increases. \\
\indent Our results suggest that the long-term bar evolution is governed by continuous secular processes instead of being permanently weakened by dynamical instabilities. We find that models with higher halo spin exhibit a more pronounced decline in bar strength, primarily during the first buckling period, which occurs over a shorter timescale. Both the DM and stellar bar regain their strength after the completion of the stellar buckling phase, exceeding their pre-buckling magnitude and eventually saturating over time. In this case, we find that buckling does not significantly reduce the strength of either bar, as both regain their strength after the buckling phase, showing a contrast with earlier studies reported by \citep{Long2014,Collier2018,collier_dark_2019}. The bar growth is governed by continuous angular momentum transfer and resonance trapping, processes that persist well beyond the buckling phase, as shown in \citep{MartinezValpuesta2006}.
\indent The difference between our results and previous studies \citep{Long2014,Collier2018,collier_dark_2019}, arises from radial variations in the fraction of retrograde halo orbits in the inner regions. In our model, a higher fraction of such orbits is present in the inner radii with no discontinuity at the halo angular momentum at $\left| L_z \right|=0$ \citep{Kataria2024}, leading to a more efficient transfer of angular momentum to the halo. The impact of the retrograde orbit fraction on the mechanism is shown in Figure 9 of \citep{kataria_effects_2022}.

\indent We also observed that the non-axisymmetric geometry of the inner region of the DM halo becomes more pronounced as the spin increases in the prograde direction.\\
 \indent Next, we analyse the evolution of the DM bar pattern speed over time. We observe that, irrespective of the initial pattern speed of the spin of the parent halo, the pattern speed of each bar saturates at a comparable level, demonstrating similar characteristics to the stellar bar \citep{kataria_effects_2022}. This convergence of pattern speeds represents a fundamental aspect of disk-halo coupling through resonant interactions.As the DM bars develop and gain strength, their pattern speed decreases. The DM bar in the halo with the highest prograde spin, $S100$, exhibits the highest initial pattern speed, which gradually decreases over time, whereas $SM100$, the DM bar, begins with the lowest initial pattern speed but shows the steepest decline.\\
\indent As time progresses, both bars develop their structural morphology, and eventually their pattern speed decreases. Consequently, the angular differences between the bars reduce, with the DM bars becoming nearly aligned with their stellar counterparts at the later stages of their evolutionary phase.\\
\indent The DM halo actively participates in the dynamical evolution of galaxies through the transfer of angular momentum between the dark matter halo and the baryonic matter. Angular momentum transfer plays the key role in galaxy evolution, as it provides the primary channel through which the dark matter and baryonic components interact and co-evolve.\\
\indent In this study, we find that the initial spin parameter, which is an intrinsic property of the DM halo, governs the evolution of the disk-halo system and consequently the transfer of angular momentum. The baryonic matter, primarily the stellar bar, interacts with the halo and consequently loses angular momentum over its evolutionary period, consequently becoming more structurally developed. In turn, the halo gains angular momentum through this interaction. A similar trend is shown by the DM bar as it is embedded inside the parent halo and formed as the dynamic interaction between the disk, primarily the stellar bar and the DM halo. The DM bar in the halo with the highest prograde spin, $S100$, exhibits a different evolutionary behavior compared to the other models, in that it loses angular momentum over time, similar to stellar bars and differing from DM bars in other models. This behavior arises because the initial angular momentum of the DM bar is much larger than the outer DM halo.\\
\indent As discussed, the DM bars develop a well-defined bar-like morphology over time, as traced by the evolution of the $b/a$ and $c/a$ axial ratios. We see that with an increase in prograde halo spin, the DM bar becomes more structurally profound by the end of its evolutionary period, while the DM bar in retrograde halo spin develops weaker structural morphology compared to its prograde counterparts. Similar to the behaviour of $b/a $ ratio, the $c/a$ ratio decreases in models with higher prograde halo spin, indicating enhanced vertical flattening at higher spins. The $SM100$ model displayed the least vertical flattening among all the models. The vertical structure of the DM bar, quantified by the $c/a$ ratio, remains nearly constant over time in all models, after the development of bar-like non-axisymmetric morphology, consistent with results from the TNG50 simulation \citep{Ash2024}.\\
\indent The formation of a non-axisymmetric DM structure contributes to the resolution of long-standing questions about whether bars can modify the density cusps predicted by CDM simulations.  Our results suggest that the structure of the inner halo becomes increasingly non-axisymmetric with higher prograde spin, potentially providing an observational test for halo spin using gravitational lensing.\\
\indent Recent JWST observations have revealed a high fraction of barred galaxies at $z > 2$ \citep{leconte2024_dxfapp,guo2025_15kw49}, questions the conventional view that bars are predominantly features of evolved, low-redshift galaxies. The results of this study suggests that galaxies forming in high-spin halos at early epochs could develop bars more rapidly, becoming one of the possible explanations for these observations \citep{Kataria2025}.

\section{Conclusion}
\label{conclusion}
In this article, we study the evolution of the dark matter bars in disk-halo systems with increasing spin parameter over a period of $9.78$ Gyr,
using N-body simulations, to understand how halo spin ($\lambda$) affects the characteristics of the DM bar. The spin parameter $(\lambda)$ in these models ranges from $0-0.1$. We also studied a retrograde spinning halo model which is counterrotating with respect to the disk. The DM bar develops as a result of DM halos' response to their interaction with the stellar bar through the transfer of angular momentum. We summarise our results below.
\begin{enumerate}[noitemsep,nolistsep]
    \item An increase in halo spin parameter in prograde direction triggers the earlier formation of the DM bar. The DM bar strength depends strongly on the initial spin of the parent DM halo. The strength increases gradually and saturates at a higher magnitude as we increase the halo spin in the prograde direction. Therefore, the DM bar in the $S100$ model demonstrates the highest DM bar strength. This behavior of the DM bar differs from that of its stellar counterpart, for which the bar strength saturates at a comparable level irrespective of the initial spin of the DM halo.
    \item The buckling of the stellar bar affects not only its own strength but also that of its dark matter counterpart. We find that the stellar disturbance leads to a decrease in the strength of the DM bar. The drop in DM bar strength is more pronounced during the first buckling phase in models that experience two buckling events in their lifetime, due to the short duration of the event. Both the stellar and DM bars regain their strength after the completion of the buckling phase and saturate at a strength level higher than that prior to buckling.
    \item The pattern speed of the DM bar saturates at a comparable level, later in its evolutionary phase, irrespective of its initial halo spin. As the DM bar starts gaining its strength, its pattern speed reduces. The spin parameter affects the initial pattern speed of the DM bar, which increases with higher prograde spin, whereas the DM bar in the retrograde model exhibits the lowest initial pattern speed. The drop in pattern speed is fastest in the retrograde model.
    \item We find that the orientation of both the bars remained mostly independent of the initial spin parameter $\lambda$. As both bars evolve and develop prominent bar-like morphology, the pattern speed of both bars decreases, which gradually decreases the angular difference between them, leading to eventual coupling.
    \item Stellar bar interacts dynamically with the DM halo through the transfer of angular momentum, eventually losing it and slowing down with time. The DM halo and the DM bar gain angular momentum during their evolutionary phase. The DM bar in $S100$ shows a reverse trend since the initial angular momentum of the DM bar is much larger than the outer DM halo.
    \item We observe that as we increase the spin of the halo in prograde direction, the inner region of the DM halo develops a more pronounced bar-like morphology, since the $b/a$ ratio decreases as we increase the spin. The retrograde model demonstrates the highest $b/a$ ratio, signifying the least pronounced bar-like morphology.
    \item The vertical flattening of the DM bar increases with higher halo spin, as reflected by a lower value of the $c/a $ ratio. The $c/a$ ratio of the DM bar for each model remains nearly constant throughout its evolutionary phase.
\end{enumerate}

\indent In summary, the evolutionary characteristics of the DM bar are strongly influenced by the initial spin of the DM halo and exhibit a significant dependence on $\lambda$. Studying the evolution of the DM bar reveals previously unexplored dynamical effects that play a significant role in the overall evolution of galaxies.

\section*{Acknowledgement}
\label{ackonwledgement}
\indent We thank the Department of Science and Technology (DST), Government of India, for providing the financial support through the INSPIRE Faculty Grant (DST/INSPIRE/03/2024/000401). We also extend our sincere thanks to the Department of S.P.A.S.E, IIT KANPUR, for providing the necessary facilities and for permitting the use of departmental resources essential to this work.  This work made use of the Gravity Supercomputer at the Department of Astronomy, Shanghai Jiao Tong University, the facilities of the Centre for High Performance Computing at Shanghai Astronomical Observatory, the facilities in the National Supercomputing Centre in Jinan, and the Param Sanganak Supercomputing facilities at IIT Kanpur. We thank referee Daniel Pfenniger for the insightful report and suggestions.

\section*{Data Availability}
We will make the data available on request.


\bibliographystyle{mnras}
\bibliography{references}

@article{kataria_effects_2022,
	title = {Effects of {Inner} {Halo} {Angular} {Momentum} on the {Peanut}/{X} {Shapes} of {Bars}},
	volume = {940},
	doi = {10.3847/1538-4357/ac9df1},
	number = {2},
	urldate = {2025-07-07},
	journal = {The Astrophysical Journal},
	author = {Kataria, Sandeep Kumar and Shen, Juntai},
	month = dec,
	year = {2022},
	pages = {175},
}

@article{collier_dark_2019,
	title = {Dark matter bars in spinning haloes},
	volume = {488},
	doi = {10.1093/mnras/stz2144},
	number = {4},
	urldate = {2025-07-07},
	journal = {Monthly Notices of the Royal Astronomical Society},
	author = {Collier, Angela and Heller, Clayton and Shlosman, Isaac},
	month = aug,
	year = {2019},
	pages = {5788--5801},
}

@article{Chen_2025,
doi = {10.3847/1538-4357/ae13a6},
url = {https://doi.org/10.3847/1538-4357/ae13a6},
year = {2025},
month = {nov},
publisher = {The American Astronomical Society},
volume = {994},
number = {1},
pages = {124},
author = {Chen, Bin-Hui and Kataria, Sandeep Kumar and Shen, Juntai and Guo, Meng},
title = {Dependency of the Bar Formation Timescale on the Halo Spin},
journal = {The Astrophysical Journal}
}

@ARTICLE{Ansar.et.al.2023,
       author = {{Ansar}, Sioree and {Kataria}, Sandeep Kumar and {Das}, Mousumi},
        title = "{Modelling dark matter halo spin using observations and simulations: application to UGC 5288}",
      journal = {\mnras},
         year = 2023,
        month = jun,
       volume = {522},
       number = {2},
        pages = {2967-2994},
          doi = {10.1093/mnras/stad1060},
archivePrefix = {arXiv},
       eprint = {2304.00724},
 primaryClass = {astro-ph.GA},
       adsurl = {https://ui.adsabs.harvard.edu/abs/2023MNRAS.522.2967A}
}

@ARTICLE{Kataria_etal_2020,
       author = {{Kataria}, Sandeep Kumar and {Das}, Mousumi and {Barway}, Sudhanshu},
        title = "{Testing a theoretical prediction for bar formation in galaxies with bulges}",
      journal = {\aap},
         year = 2020,
        month = aug,
       volume = {640},
          eid = {A14},
        pages = {A14},
          doi = {10.1051/0004-6361/202037527},
archivePrefix = {arXiv},
       eprint = {2006.05870},
 primaryClass = {astro-ph.GA},
       adsurl = {https://ui.adsabs.harvard.edu/abs/2020A&A...640A..14K}
}

@ARTICLE{Kataria.Das.2018,
       author = {{Kataria}, Sandeep Kumar and {Das}, Mousumi},
        title = "{A study of the effect of bulges on bar formation in disc galaxies}",
      journal = {\mnras},
         year = 2018,
        month = apr,
       volume = {475},
       number = {2},
        pages = {1653-1664},
          doi = {10.1093/mnras/stx3279},
       adsurl = {https://ui.adsabs.harvard.edu/abs/2018MNRAS.475.1653K}
}

@ARTICLE{KumarA.et.al.2022,
       author = {{Kumar}, Ankit and {Das}, Mousumi and {Kataria}, Sandeep Kumar},
        title = "{The effect of dark matter halo shape on bar buckling and boxy/peanut bulges}",
      journal = {\mnras},
         year = 2022,
        month = jan,
       volume = {509},
       number = {1},
        pages = {1262-1268},
          doi = {10.1093/mnras/stab3019},
archivePrefix = {arXiv},
       eprint = {2110.08165},
 primaryClass = {astro-ph.GA},
       adsurl = {https://ui.adsabs.harvard.edu/abs/2022MNRAS.509.1262K}
}

@ARTICLE{Kataria.Das.2019,
       author = {{Kataria}, Sandeep Kumar and {Das}, Mousumi},
        title = "{The Effect of Bulge Mass on Bar Pattern Speed in Disk Galaxies}",
      journal = {\apj},
         year = 2019,
        month = nov,
       volume = {886},
       number = {1},
          eid = {43},
        pages = {43},
          doi = {10.3847/1538-4357/ab48f7},
archivePrefix = {arXiv},
       eprint = {1910.03967},
 primaryClass = {astro-ph.GA},
       adsurl = {https://ui.adsabs.harvard.edu/abs/2019ApJ...886...43K}
}

@ARTICLE{Kataria.2024,
       author = {{Kataria}, Sandeep Kumar},
        title = "{How do the successive buckling events affect a galaxy bar and stellar disc? Potential observable signatures for spotting the buckling action - I}",
      journal = {\mnras},
         year = 2024,
        month = nov,
       volume = {534},
       number = {4},
        pages = {3565-3575},
          doi = {10.1093/mnras/stae2311},
archivePrefix = {arXiv},
       eprint = {2407.04113},
 primaryClass = {astro-ph.GA},
       adsurl = {https://ui.adsabs.harvard.edu/abs/2024MNRAS.534.3565K}
}

@article{Kumar.Kataria.2022,
    author = {Kumar, Ankit and Kataria, Sandeep Kumar},
    title = {Growth of disc-like pseudo-bulges in SDSS DR7 since z = 0.1},
    journal = {Monthly Notices of the Royal Astronomical Society},
    volume = {514},
    number = {2},
    pages = {2497-2512},
    year = {2022},
    month = {05},
    issn = {0035-8711},
    doi = {10.1093/mnras/stac1487},
    url = {https://doi.org/10.1093/mnras/stac1487},
    eprint = {https://academic.oup.com/mnras/article-pdf/514/2/2497/44137472/stac1487.pdf},
}

@ARTICLE{Kataria.Vivek.2024,
       author = {{Kataria}, Sandeep Kumar and {Vivek}, M.},
        title = "{How does the presence of bar affects the fueling of supermassive black holes? An IllustrisTNG100 perspective}",
      journal = {\mnras},
         year = 2024,
        month = jan,
       volume = {527},
       number = {2},
        pages = {3366-3380},
          doi = {10.1093/mnras/stad3383},
archivePrefix = {arXiv},
       eprint = {2311.00040},
 primaryClass = {astro-ph.GA},
       adsurl = {https://ui.adsabs.harvard.edu/abs/2024MNRAS.527.3366K}
}

@ARTICLE{2015MNRAS.446..521S,
       author = {{Schaye}, Joop and {Crain}, Robert A. and {Bower}, Richard G. and {Furlong}, Michelle and {Schaller}, Matthieu and {Theuns}, Tom and {Dalla Vecchia}, Claudio and {Frenk}, Carlos S. and {McCarthy}, I.~G. and {Helly}, John C. and {Jenkins}, Adrian and {Rosas-Guevara}, Y.~M. and {White}, Simon D.~M. and {Baes}, Maarten and {Booth}, C.~M. and {Camps}, Peter and {Navarro}, Julio F. and {Qu}, Yan and {Rahmati}, Alireza and {Sawala}, Till and {Thomas}, Peter A. and {Trayford}, James},
        title = "{The EAGLE project: simulating the evolution and assembly of galaxies and their environments}",
      journal = {\mnras},
         year = 2015,
        month = jan,
       volume = {446},
       number = {1},
        pages = {521-554},
          doi = {10.1093/mnras/stu2058},
archivePrefix = {arXiv},
       eprint = {1407.7040},
 primaryClass = {astro-ph.GA},
       adsurl = {https://ui.adsabs.harvard.edu/abs/2015MNRAS.446..521S}
}

@ARTICLE{2005Natur.435..629S,
       author = {{Springel}, Volker and {White}, Simon D.~M. and {Jenkins}, Adrian and {Frenk}, Carlos S. and {Yoshida}, Naoki and {Gao}, Liang and {Navarro}, Julio and {Thacker}, Robert and {Croton}, Darren and {Helly}, John and {Peacock}, John A. and {Cole}, Shaun and {Thomas}, Peter and {Couchman}, Hugh and {Evrard}, August and {Colberg}, J{\"o}rg and {Pearce}, Frazer},
        title = "{Simulations of the formation, evolution and clustering of galaxies and quasars}",
      journal = {\nat},
         year = 2005,
        month = jun,
       volume = {435},
       number = {7042},
        pages = {629-636},
          doi = {10.1038/nature03597},
archivePrefix = {arXiv},
       eprint = {astro-ph/0504097},
 primaryClass = {astro-ph},
       adsurl = {https://ui.adsabs.harvard.edu/abs/2005Natur.435..629S}
}

@article{Long2014,
  title = {SECULAR DAMPING OF STELLAR BARS IN SPINNING DARK MATTER HALOS},
  volume = {783},
  ISSN = {2041-8213},
  url = {http://dx.doi.org/10.1088/2041-8205/783/1/L18},
  DOI = {10.1088/2041-8205/783/1/l18},
  number = {1},
  journal = {The Astrophysical Journal},
  publisher = {American Astronomical Society},
  author = {Long,  Stacy and Shlosman,  Isaac and Heller,  Clayton},
  year = {2014},
  month = feb,
  pages = {L18}
}

@article{Collier2018,
  title = {What makes the family of barred disc galaxies so rich: damping stellar bars in spinning haloes},
  volume = {476},
  ISSN = {1365-2966},
  url = {http://dx.doi.org/10.1093/mnras/sty270},
  DOI = {10.1093/mnras/sty270},
  number = {1},
  journal = {Monthly Notices of the Royal Astronomical Society},
  publisher = {Oxford University Press (OUP)},
  author = {Collier,  Angela and Shlosman,  Isaac and Heller,  Clayton},
  year = {2018},
  month = feb,
  pages = {1331–1344}
}

@ARTICLE{2006ApJ...648..807B,
       author = {{Berentzen}, Ingo and {Shlosman}, Isaac},
        title = "{Growing Live Disks within Cosmologically Assembling Asymmetric Halos: Washing Out the Halo Prolateness}",
      journal = {\apj},
         year = 2006,
        month = sep,
       volume = {648},
       number = {2},
        pages = {807-819},
          doi = {10.1086/506016},
archivePrefix = {arXiv},
       eprint = {astro-ph/0603487},
 primaryClass = {astro-ph},
       adsurl = {https://ui.adsabs.harvard.edu/abs/2006ApJ...648..807B}
}

@ARTICLE{2013MNRAS.429.1949A,
       author = {{Athanassoula}, E. and {Machado}, Rubens E.~G. and {Rodionov}, S.~A.},
        title = "{Bar formation and evolution in disc galaxies with gas and a triaxial halo: morphology, bar strength and halo properties}",
      journal = {\mnras},
         year = 2013,
        month = mar,
       volume = {429},
       number = {3},
        pages = {1949-1969},
          doi = {10.1093/mnras/sts452},
archivePrefix = {arXiv},
       eprint = {1211.6754},
 primaryClass = {astro-ph.CO},
       adsurl = {https://ui.adsabs.harvard.edu/abs/2013MNRAS.429.1949A}
}

@ARTICLE{2007MNRAS.377.1569A,
       author = {{Athanassoula}, E.},
        title = "{A bar in the inner halo of barred galaxies - I. Structure and kinematics of a representative model}",
      journal = {\mnras},
         year = 2007,
        month = jun,
       volume = {377},
       number = {4},
        pages = {1569-1578},
          doi = {10.1111/j.1365-2966.2007.11711.x},
archivePrefix = {arXiv},
       eprint = {astro-ph/0703184},
 primaryClass = {astro-ph},
       adsurl = {https://ui.adsabs.harvard.edu/abs/2007MNRAS.377.1569A}
}

@ARTICLE{2005NYASA1045..168A,
       author = {{Athanassoula}, E.},
        title = "{On Bars and Haloes: Their Interaction and Their Orbital Structure}",
      journal = {Annals of the New York Academy of Sciences},
         year = 2005,
        month = jan,
       volume = {1045},
        pages = {168},
          doi = {10.1196/annals.1350.013},
archivePrefix = {arXiv},
       eprint = {astro-ph/0504664},
 primaryClass = {astro-ph},
       adsurl = {https://ui.adsabs.harvard.edu/abs/2005NYASA1045..168A}
}

@ARTICLE{2006ApJ...644..687C,
       author = {{Col{\'\i}n}, Pedro and {Valenzuela}, O. and {Klypin}, A.},
        title = "{Bars and Cold Dark Matter Halos}",
      journal = {\apj},
         year = 2006,
        month = jun,
       volume = {644},
       number = {2},
        pages = {687-700},
          doi = {10.1086/503791},
archivePrefix = {arXiv},
       eprint = {astro-ph/0506627},
 primaryClass = {astro-ph},
       adsurl = {https://ui.adsabs.harvard.edu/abs/2006ApJ...644..687C}
}

@ARTICLE{2021ApJ...915...23C,
       author = {{Collier}, Angela and {Madigan}, Ann-Marie},
        title = "{The Coupling of Galactic Dark Matter Halos with Stellar Bars}",
      journal = {\apj},
         year = 2021,
        month = jul,
       volume = {915},
       number = {1},
          eid = {23},
        pages = {23},
          doi = {10.3847/1538-4357/ac004d},
archivePrefix = {arXiv},
       eprint = {2105.04698},
 primaryClass = {astro-ph.GA},
       adsurl = {https://ui.adsabs.harvard.edu/abs/2021ApJ...915...23C}
}

@ARTICLE{2024Galax..12...27M,
       author = {{Marostica}, Daniel A. and {Machado}, Rubens E.~G. and {Athanassoula}, E. and {Manos}, T.},
        title = "{The Response of the Inner Dark Matter Halo to Stellar Bars}",
      journal = {Galaxies},
         year = 2024,
        month = may,
       volume = {12},
       number = {3},
          eid = {27},
        pages = {27},
          doi = {10.3390/galaxies12030027},
archivePrefix = {arXiv},
       eprint = {2405.17128},
 primaryClass = {astro-ph.GA},
       adsurl = {https://ui.adsabs.harvard.edu/abs/2024Galax..12...27M}
}

@article{Saha2013,
  title = {Spinning dark matter haloes promote bar formation},
  volume = {434},
  ISSN = {1365-2966},
  url = {http://dx.doi.org/10.1093/mnras/stt1088},
  DOI = {10.1093/mnras/stt1088},
  number = {2},
  journal = {Monthly Notices of the Royal Astronomical Society},
  publisher = {Oxford University Press (OUP)},
  author = {Saha,  Kanak and Naab,  Thorsten},
  year = {2013},
  month = jul,
  pages = {1287–1299}
}

@article{VillaVargas2009,
  title = {DARK MATTER HALOS AND EVOLUTION OF BARS IN DISK GALAXIES: COLLISIONLESS MODELS REVISITED},
  volume = {707},
  ISSN = {1538-4357},
  url = {http://dx.doi.org/10.1088/0004-637X/707/1/218},
  DOI = {10.1088/0004-637x/707/1/218},
  number = {1},
  journal = {The Astrophysical Journal},
  publisher = {American Astronomical Society},
  author = {Villa-Vargas,  Jorge and Shlosman,  Isaac and Heller,  Clayton},
  year = {2009},
  month = nov,
  pages = {218–232}
}

@article{VillaVargas2010,
  title = {DARK MATTER HALOS AND EVOLUTION OF BARS IN DISK GALAXIES: VARYING GAS FRACTION AND GAS SPATIAL RESOLUTION},
  volume = {719},
  ISSN = {1538-4357},
  url = {http://dx.doi.org/10.1088/0004-637X/719/2/1470},
  DOI = {10.1088/0004-637x/719/2/1470},
  number = {2},
  journal = {The Astrophysical Journal},
  publisher = {American Astronomical Society},
  author = {Villa-Vargas,  Jorge and Shlosman,  Isaac and Heller,  Clayton},
  year = {2010},
  month = jul,
  pages = {1470–1480}
}

@article{Athanassoula2003,
  title = {What determines the strength and the slowdown rate of bars?},
  volume = {341},
  ISSN = {1365-2966},
  url = {http://dx.doi.org/10.1046/j.1365-8711.2003.06473.x},
  DOI = {10.1046/j.1365-8711.2003.06473.x},
  number = {4},
  journal = {Monthly Notices of the Royal Astronomical Society},
  publisher = {Oxford University Press (OUP)},
  author = {Athanassoula,  E.},
  year = {2003},
  month = jun,
  pages = {1179–1198}
}

@article{MartinezValpuesta2006,
  title = {Evolution of Stellar Bars in Live Axisymmetric Halos: Recurrent Buckling and Secular Growth},
  volume = {637},
  ISSN = {1538-4357},
  url = {http://dx.doi.org/10.1086/498338},
  DOI = {10.1086/498338},
  number = {1},
  journal = {The Astrophysical Journal},
  publisher = {American Astronomical Society},
  author = {Martinez‐Valpuesta,  Inma and Shlosman,  Isaac and Heller,  Clayton},
  year = {2006},
  month = jan,
  pages = {214–226}
}

@article{Berentzen2007,
  title = {Gas Feedback on Stellar Bar Evolution},
  volume = {666},
  ISSN = {1538-4357},
  url = {http://dx.doi.org/10.1086/520531},
  DOI = {10.1086/520531},
  number = {1},
  journal = {The Astrophysical Journal},
  publisher = {American Astronomical Society},
  author = {Berentzen,  Ingo and Shlosman,  Isaac and Martinez‐Valpuesta,  Inma and Heller,  Clayton H.},
  year = {2007},
  month = sep,
  pages = {189–200}
}

@article{Ash2024,
  title = {Stellar Bars Form Dark Matter Counterparts in TNG50},
  volume = {976},
  ISSN = {1538-4357},
  url = {http://dx.doi.org/10.3847/1538-4357/ad863a},
  DOI = {10.3847/1538-4357/ad863a},
  number = {2},
  journal = {The Astrophysical Journal},
  publisher = {American Astronomical Society},
  author = {Ash,  Neil and Valluri,  Monica and Chen,  Yingtian and Bell,  Eric F.},
  year = {2024},
  month = nov,
  pages = {189}
}

@article{Athanassoula2005,
  title = {On the nature of bulges in general and of box/peanut bulges in particular: input fromN-body simulations},
  volume = {358},
  ISSN = {1365-2966},
  url = {http://dx.doi.org/10.1111/j.1365-2966.2005.08872.x},
  DOI = {10.1111/j.1365-2966.2005.08872.x},
  number = {4},
  journal = {Monthly Notices of the Royal Astronomical Society},
  publisher = {Oxford University Press (OUP)},
  author = {Athanassoula,  E.},
  year = {2005},
  month = apr,
  pages = {1477–1488}
}

@article{Dubinski2009,
  title = {ANATOMY OF THE BAR INSTABILITY IN CUSPY DARK MATTER HALOS},
  volume = {697},
  ISSN = {1538-4357},
  url = {http://dx.doi.org/10.1088/0004-637X/697/1/293},
  DOI = {10.1088/0004-637x/697/1/293},
  number = {1},
  journal = {The Astrophysical Journal},
  publisher = {American Astronomical Society},
  author = {Dubinski,  John and Berentzen,  Ingo and Shlosman,  Isaac},
  year = {2009},
  month = may,
  pages = {293–310}
}

@ARTICLE{2001ApJ...555..240B,
       author = {{Bullock}, J.~S. and {Dekel}, A. and {Kolatt}, T.~S. and {Kravtsov}, A.~V. and {Klypin}, A.~A. and {Porciani}, C. and {Primack}, J.~R.},
        title = "{A Universal Angular Momentum Profile for Galactic Halos}",
      journal = {\apj},
         year = 2001,
        month = jul,
       volume = {555},
       number = {1},
        pages = {240-257},
          doi = {10.1086/321477},
archivePrefix = {arXiv},
       eprint = {astro-ph/0011001},
 primaryClass = {astro-ph},
       adsurl = {https://ui.adsabs.harvard.edu/abs/2001ApJ...555..240B}
}

@article{Eskridge2000,
  title = {The Frequency of Barred Spiral Galaxies in the Near-Infrared},
  volume = {119},
  ISSN = {0004-6256},
  url = {http://dx.doi.org/10.1086/301203},
  DOI = {10.1086/301203},
  number = {2},
  journal = {The Astronomical Journal},
  publisher = {American Astronomical Society},
  author = {Eskridge,  Paul B. and Frogel,  Jay A. and Pogge,  Richard W. and Quillen,  Alice C. and Davies,  Roger L. and DePoy,  D. L. and Houdashelt,  Mark L. and Kuchinski,  Leslie E. and Ramírez,  Solange V. and Sellgren,  K. and Terndrup,  Donald M. and Tiede,  Glenn P.},
  year = {2000},
  month = feb,
  pages = {536–544}
}

@ARTICLE{2008ApJ...675.1194B,
       author = {{Barazza}, Fabio D. and {Jogee}, Shardha and {Marinova}, Irina},
        title = "{Bars in Disk-dominated and Bulge-dominated Galaxies at z \raisebox{-0.5ex}\textasciitilde 0: New Insights from \raisebox{-0.5ex}\textasciitilde3600 SDSS Galaxies}",
      journal = {\apj},
         year = 2008,
        month = mar,
       volume = {675},
       number = {2},
        pages = {1194-1212},
          doi = {10.1086/526510},
archivePrefix = {arXiv},
       eprint = {0710.4674},
 primaryClass = {astro-ph},
       adsurl = {https://ui.adsabs.harvard.edu/abs/2008ApJ...675.1194B}
}

@ARTICLE{2009A&A...495..491A,
       author = {{Aguerri}, J.~A.~L. and {M{\'e}ndez-Abreu}, J. and {Corsini}, E.~M.},
        title = "{The population of barred galaxies in the local universe. I. Detection and characterisation of bars}",
      journal = {\aap},
         year = 2009,
        month = feb,
       volume = {495},
       number = {2},
        pages = {491-504},
          doi = {10.1051/0004-6361:200810931},
archivePrefix = {arXiv},
       eprint = {0901.2346},
 primaryClass = {astro-ph.GA},
       adsurl = {https://ui.adsabs.harvard.edu/abs/2009A&A...495..491A}
}

@ARTICLE{2010ApJ...714L.260N,
       author = {{Nair}, Preethi B. and {Abraham}, Roberto G.},
        title = "{On the Fraction of Barred Spiral Galaxies}",
      journal = {\apjl},
         year = 2010,
        month = may,
       volume = {714},
       number = {2},
        pages = {L260-L264},
          doi = {10.1088/2041-8205/714/2/L260},
archivePrefix = {arXiv},
       eprint = {1004.0684},
 primaryClass = {astro-ph.CO},
       adsurl = {https://ui.adsabs.harvard.edu/abs/2010ApJ...714L.260N}
}

@ARTICLE{2015ApJS..217...32B,
       author = {{Buta}, Ronald J. and {Sheth}, Kartik and {Athanassoula}, E. and {Bosma}, A. and {Knapen}, Johan H. and {Laurikainen}, Eija and {Salo}, Heikki and {Elmegreen}, Debra and {Ho}, Luis C. and {Zaritsky}, Dennis and {Courtois}, Helene and {Hinz}, Joannah L. and {Mu{\~n}oz-Mateos}, Juan-Carlos and {Kim}, Taehyun and {Regan}, Michael W. and {Gadotti}, Dimitri A. and {Gil de Paz}, Armando and {Laine}, Jarkko and {Men{\'e}ndez-Delmestre}, Kar{\'\i}n and {Comer{\'o}n}, S{\'e}bastien and {Erroz Ferrer}, Santiago and {Seibert}, Mark and {Mizusawa}, Trisha and {Holwerda}, Benne and {Madore}, Barry F.},
        title = "{A Classical Morphological Analysis of Galaxies in the Spitzer Survey of Stellar Structure in Galaxies (S4G)}",
      journal = {\apjs},
         year = 2015,
        month = apr,
       volume = {217},
       number = {2},
          eid = {32},
        pages = {32},
          doi = {10.1088/0067-0049/217/2/32},
archivePrefix = {arXiv},
       eprint = {1501.00454},
 primaryClass = {astro-ph.GA},
       adsurl = {https://ui.adsabs.harvard.edu/abs/2015ApJS..217...32B}
}

@ARTICLE{2018MNRAS.474.5372E,
       author = {{Erwin}, Peter},
        title = "{The dependence of bar frequency on galaxy mass, colour, and gas content - and angular resolution - in the local universe}",
      journal = {\mnras},
         year = 2018,
        month = mar,
       volume = {474},
       number = {4},
        pages = {5372-5392},
          doi = {10.1093/mnras/stx3117},
archivePrefix = {arXiv},
       eprint = {1711.04867},
 primaryClass = {astro-ph.GA},
       adsurl = {https://ui.adsabs.harvard.edu/abs/2018MNRAS.474.5372E}
}

@ARTICLE{2003MNRAS.341.1179A,
       author = {{Athanassoula}, E.},
        title = "{What determines the strength and the slowdown rate of bars?}",
      journal = {\mnras},
         year = 2003,
        month = jun,
       volume = {341},
       number = {4},
        pages = {1179-1198},
          doi = {10.1046/j.1365-8711.2003.06473.x},
archivePrefix = {arXiv},
       eprint = {astro-ph/0302519},
 primaryClass = {astro-ph},
       adsurl = {https://ui.adsabs.harvard.edu/abs/2003MNRAS.341.1179A}
}

@ARTICLE{2005ApJ...632..217S,
       author = {{Sheth}, Kartik and {Vogel}, Stuart N. and {Regan}, Michael W. and {Thornley}, Michele D. and {Teuben}, Peter J.},
        title = "{Secular Evolution via Bar-driven Gas Inflow: Results from BIMA SONG}",
      journal = {\apj},
         year = 2005,
        month = oct,
       volume = {632},
       number = {1},
        pages = {217-226},
          doi = {10.1086/432409},
archivePrefix = {arXiv},
       eprint = {astro-ph/0505393},
 primaryClass = {astro-ph},
       adsurl = {https://ui.adsabs.harvard.edu/abs/2005ApJ...632..217S}
}

@ARTICLE{2006ApJ...645..209D,
       author = {{Debattista}, Victor P. and {Mayer}, Lucio and {Carollo}, C. Marcella and {Moore}, Ben and {Wadsley}, James and {Quinn}, Thomas},
        title = "{The Secular Evolution of Disk Structural Parameters}",
      journal = {\apj},
         year = 2006,
        month = jul,
       volume = {645},
       number = {1},
        pages = {209-227},
          doi = {10.1086/504147},
archivePrefix = {arXiv},
       eprint = {astro-ph/0509310},
 primaryClass = {astro-ph},
       adsurl = {https://ui.adsabs.harvard.edu/abs/2006ApJ...645..209D}
}

@inbook{Athanassoula_2013, place={Cambridge}, series={Canary Islands Winter School of Astrophysics}, title={Bars and secular evolution in disk galaxies: Theoretical input}, booktitle={Secular Evolution of Galaxies}, publisher={Cambridge University Press}, author={Athanassoula, E.}, editor={Falcón-Barroso, Jesús and Knapen, Johan H.Editors}, year={2013}, pages={305–352}, collection={Canary Islands Winter School of Astrophysics}}

@ARTICLE{2013A&A...553A.102D,
       author = {{Di Matteo}, P. and {Haywood}, M. and {Combes}, F. and {Semelin}, B. and {Snaith}, O.~N.},
        title = "{Signatures of radial migration in barred galaxies: Azimuthal variations in the metallicity distribution of old stars}",
      journal = {\aap},
         year = 2013,
        month = may,
       volume = {553},
          eid = {A102},
        pages = {A102},
          doi = {10.1051/0004-6361/201220539},
archivePrefix = {arXiv},
       eprint = {1301.2545},
 primaryClass = {astro-ph.GA},
       adsurl = {https://ui.adsabs.harvard.edu/abs/2013A&A...553A.102D}
}

@ARTICLE{2014RvMP...86....1S,
       author = {{Sellwood}, J.~A.},
        title = "{Secular evolution in disk galaxies}",
      journal = {Reviews of Modern Physics},
         year = 2014,
        month = jan,
       volume = {86},
       number = {1},
        pages = {1-46},
          doi = {10.1103/RevModPhys.86.1},
archivePrefix = {arXiv},
       eprint = {1310.0403},
 primaryClass = {astro-ph.GA},
       adsurl = {https://ui.adsabs.harvard.edu/abs/2014RvMP...86....1S}
}

@ARTICLE{2023MNRAS.518.1002R,
       author = {{Romeo}, Alessandro B. and {Agertz}, Oscar and {Renaud}, Florent},
        title = "{The specific angular momentum of disc galaxies and its connection with galaxy morphology, bar structure, and disc gravitational instability}",
      journal = {\mnras},
         year = 2023,
        month = jan,
       volume = {518},
       number = {1},
        pages = {1002-1021},
          doi = {10.1093/mnras/stac3074},
archivePrefix = {arXiv},
       eprint = {2204.02695},
 primaryClass = {astro-ph.GA},
       adsurl = {https://ui.adsabs.harvard.edu/abs/2023MNRAS.518.1002R}
}

@ARTICLE{1969JCoPh...4..306H,
       author = {{Hohl}, Frank and {Hockney}, R.~W.},
        title = "{A Computer Model of Disks of Stars}",
      journal = {Journal of Computational Physics},
         year = 1969,
        month = oct,
       volume = {4},
        pages = {306},
          doi = {10.1016/0021-9991(69)90002-3},
       adsurl = {https://ui.adsabs.harvard.edu/abs/1969JCoPh...4..306H}
}

@ARTICLE{1970ApJ...161..903M,
       author = {{Miller}, R.~H. and {Prendergast}, K.~H. and {Quirk}, William J.},
        title = "{Numerical Experiments on Spiral Structure}",
      journal = {\apj},
         year = 1970,
        month = sep,
       volume = {161},
        pages = {903},
          doi = {10.1086/150593},
       adsurl = {https://ui.adsabs.harvard.edu/abs/1970ApJ...161..903M}
}

@ARTICLE{1981A&A....96..164C,
       author = {{Combes}, F. and {Sanders}, R.~H.},
        title = "{Formation and properties of persisting stellar bars.}",
      journal = {\aap},
         year = 1981,
        month = mar,
       volume = {96},
        pages = {164-173},
       adsurl = {https://ui.adsabs.harvard.edu/abs/1981A&A....96..164C}
}

@ARTICLE{1992ApJ...400...80H,
       author = {{Hernquist}, Lars and {Weinberg}, Martin D.},
        title = "{Bar-Spheroid Interaction in Galaxies}",
      journal = {\apj},
         year = 1992,
        month = nov,
       volume = {400},
        pages = {80},
          doi = {10.1086/171975},
       adsurl = {https://ui.adsabs.harvard.edu/abs/1992ApJ...400...80H}
}

@ARTICLE{2000ApJ...543..704D,
       author = {{Debattista}, Victor P. and {Sellwood}, J.~A.},
        title = "{Constraints from Dynamical Friction on the Dark Matter Content of Barred Galaxies}",
      journal = {\apj},
         year = 2000,
        month = nov,
       volume = {543},
       number = {2},
        pages = {704-721},
          doi = {10.1086/317148},
archivePrefix = {arXiv},
       eprint = {astro-ph/0006275},
 primaryClass = {astro-ph},
       adsurl = {https://ui.adsabs.harvard.edu/abs/2000ApJ...543..704D}
}

@ARTICLE{2002ApJ...569L..83A,
       author = {{Athanassoula}, E.},
        title = "{Bar-Halo Interaction and Bar Growth}",
      journal = {\apjl},
         year = 2002,
        month = apr,
       volume = {569},
       number = {2},
        pages = {L83-L86},
          doi = {10.1086/340784},
archivePrefix = {arXiv},
       eprint = {astro-ph/0203368},
 primaryClass = {astro-ph},
       adsurl = {https://ui.adsabs.harvard.edu/abs/2002ApJ...569L..83A}
}

@ARTICLE{2002MNRAS.330...35A,
       author = {{Athanassoula}, E. and {Misiriotis}, A.},
        title = "{Morphology, photometry and kinematics of N -body bars - I. Three models with different halo central concentrations}",
      journal = {\mnras},
         year = 2002,
        month = feb,
       volume = {330},
       number = {1},
        pages = {35-52},
          doi = {10.1046/j.1365-8711.2002.05028.x},
archivePrefix = {arXiv},
       eprint = {astro-ph/0111449},
 primaryClass = {astro-ph},
       adsurl = {https://ui.adsabs.harvard.edu/abs/2002MNRAS.330...35A}
}

@ARTICLE{2003MNRAS.346..251O,
       author = {{O'Neill}, J.~K. and {Dubinski}, John},
        title = "{Detailed comparison of the structures and kinematics of simulated and observed barred galaxies}",
      journal = {\mnras},
         year = 2003,
        month = nov,
       volume = {346},
       number = {1},
        pages = {251-264},
          doi = {10.1046/j.1365-2966.2003.07085.x},
archivePrefix = {arXiv},
       eprint = {astro-ph/0305169},
 primaryClass = {astro-ph},
       adsurl = {https://ui.adsabs.harvard.edu/abs/2003MNRAS.346..251O}
}

@ARTICLE{2005MNRAS.363..991H,
       author = {{Holley-Bockelmann}, K. and {Weinberg}, M. and {Katz}, N.},
        title = "{Bar-induced evolution of dark matter cusps}",
      journal = {\mnras},
         year = 2005,
        month = nov,
       volume = {363},
       number = {3},
        pages = {991-1007},
          doi = {10.1111/j.1365-2966.2005.09501.x},
archivePrefix = {arXiv},
       eprint = {astro-ph/0306374},
 primaryClass = {astro-ph},
       adsurl = {https://ui.adsabs.harvard.edu/abs/2005MNRAS.363..991H}
}

@ARTICLE{2006ApJ...637..214M,
       author = {{Martinez-Valpuesta}, Inma and {Shlosman}, Isaac and {Heller}, Clayton},
        title = "{Evolution of Stellar Bars in Live Axisymmetric Halos: Recurrent Buckling and Secular Growth}",
      journal = {\apj},
         year = 2006,
        month = jan,
       volume = {637},
       number = {1},
        pages = {214-226},
          doi = {10.1086/498338},
archivePrefix = {arXiv},
       eprint = {astro-ph/0507219},
 primaryClass = {astro-ph},
       adsurl = {https://ui.adsabs.harvard.edu/abs/2006ApJ...637..214M}
}

@ARTICLE{2007MNRAS.375..460W,
       author = {{Weinberg}, Martin D. and {Katz}, Neal},
        title = "{The bar-halo interaction - II. Secular evolution and the religion of N-body simulations}",
      journal = {\mnras},
         year = 2007,
        month = feb,
       volume = {375},
       number = {2},
        pages = {460-476},
          doi = {10.1111/j.1365-2966.2006.11307.x},
archivePrefix = {arXiv},
       eprint = {astro-ph/0601138},
 primaryClass = {astro-ph},
       adsurl = {https://ui.adsabs.harvard.edu/abs/2007MNRAS.375..460W}
}

@ARTICLE{2009ApJ...697..293D,
       author = {{Dubinski}, John and {Berentzen}, Ingo and {Shlosman}, Isaac},
        title = "{Anatomy of the Bar Instability in Cuspy Dark Matter Halos}",
      journal = {\apj},
         year = 2009,
        month = may,
       volume = {697},
       number = {1},
        pages = {293-310},
          doi = {10.1088/0004-637X/697/1/293},
archivePrefix = {arXiv},
       eprint = {0810.4925},
 primaryClass = {astro-ph},
       adsurl = {https://ui.adsabs.harvard.edu/abs/2009ApJ...697..293D}
}

@ARTICLE{2005MNRAS.361..776S,
       author = {{Springel}, Volker and {Di Matteo}, Tiziana and {Hernquist}, Lars},
        title = "{Modelling feedback from stars and black holes in galaxy mergers}",
      journal = {\mnras},
         year = 2005,
        month = aug,
       volume = {361},
       number = {3},
        pages = {776-794},
          doi = {10.1111/j.1365-2966.2005.09238.x},
archivePrefix = {arXiv},
       eprint = {astro-ph/0411108},
 primaryClass = {astro-ph},
       adsurl = {https://ui.adsabs.harvard.edu/abs/2005MNRAS.361..776S}
}

@ARTICLE{Chiba2024-bh,
  title     = "Origin of reduced dynamical friction by dark matter haloes with
               net prograde rotation",
  author    = "Chiba, Rimpei and Kataria, Sandeep Kumar",
  journal   = "Mon. Not. R. Astron. Soc.",
  publisher = "Oxford University Press (OUP)",
  volume    =  528,
  number    =  3,
  pages     = "4115--4124",
  month     =  feb,
  year      =  2024,
  copyright = "https://creativecommons.org/licenses/by/4.0/",
  language  = "en"
}

@article{Kataria2024,
  title = {Importance of Initial Condition on Bar Secular Evolution: Role of Halo Angular Momentum Distribution Discontinuity},
  volume = {970},
  ISSN = {1538-4357},
  url = {http://dx.doi.org/10.3847/1538-4357/ad5b58},
  DOI = {10.3847/1538-4357/ad5b58},
  number = {1},
  journal = {The Astrophysical Journal},
  publisher = {American Astronomical Society},
  author = {Kataria,  Sandeep Kumar and Shen,  Juntai},
  year = {2024},
  month = jul,
  pages = {45}
}

@misc{Kataria2025,
  doi = {10.48550/ARXIV.2512.21632},
  url = {https://arxiv.org/abs/2512.21632},
  author = {Kataria,  Sandeep Kumar},
  title = {Can A Kinematically Hot and Thick Disk Form A Bar? : Role of Highly Spinning Dark Matter Halos},
  publisher = {arXiv},
  year = {2025},
  copyright = {arXiv.org perpetual,  non-exclusive license}
}

@misc{ansar2024stellarbardark,
      title={The stellar bar - dark matter halo connection in the TNG50 simulations}, 
      author={Sioree Ansar and Mousumi Das},
      year={2024},
      eprint={2311.11998},
      archivePrefix={arXiv},
      primaryClass={astro-ph.GA},
      url={https://arxiv.org/abs/2311.11998}, 
}

@article{leconte2024_dxfapp,
  title={A <i>JWST</i> investigation into the bar fraction at redshifts 1 ≤ <i>z</i> ≤ 3},
  author={Le Conte, Zoe A and Gadotti, Dimitri A and Ferreira, Leonardo and Conselice, Christopher J and de Sá-Freitas, Camila and Kim, Taehyun and Neumann, Justus and Fragkoudi, Francesca and Athanassoula, E and Adams, Nathan J},
  journal={Monthly Notices of the Royal Astronomical Society},
  year={2024},
  volume={530},
  number={2},
  pages={1984-2000},
  doi={10.1093/mnras/stae921},
  publisher={Oxford University Press (OUP)},
  url={https://doi.org/10.1093/mnras/stae921}
}

@article{Tahmasebzadeh_2024,
doi = {10.3847/1538-4357/ad77c8},
url = {https://doi.org/10.3847/1538-4357/ad77c8},
year = {2024},
month = {oct},
publisher = {The American Astronomical Society},
volume = {975},
number = {1},
pages = {120},
author = {Tahmasebzadeh, Behzad and Dattathri, Shashank and Valluri, Monica and Shen, Juntai and Zhu, Ling and Wheeler, Vance and Gerhard, Ortwin and Kataria, Sandeep Kumar and Beraldo e Silva, Leandro and Daniel, Kathryne J.},
title = {Orbital Support and Evolution of CX/OX Structures in Boxy/Peanut Bars},
journal = {The Astrophysical Journal}
}

@article{guo2025_15kw49,
  title={The Abundance and Properties of Barred Galaxies out to <i>z</i> ∼ 4 Using JWST CEERS Data},
  author={Guo, Yuchen and Jogee, Shardha and Wise, Eden and Pritchett, Keith and McGrath, Elizabeth J. and Finkelstein, Steven L. and Iyer, Kartheik G. and Haro, Pablo Arrabal and Bagley, Micaela B. and Dickinson, Mark},
  journal={The Astrophysical Journal},
  year={2025},
  volume={985},
  number={2},
  pages={181},
  doi={10.3847/1538-4357/adc8a7},
  publisher={American Astronomical Society},
  url={https://doi.org/10.3847/1538-4357/adc8a7}
}

@ARTICLE{1991A&A...252...75P,
       author = {{Pfenniger}, D. and {Friedli}, D.},
        title = "{Structure and dynamics of 3D N-body barred galaxies.}",
      journal = {\aap},
         year = 1991,
        month = dec,
       volume = {252},
        pages = {75-93},
       adsurl = {https://ui.adsabs.harvard.edu/abs/1991A&A...252...75P}
}

@ARTICLE{1991Natur.352..411R,
       author = {{Raha}, N. and {Sellwood}, J.~A. and {James}, R.~A. and {Kahn}, F.~D.},
        title = "{A dynamical instability of bars in disk galaxies}",
      journal = {\nat},
         year = 1991,
        month = aug,
       volume = {352},
       number = {6334},
        pages = {411-412},
          doi = {10.1038/352411a0},
       adsurl = {https://ui.adsabs.harvard.edu/abs/1991Natur.352..411R}
}

@ARTICLE{2019A&A...624A..37L,
       author = {{{\L}okas}, Ewa L.},
        title = "{Buckling instability in tidally induced galactic bars}",
      journal = {\aap},
         year = 2019,
        month = apr,
       volume = {624},
          eid = {A37},
        pages = {A37},
          doi = {10.1051/0004-6361/201935011},
archivePrefix = {arXiv},
       eprint = {1902.07103},
 primaryClass = {astro-ph.GA},
       adsurl = {https://ui.adsabs.harvard.edu/abs/2019A&A...624A..37L}
}

@ARTICLE{1990A&A...233...82C,
       author = {{Combes}, F. and {Debbasch}, F. and {Friedli}, D. and {Pfenniger}, D.},
        title = "{Box and peanut shapes generated by stellar bars.}",
      journal = {\aap},
         year = 1990,
        month = jul,
       volume = {233},
        pages = {82},
       adsurl = {https://ui.adsabs.harvard.edu/abs/1990A&A...233...82C}
}

@ARTICLE{1998MNRAS.300...49B,
       author = {{Berentzen}, I. and {Heller}, C.~H. and {Shlosman}, I. and {Fricke}, K.~J.},
        title = "{Gas-driven evolution of stellar orbits in barred galaxies}",
      journal = {\mnras},
         year = 1998,
        month = oct,
       volume = {300},
       number = {1},
        pages = {49-63},
          doi = {10.1046/j.1365-8711.1998.01836.x},
archivePrefix = {arXiv},
       eprint = {astro-ph/9806138},
 primaryClass = {astro-ph},
       adsurl = {https://ui.adsabs.harvard.edu/abs/1998MNRAS.300...49B}
}

@ARTICLE{2002MNRAS.337..578P,
       author = {{Patsis}, P.~A. and {Skokos}, Ch. and {Athanassoula}, E.},
        title = "{Orbital dynamics of three-dimensional bars - III. Boxy/peanut edge-on profiles}",
      journal = {\mnras},
         year = 2002,
        month = dec,
       volume = {337},
       number = {2},
        pages = {578-596},
          doi = {10.1046/j.1365-8711.2002.05943.x},
       adsurl = {https://ui.adsabs.harvard.edu/abs/2002MNRAS.337..578P}
}

@ARTICLE{1986AJ.....91...65J,
       author = {{Jarvis}, B.~J.},
        title = "{A search for box- and peanut-shaped bulges.}",
      journal = {\aj},
         year = 1986,
        month = jan,
       volume = {91},
        pages = {65-69},
          doi = {10.1086/113980},
       adsurl = {https://ui.adsabs.harvard.edu/abs/1986AJ.....91...65J}
}

@ARTICLE{1987MNRAS.229..691S,
       author = {{Shaw}, Martin A.},
        title = "{The nature of 'box' and 'peanut' shaped galactic bulges.}",
      journal = {\mnras},
         year = 1987,
        month = dec,
       volume = {229},
        pages = {691-706},
          doi = {10.1093/mnras/229.4.691},
       adsurl = {https://ui.adsabs.harvard.edu/abs/1987MNRAS.229..691S}
}

@ARTICLE{1999AJ....118..126B,
       author = {{Bureau}, M. and {Freeman}, K.~C.},
        title = "{The Nature of Boxy/Peanut-Shaped Bulges in Spiral Galaxies}",
      journal = {\aj},
         year = 1999,
        month = jul,
       volume = {118},
       number = {1},
        pages = {126-138},
          doi = {10.1086/300922},
archivePrefix = {arXiv},
       eprint = {astro-ph/9904015},
 primaryClass = {astro-ph},
       adsurl = {https://ui.adsabs.harvard.edu/abs/1999AJ....118..126B}
}

@ARTICLE{1999A&A...345L..47M,
       author = {{Merrifield}, Michael R. and {Kuijken}, Konrad},
        title = "{Hidden bars and boxy bulges}",
      journal = {\aap},
         year = 1999,
        month = may,
       volume = {345},
        pages = {L47-L50},
          doi = {10.48550/arXiv.astro-ph/9904158},
archivePrefix = {arXiv},
       eprint = {astro-ph/9904158},
 primaryClass = {astro-ph},
       adsurl = {https://ui.adsabs.harvard.edu/abs/1999A&A...345L..47M}
}

@article{10.1093/mnras/stt1972,
    author = {Quillen, Alice C. and Minchev, Ivan and Sharma, Sanjib and Qin, Yu-Jing and Di Matteo, Paola},
    title = {A vertical resonance heating model for X- or peanut-shaped galactic bulges},
    journal = {Monthly Notices of the Royal Astronomical Society},
    volume = {437},
    number = {2},
    pages = {1284-1307},
    year = {2013},
    month = {11},
        issn = {0035-8711},
    doi = {10.1093/mnras/stt1972},
    url = {https://doi.org/10.1093/mnras/stt1972},
    eprint = {https://academic.oup.com/mnras/article-pdf/437/2/1284/3845904/stt1972.pdf},
}

@misc{https://doi.org/10.48550/arxiv.2409.03746,
  doi = {10.48550/ARXIV.2409.03746},
  url = {https://arxiv.org/abs/2409.03746},
  author = {Tahmasebzadeh,  Behzad and Dattathri,  Shashank and Valluri,  Monica and Shen,  Juntai and Zhu,  Ling and Wheeler,  Vance and Gerhard,  Ortwin and Kataria,  Sandeep Kumar and Silva,  Leandro Beraldo e and Daniel,  Kathryne J.},
  title = {Orbital Support and Evolution of CX/OX Structures in Boxy/Peanut Bars},
  publisher = {arXiv},
  year = {2024},
  copyright = {arXiv.org perpetual,  non-exclusive license}
}

@article{10.1093/mnras/stad2799,
    author = {Li, Xingchen and Shlosman, Isaac and Heller, Clayton and Pfenniger, Daniel},
    title = {Stellar bars in spinning haloes: delayed buckling and absence of slowdown},
    journal = {Monthly Notices of the Royal Astronomical Society},
    volume = {526},
    number = {2},
    pages = {1972-1986},
    year = {2023},
    month = {12},
    issn = {0035-8711},
    doi = {10.1093/mnras/stad2799},
    url = {https://doi.org/10.1093/mnras/stad2799},
    eprint = {https://academic.oup.com/mnras/article-pdf/526/2/1972/51820827/stad2799.pdf},
}

\section*{Appendix}
\phantomsection
\label{app:appendix}

\begin{figure*}
\centering

\begin{minipage}{0.48\textwidth}
\centering
\includegraphics[width=\linewidth]{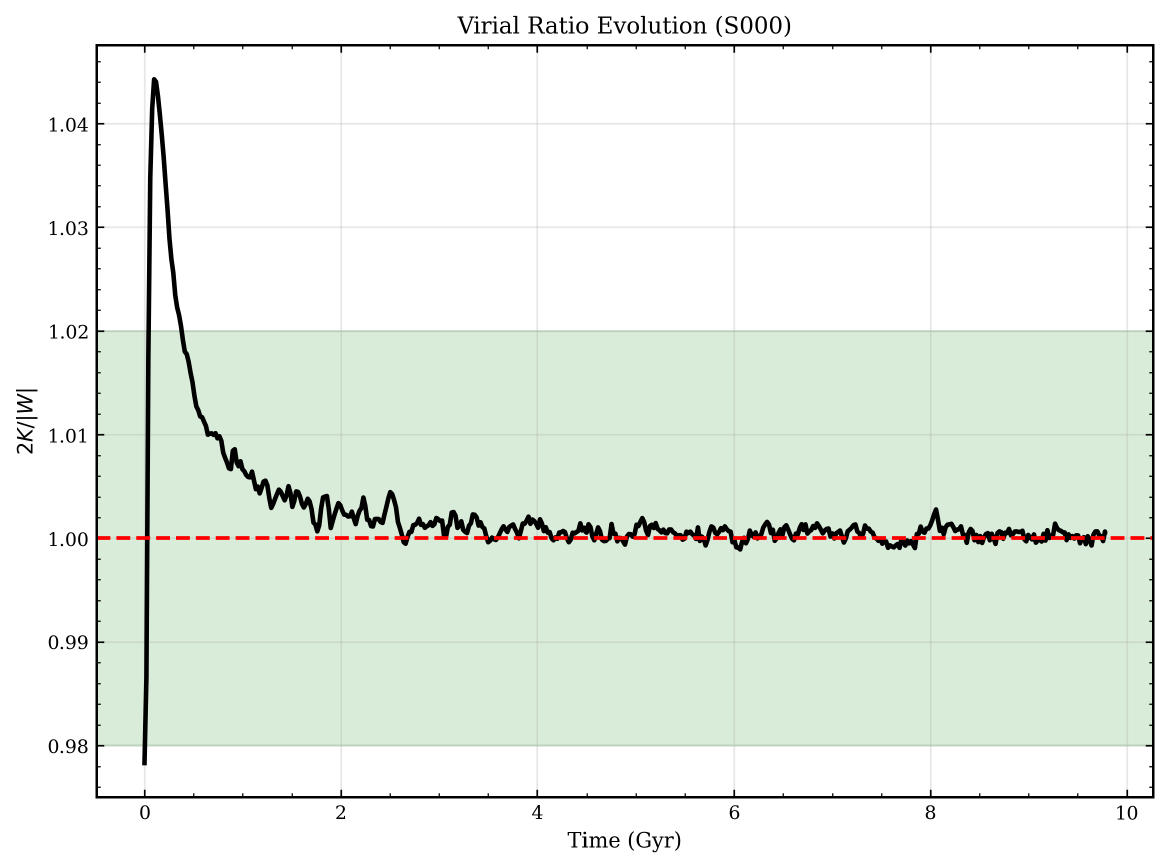}

(a) $S000$
\end{minipage}
\hfill
\begin{minipage}{0.48\textwidth}
\centering
\includegraphics[width=\linewidth]{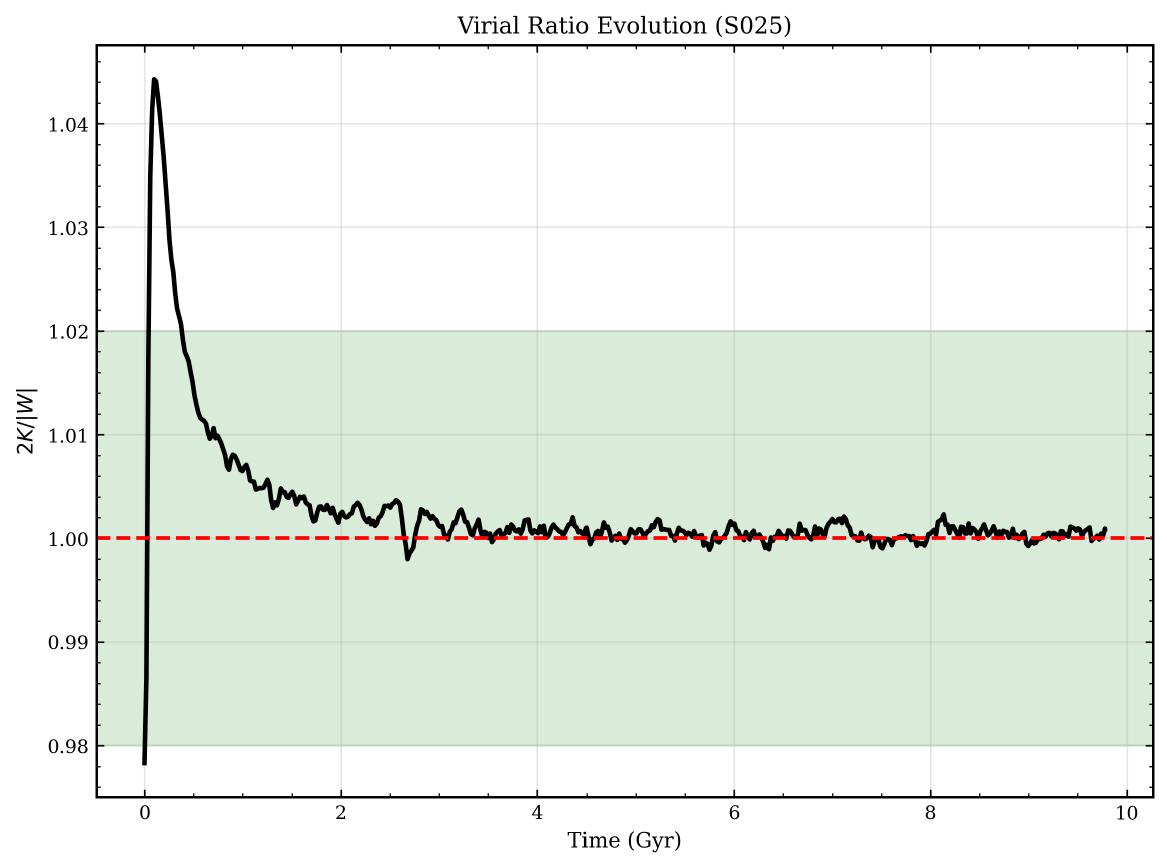}

(b) $S025$
\end{minipage}

\vspace{0.3cm}

\begin{minipage}{0.48\textwidth}
\centering
\includegraphics[width=\linewidth]{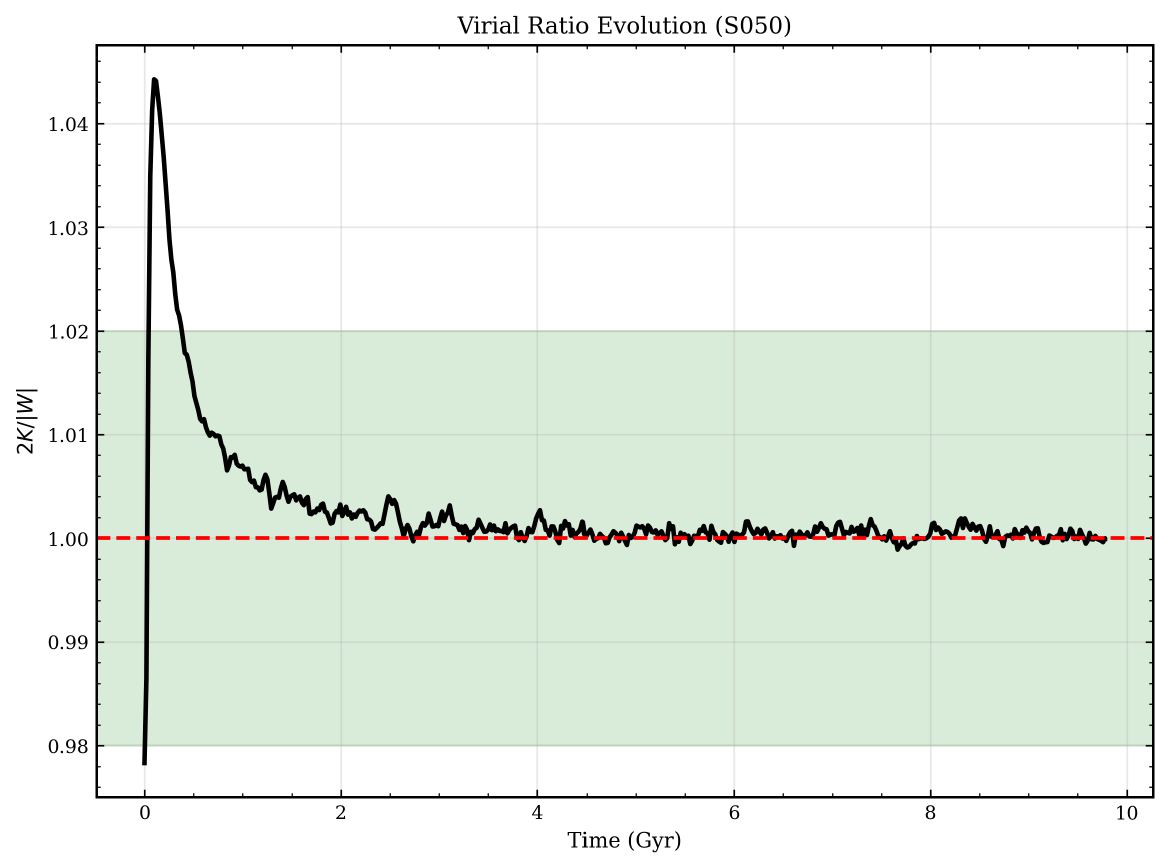}

(c) $S050$
\end{minipage}
\hfill
\begin{minipage}{0.48\textwidth}
\centering
\includegraphics[width=\linewidth]{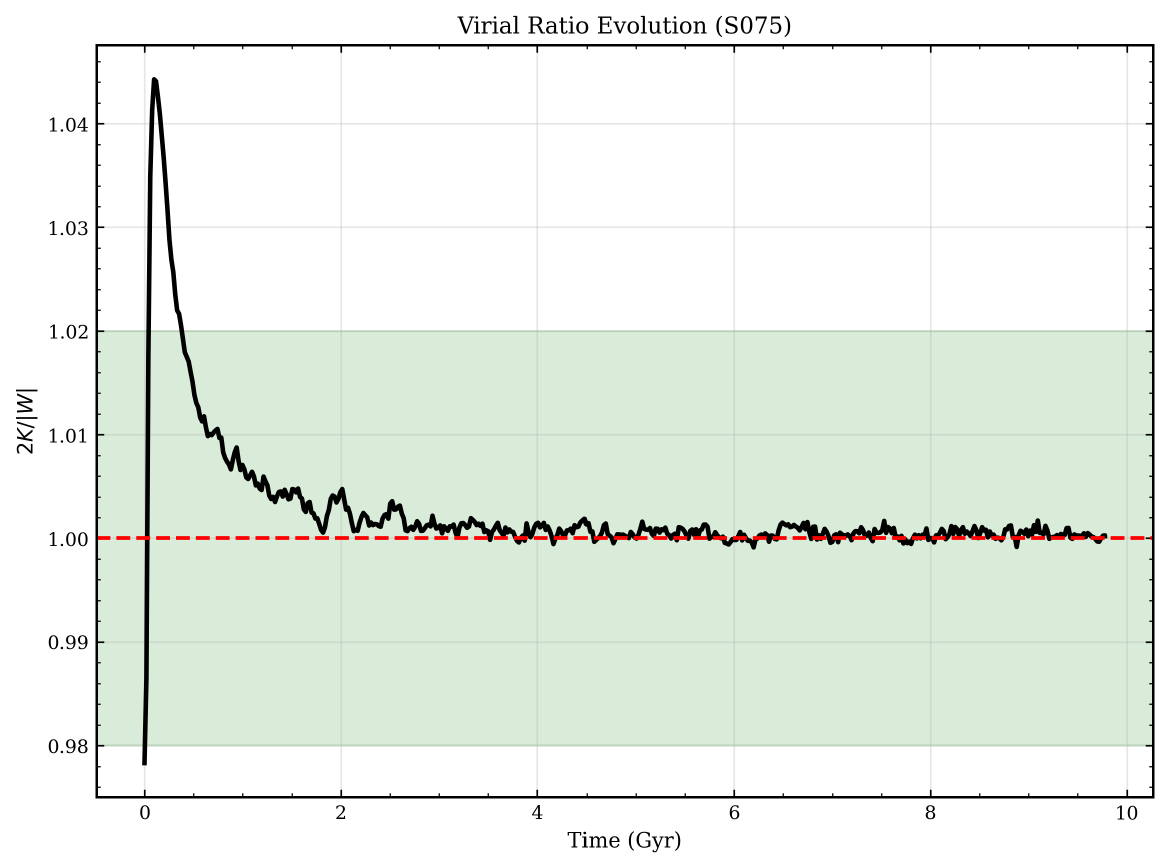}

(d) $S075$
\end{minipage}

\vspace{0.3cm}

\begin{minipage}{0.48\textwidth}
\centering
\includegraphics[width=\linewidth]{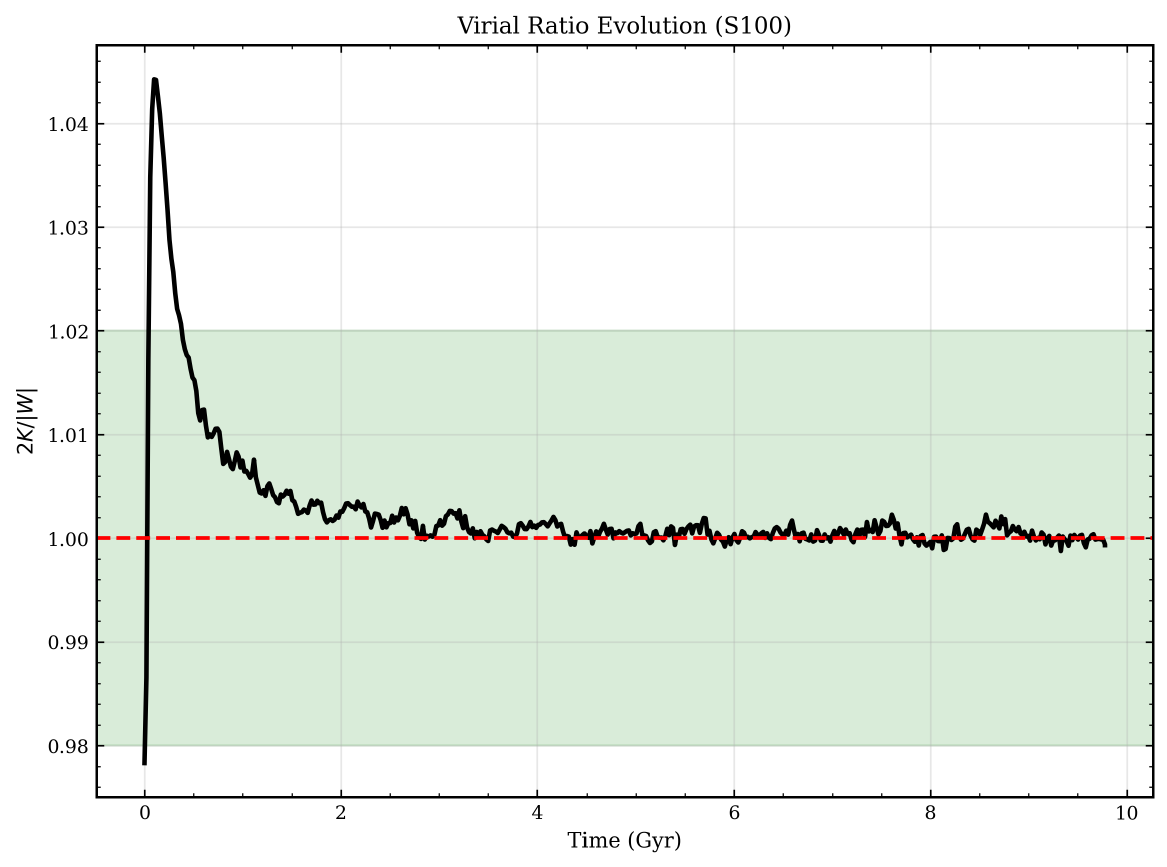}

(e) $S100$
\end{minipage}
\hfill
\begin{minipage}{0.48\textwidth}
\centering
\includegraphics[width=\linewidth]{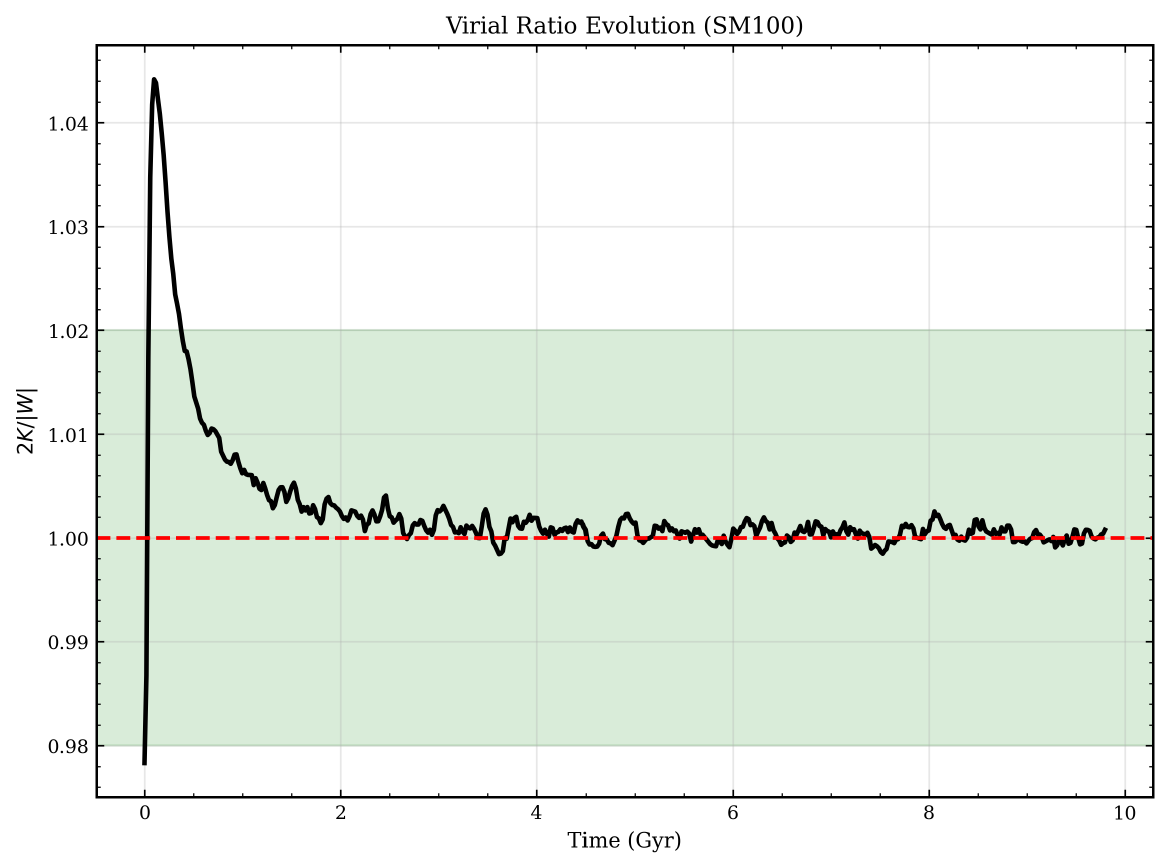}

(f) $SM100$
\end{minipage}

\caption{Evolution of the Virial Ratio throughout the simulation.}
\label{fig:evolution of virial ratio}

\end{figure*}

\begin{figure*}
\centering
\begin{minipage}{0.48\textwidth}
\centering
\includegraphics[width=\linewidth]{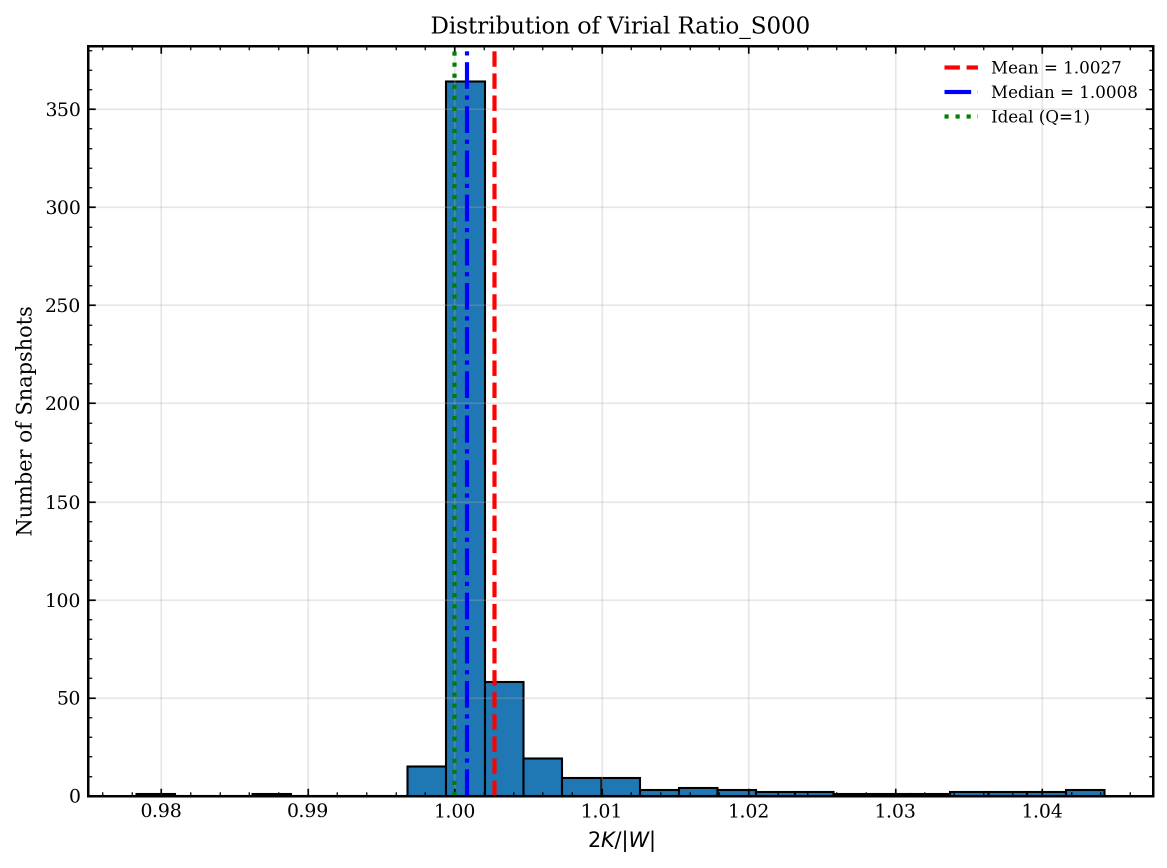}

(a) $S000$
\end{minipage}
\hfill
\begin{minipage}{0.48\textwidth}
\centering
\includegraphics[width=\linewidth]{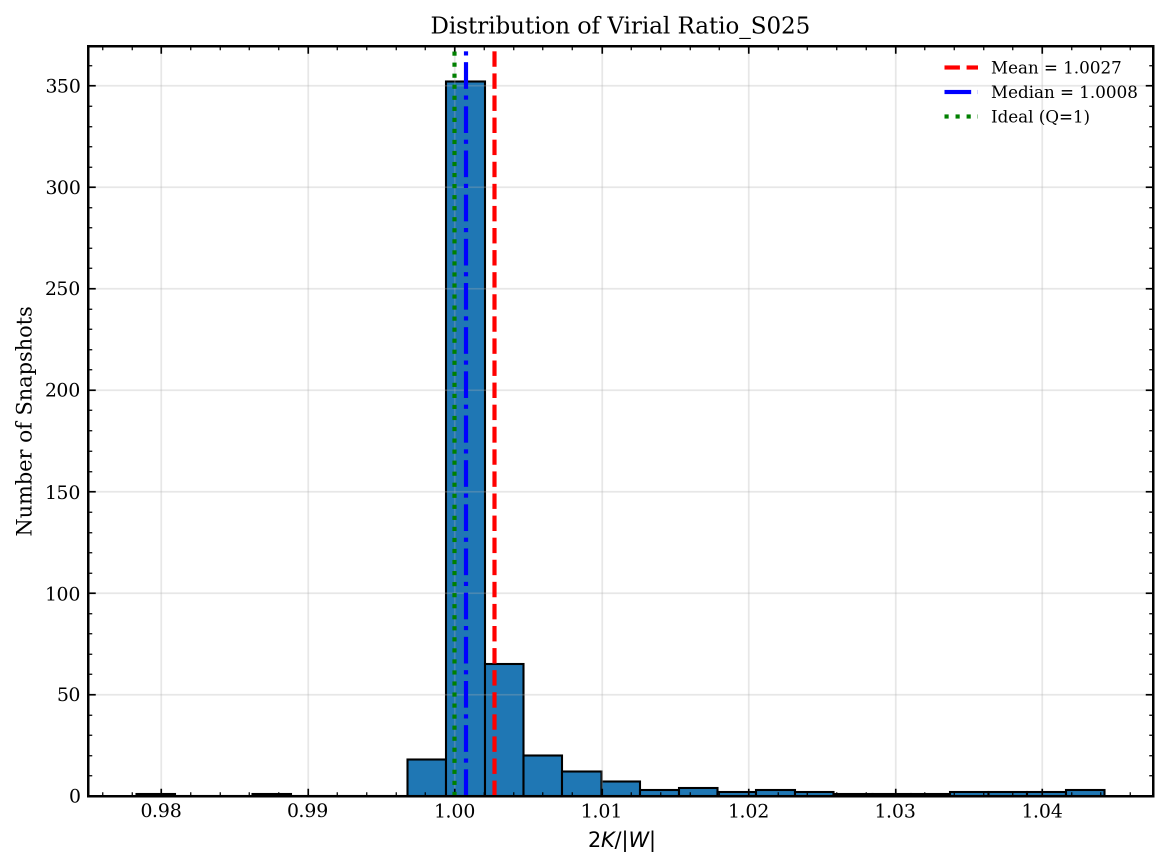}

(b) $S025$
\end{minipage}

\vspace{0.3cm}

\begin{minipage}{0.48\textwidth}
\centering
\includegraphics[width=\linewidth]{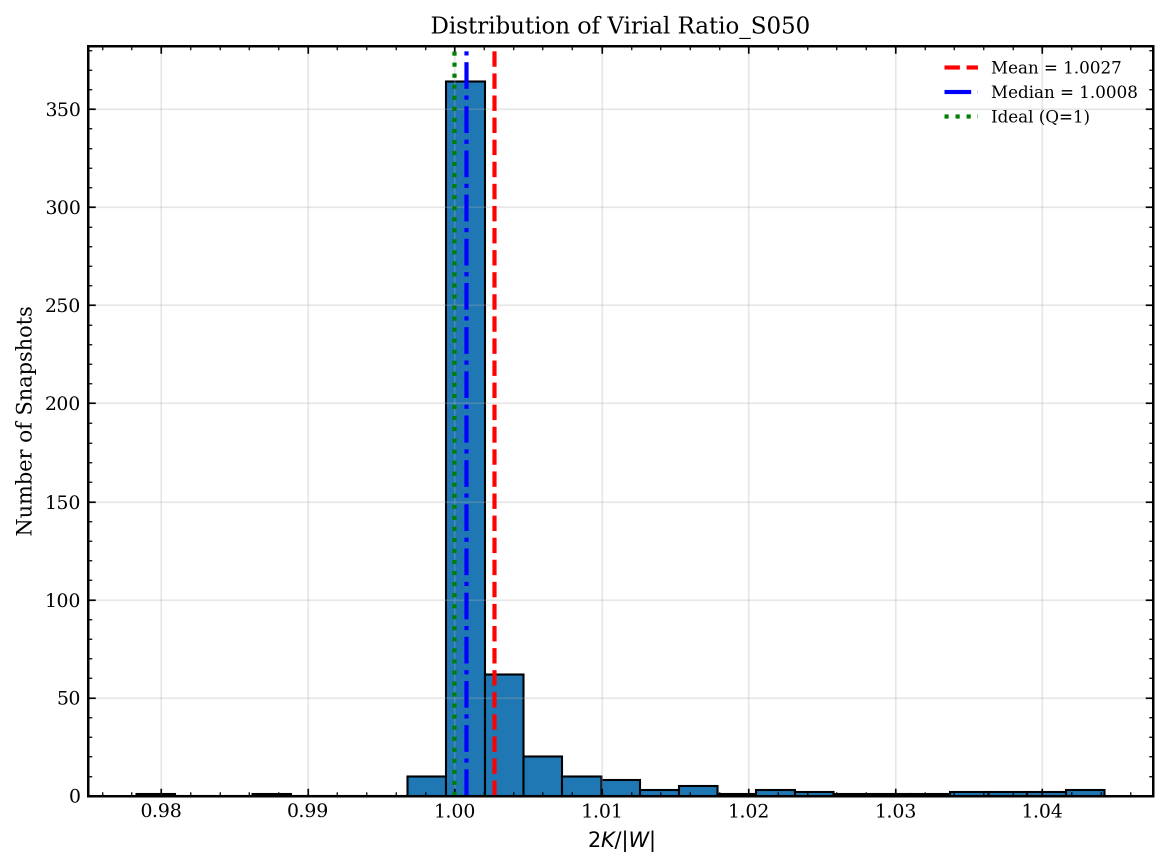}

(c) $S050$
\end{minipage}
\hfill
\begin{minipage}{0.48\textwidth}
\centering
\includegraphics[width=\linewidth]{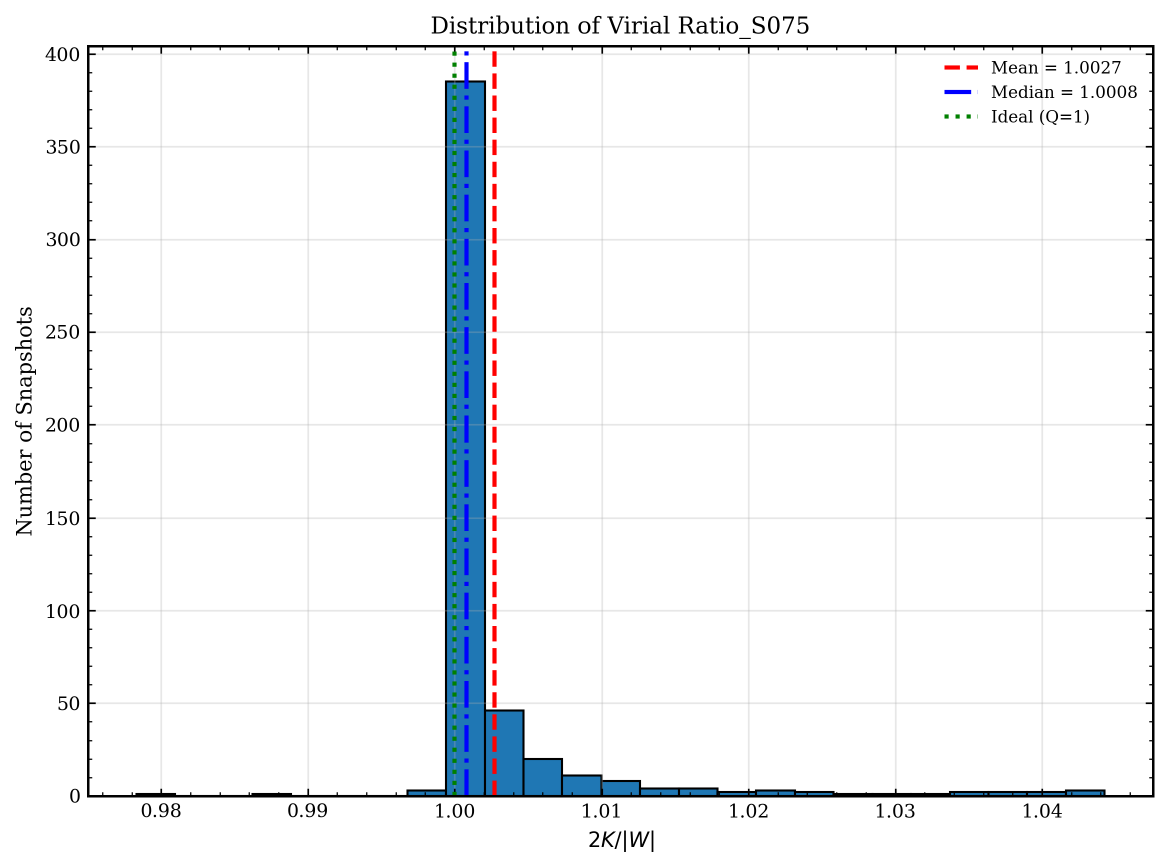}

(d) $S075$
\end{minipage}

\vspace{0.3cm}

\begin{minipage}{0.48\textwidth}
\centering
\includegraphics[width=\linewidth]{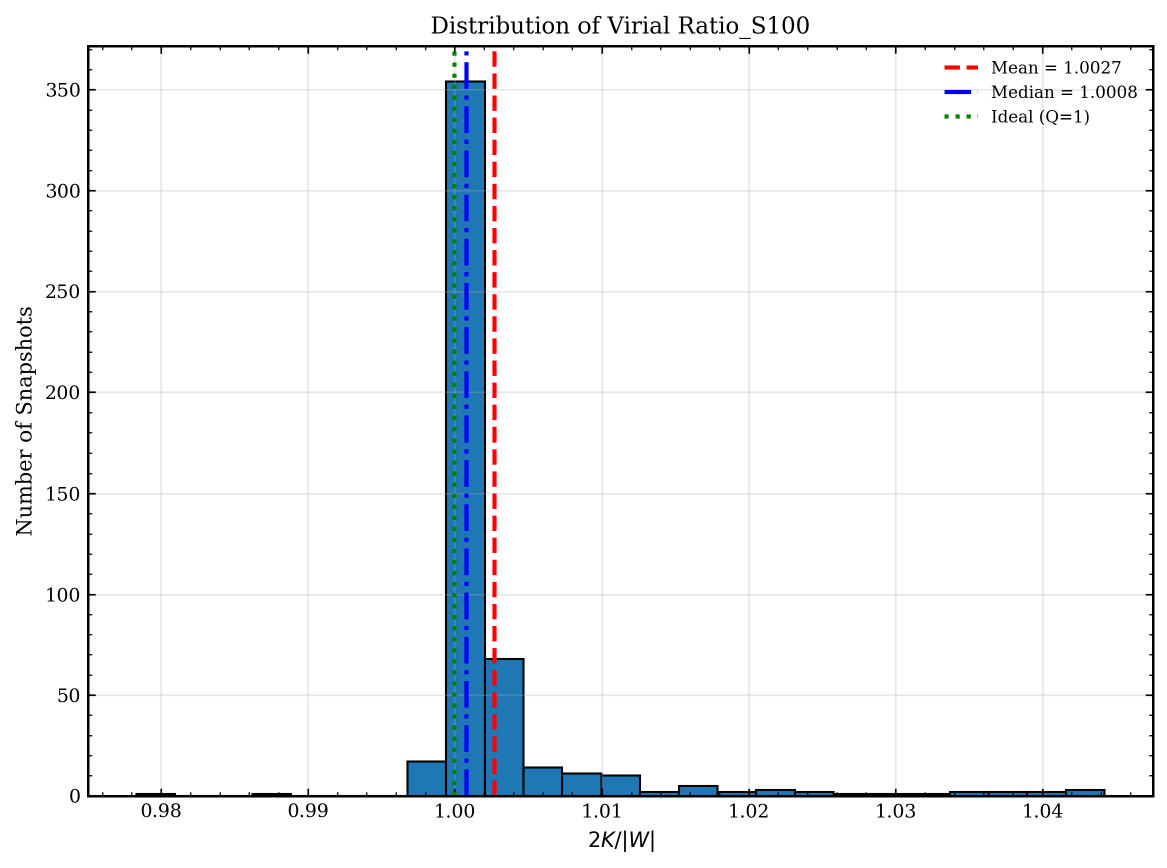}

(e) $S100$
\end{minipage}
\hfill
\begin{minipage}{0.48\textwidth}
\centering
\includegraphics[width=\linewidth]{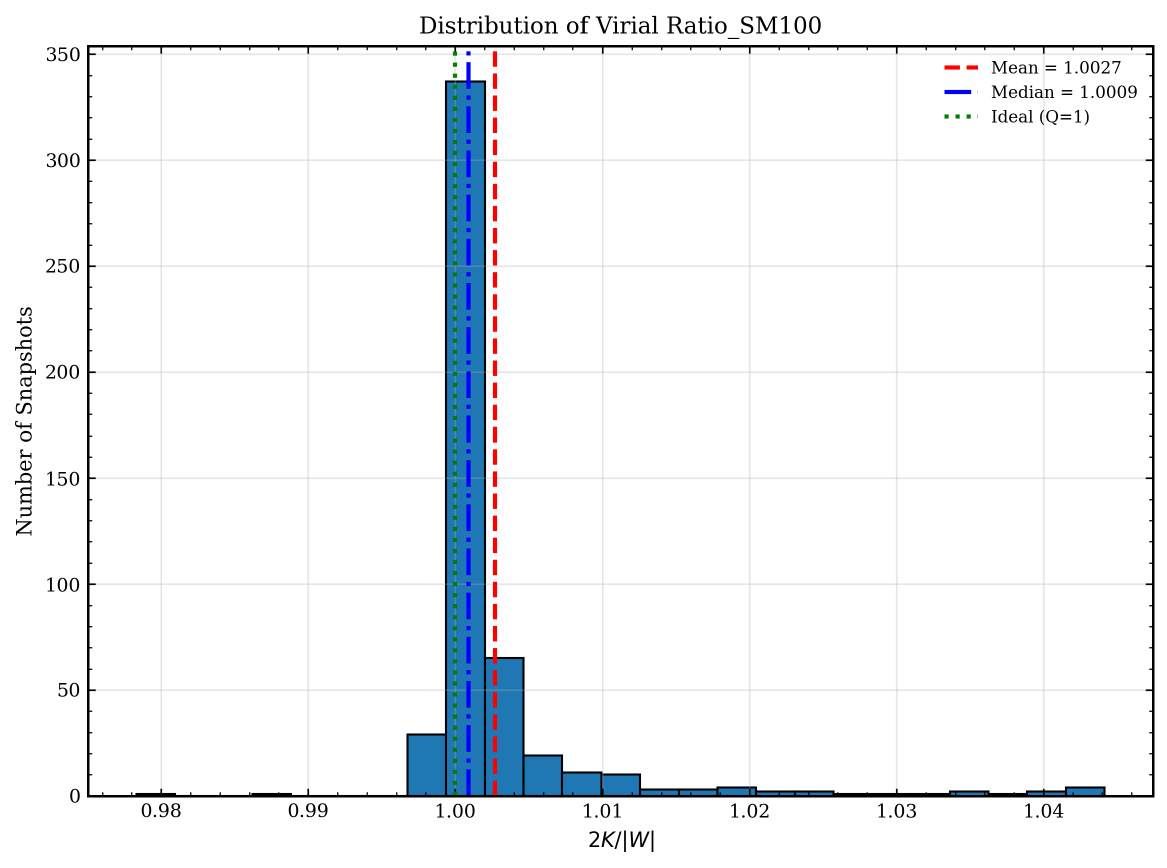}

(f) $SM100$
\end{minipage}
\caption{Distribution of the Virial Ratio throughout the simulation.}
\label{fig:Distribution of the Virial Ratio throughout the simulation.}
\end{figure*}

\begin{figure*}
\centering
\begin{minipage}{0.48\textwidth}
\centering
\includegraphics[width=\linewidth]{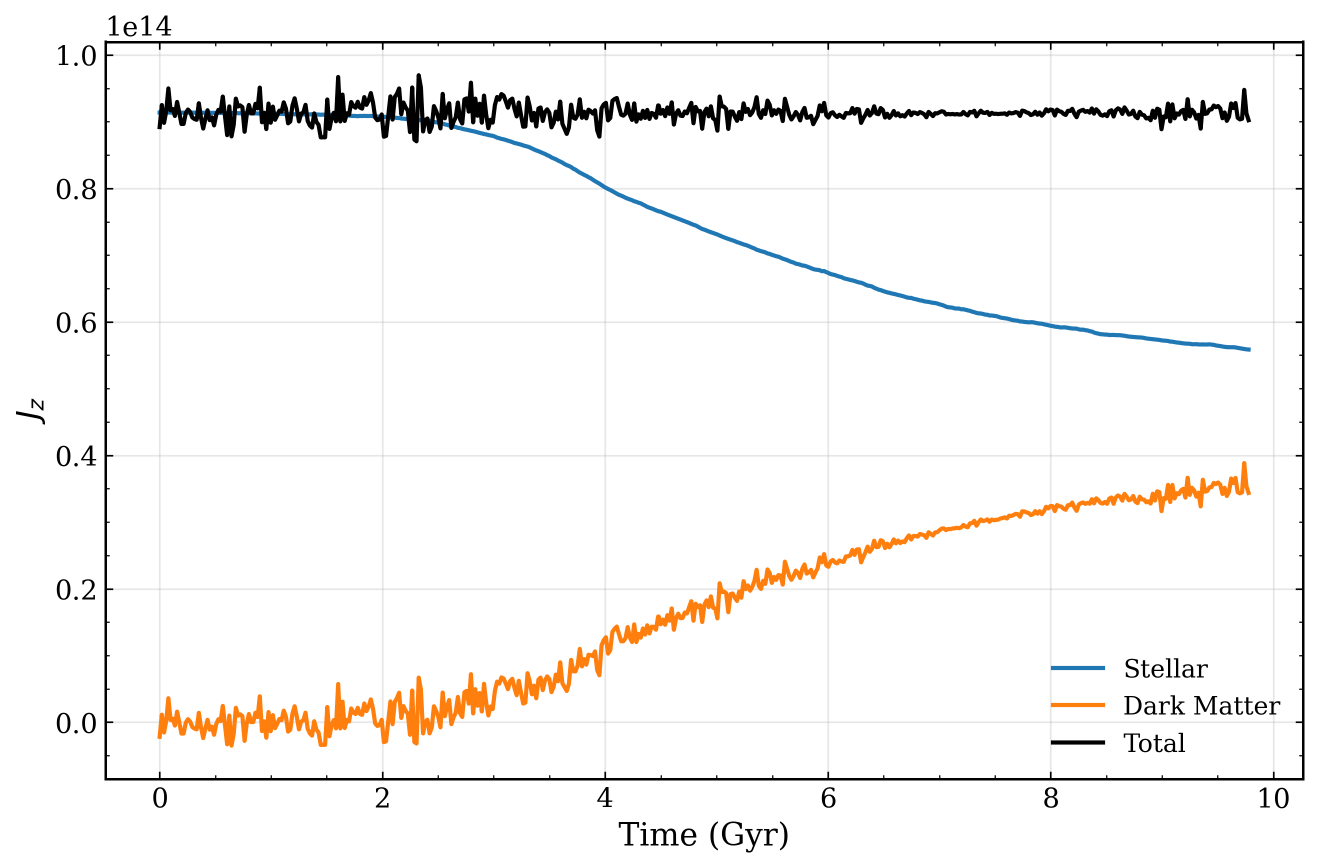}

(a) $S000$
\end{minipage}
\hfill
\begin{minipage}{0.48\textwidth}
\centering
\includegraphics[width=\linewidth]{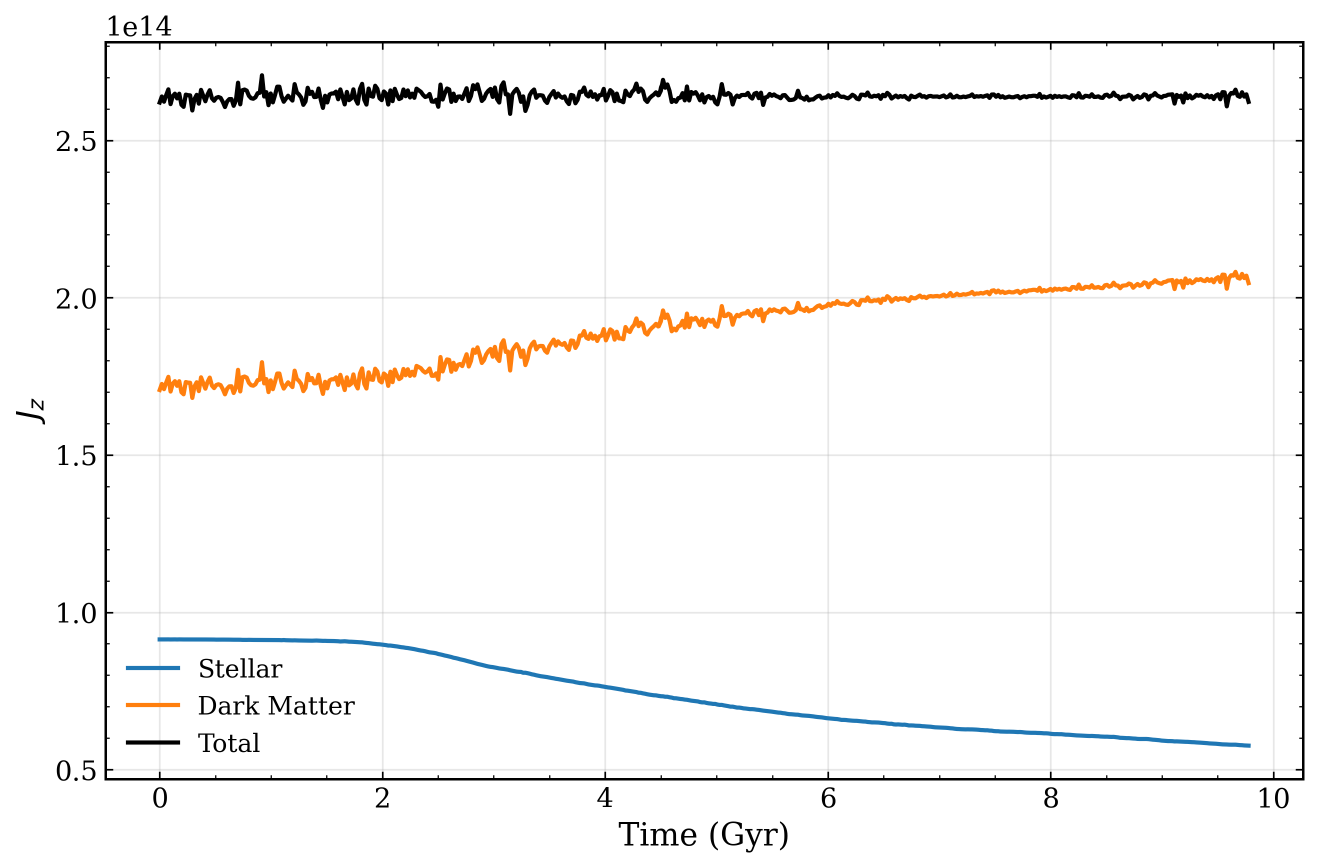}

(b) $S025$
\end{minipage}

\vspace{0.3cm}

\begin{minipage}{0.48\textwidth}
\centering
\includegraphics[width=\linewidth]{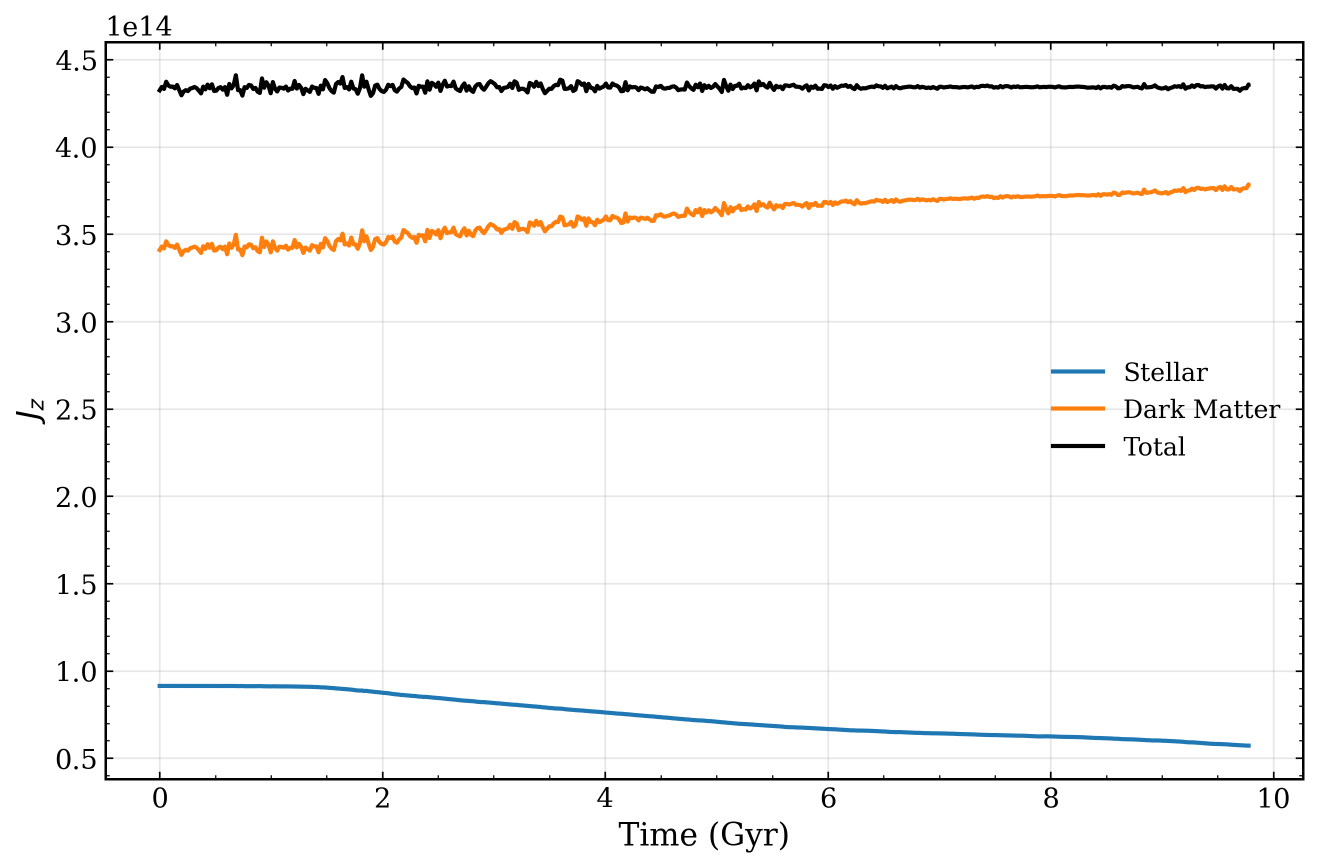}

(c) $S050$
\end{minipage}
\hfill
\begin{minipage}{0.48\textwidth}
\centering
\includegraphics[width=\linewidth]{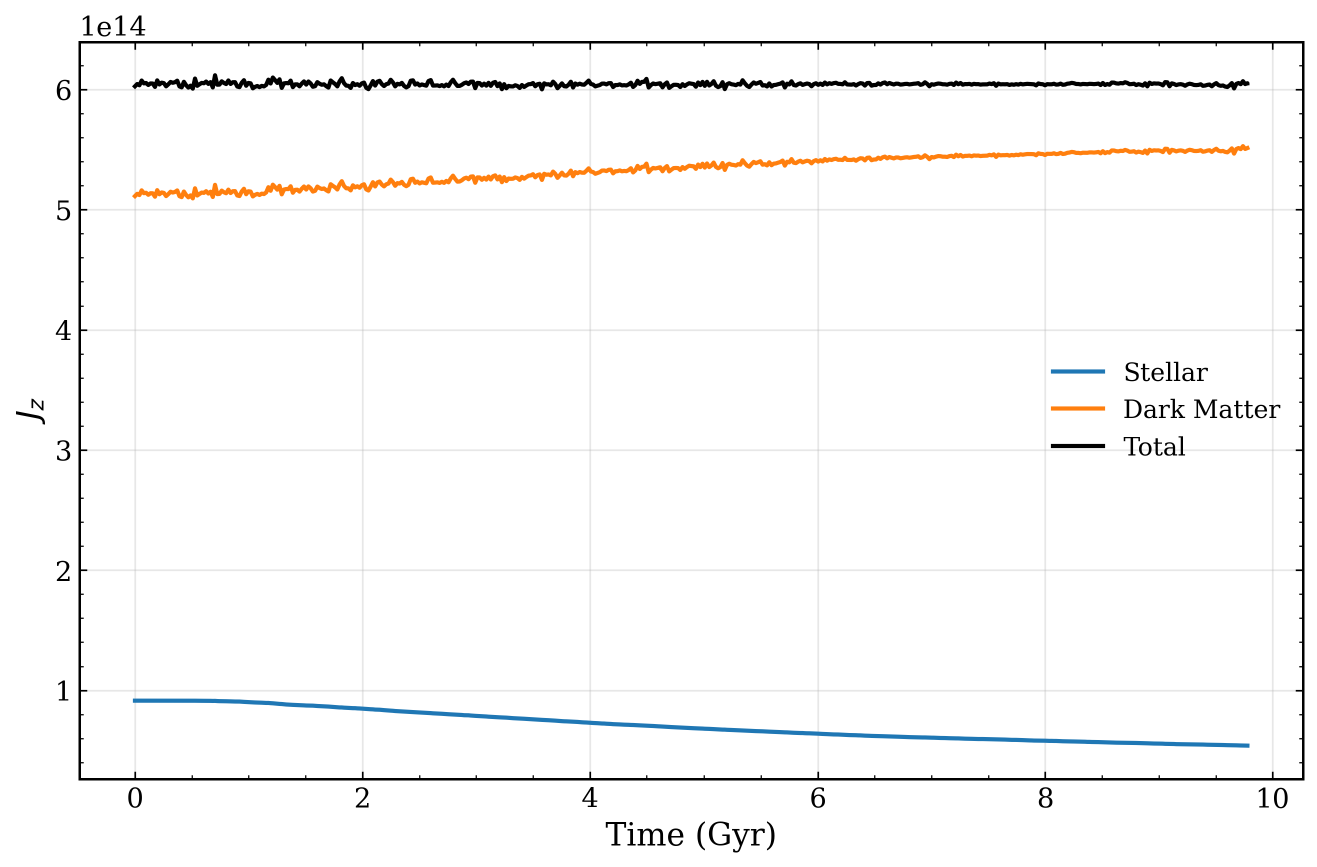}

(d) $S075$
\end{minipage}

\vspace{0.3cm}

\begin{minipage}{0.48\textwidth}
\centering
\includegraphics[width=\linewidth]{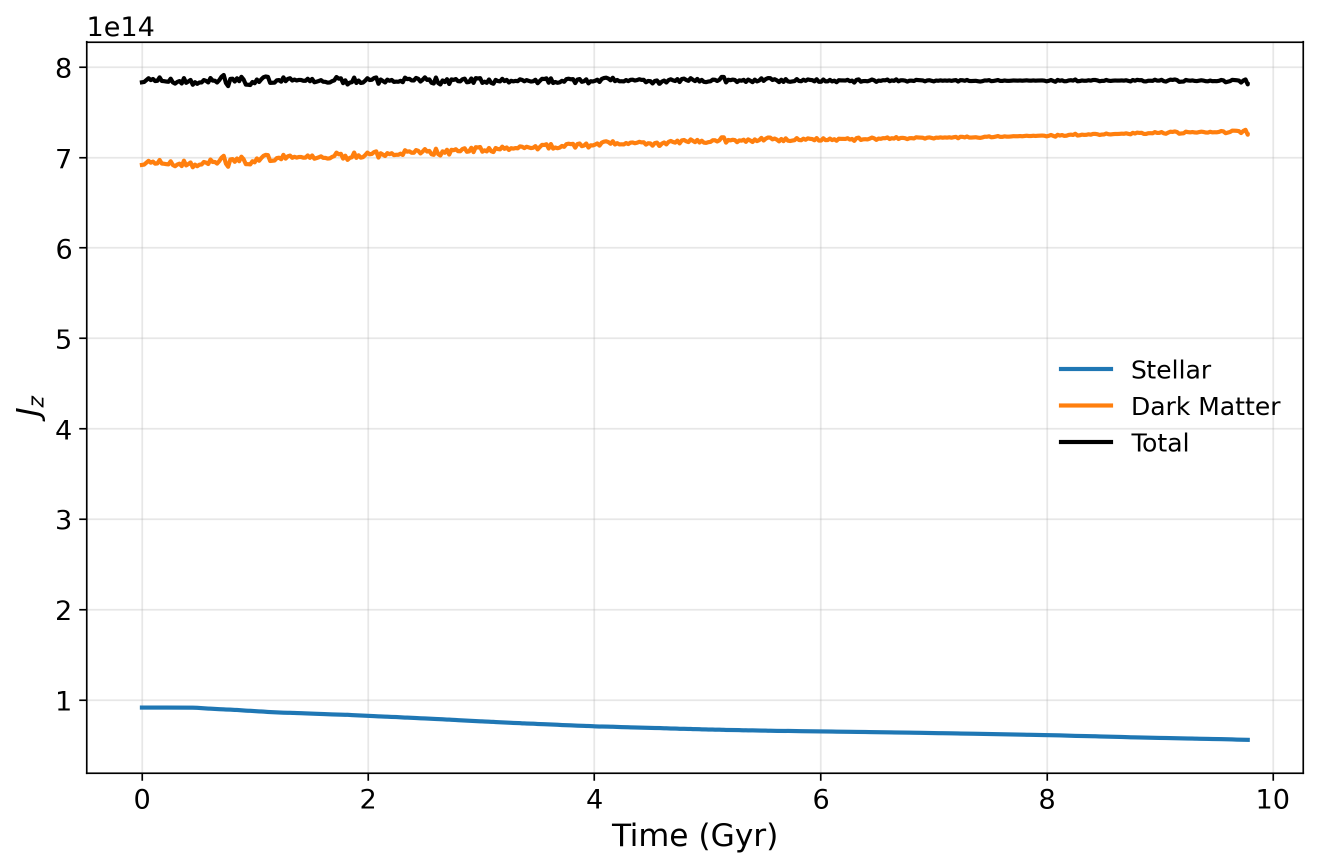}

(e) $S100$
\end{minipage}
\hfill
\begin{minipage}{0.48\textwidth}
\centering
\includegraphics[width=\linewidth]{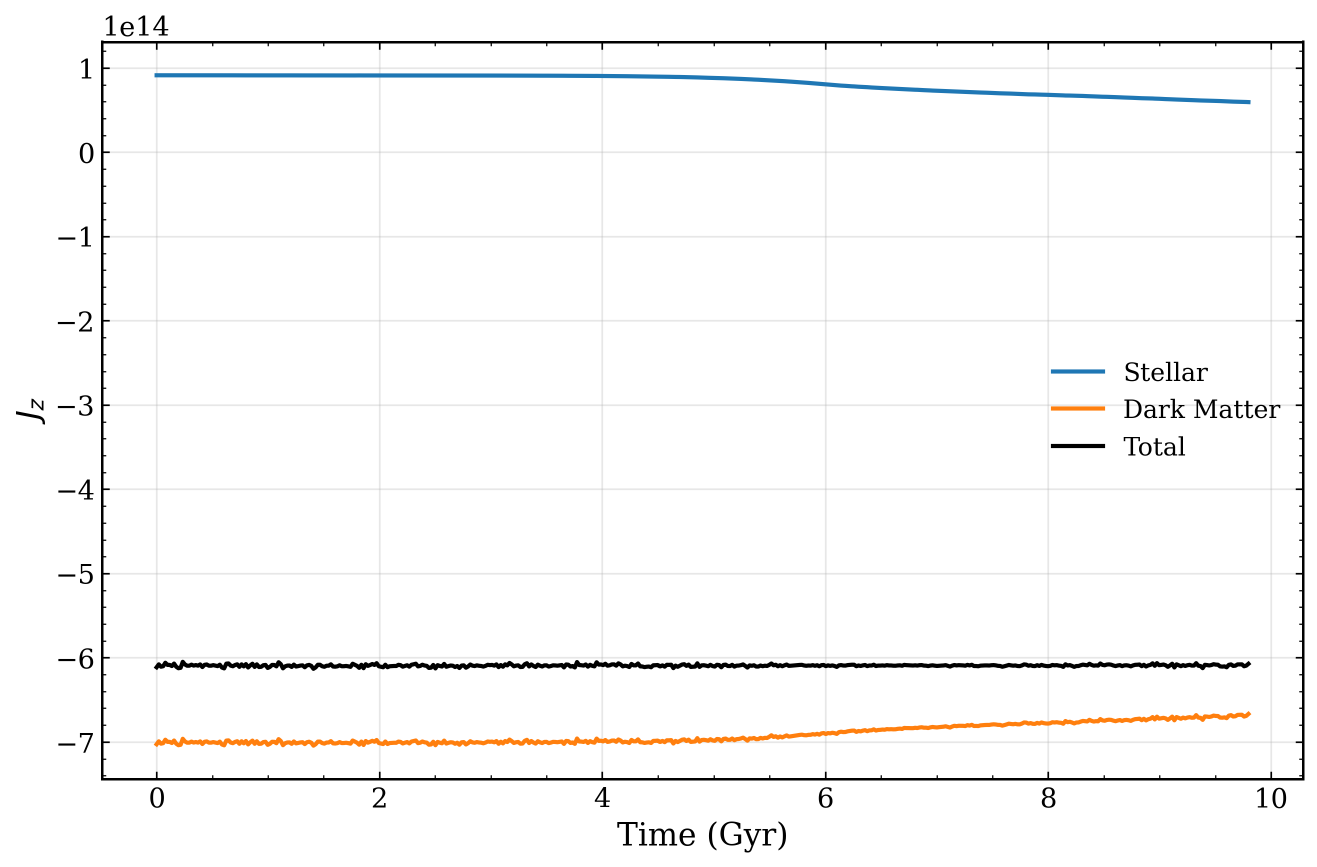}

(f) $SM100$
\end{minipage}
\caption{Total Angular Momentum Conservation of different spin halo models.}
\label{fig:Total Angular Momentum Conservation of different spin halo models.}
\end{figure*}

\indent In this \hyperref[app:appendix]{Appendix}, we show the numerical stability of the simulations. The virial ratio, $Q = 2K/|W|$, throughout the simulation remains close to unity as shown in Figure \ref{fig:evolution of virial ratio}, indicating that the galaxy remains in dynamical equilibrium. The virial ratio analysis indicates that there is no fast-growing numerical instability developing in the core. \\
\indent Figure \ref{fig:evolution of virial ratio} demonstrates that the system exhibits a brief transient during the first $\sim 0.5$--$1\,\mathrm{Gyr}$, where $Q$ reaches a maximum value of approximately $1.044$ before rapidly relaxing toward unity. Subsequently, the virial ratio remains extremely close to the equilibrium value, fluctuating around $Q \simeq 1$ for the remainder of the $10\,\mathrm{Gyr}$ evolution.\\
\indent Figure \ref{fig:Distribution of the Virial Ratio throughout the simulation.} represents that the overwhelming majority of snapshots are concentrated around $Q = 1$, with a mean value of $1.0027$ and a median value of $1.0008$, confirming that the galaxy remains very close to virial equilibrium throughout its evolution.\\
\indent As shown in Figure \ref{fig:Total Angular Momentum Conservation of different spin halo models.}, the total angular momentum, $J_z$, remains remarkably well conserved throughout the simulation for all the models. The total $J_z$ is nearly constant over the entire $10\,\mathrm{Gyr}$ evolution, exhibiting only negligible numerical fluctuations. This demonstrates that the simulation conserves angular momentum to a high degree of accuracy.\\
\indent The diagnostics indicate that the initial variations in the component of angular momenta are not indicative of a fast-growing core instability. Instead, they arise from the brief initial relaxation of the finite-$N$ realisation generated by \textsc{galic}, as the discrete $N$-body system adjusts to the self-consistent gravitational potential evolved with \textsc{gadget}. This transient reflects the finite sampling of the equilibrium distribution function rather than the development of any physical or numerical instability. During this short settling phase, angular momentum is redistributed among the stellar bar, the dark matter bar, and the halo, while the total angular momentum is conserved to within numerical precision. Concurrently, the system rapidly relaxes toward virial equilibrium. After this initial transient ($\lesssim 1\,\mathrm{Gyr}$), both the virial ratio and the total angular momentum remain stable throughout the remainder of the simulation, demonstrating that no fast-growing numerical instability develops in the core.

\bsp
\label{lastpage}
\end{document}